\documentclass[12pt,preprint]{aastex}
\usepackage{psfig}
\def\HI {H\kern0.1em{\sc i}} 
\def\radm {rad m$^{-2}$} 

\def\dg{$^{\circ}$}
\def\l2{$\lambda^2$}

\begin{document}
\title{~~\\ ~~\\ A View through Faraday's Fog 2:\\
Parsec Scale Rotation Measures in 40 AGN}
\shorttitle{Faraday's Fog 2}
\shortauthors{Zavala \& Taylor}
\author{R. T. Zavala\altaffilmark{1,2,3} \& G. B. Taylor\altaffilmark{1}}
\email{bzavala@nofs.navy.mil, gtaylor@nrao.edu}
\altaffiltext{1}{National Radio Astronomy Observatory, P.O. Box 0, 
Socorro, NM 87801}
\altaffiltext{2}{Department of Astronomy, New Mexico State University, 
MSC 4500 P.O. Box 30001, Las Cruces, NM 88003-8001}
\altaffiltext{2}{Present address: U. S. Naval Observatory Flagstaff Station, 
P.O. Box 1149 Flagstaff, AZ 86002-1149}
\slugcomment{Accepted to the Astrophysical Journal}

\begin{abstract}
Results from a survey of the parsec scale Faraday rotation 
measure properties for 40 quasars, radio galaxies and BL Lac
objects are presented. Core rotation measures for quasars vary 
from approximately 500 to several thousand \radm\,. Quasar jets have 
rotation measures which are typically 500 \radm\ or less. The cores and 
jets of the BL Lac objects have rotation measures similar to 
those found in quasar jets. The jets of radio galaxies exhibit a range
of rotation measures from a few hundred \radm\ to almost 10,000 \radm\ 
for the jet of M87. Radio galaxy cores are generally depolarized, and 
only one of four radio galaxies (3C\,120) has a detectable 
rotation measure in the core. Several potential identities for the 
foreground Faraday screen are considered and we believe the most
promising candidate for all the AGN types considered is a screen in 
close proximity to the jet. This constrains the path length to 
approximately 10 parsecs, and magnetic field strengths of approximately 1
$\mu$Gauss can account for the observed rotation measures. 
For 27 out of 34 quasars and BL Lacs their optically thick cores have good 
agreement to a \l2\ law. This requires the different $\tau = 1$ surfaces to 
have the same intrinsic polarization angle independent of frequency and 
distance from the black hole.
\end{abstract}


\section{Introduction}

The first rotation measure (RM) towards an extragalactic radio source 
was published by \citet{cp62}. They discussed the potential for such 
measurements as a probe of the galactic Faraday screen. \citet{war} 
analyzed radio polarization monitoring observations of compact extragalactic 
sources for signs of Faraday rotation. Wardle suggested that the 
combined Faraday rotation from our galaxy and the host object were 
relatively small. Observations on arcsecond scales of 555 steep spectrum 
sources by \citet{sim}, and of flat spectrum sources by \citet{rj} and 
\citet{rusk} all confirmed this result. An expectation was established using 
these observations that the unresolved parsec scale cores of AGN 
would similarly show negligible Faraday rotation. It was not until 
simultaneous, multi-frequency polarimetry became available with the 
Very Long Baseline Array (VLBA\footnote{The National Radio Astronomy 
Observatory is operated by Associated Universities, Inc., under 
cooperative agreement with the National Science Foundation.}) that 
these expectations were proven false. Extreme parsec scale rest frame 
rotation measures were first reported for OQ 172 \citep[40000 \radm\ ]{udom} 
and 3C\,138 \citep[5300 \radm\ ]{138}. Such RMs show that interpretations 
of polarization observations on parsec scales in AGN requires simultaneous 
determination of the rotation measure. Without knowledge of the rotation 
measure in a source the orientations of polarization vectors in the 
relativistic jets of AGN is uncertain. For example, a 
rotation measure of 250 \radm\ will change the intrinsic polarization 
angle of a source by 25\dg\ at 8 GHz. The existence of RMs in quasar 
cores of 1000 \radm\ or more \citep{tay98,tay00} and time variability 
of RMs in quasar cores \citep{zt01} shows how essential knowledge of the 
rotation measure is for the correct interpretation of the observed 
polarization.

Michael Faraday first observed what we now refer to as Faraday rotation
when he passed polarized light through glass in the presence of a 
magnetic field \citep{farad}. He correctly surmised that this 
observation hinted at the connection between electric and magnetic 
fields and light. Light subject to Faraday rotation will have its 
intrinsic polarization angle $\chi_0$ rotated to an observed angle  
$\chi$ by 

\begin{equation}
\chi = \chi_{0} + \rm{RM}{\lambda}^{2}
\end{equation}

\noindent where $\lambda$ is the observed wavelength.  The 
linear relationship to \l2 is the characteristic signature 
of Faraday rotation. The slope of the line is known as the      
Rotation Measure (RM) and depends linearly on the electron 
density $\rm{n_{e}}$, the net line of sight magnetic field 
$\rm{B_{\parallel}}$, and path length $\rm{dl}$ through the plasma. 
Using units of cm$^{-3}$, mG and parsecs the rotation measure 
is given by:

\begin{equation}
\rm{RM = 812\int{n_{e}B_{\parallel}\,dl}\quad rad\ m^{-2}}.
\end{equation}

\noindent A suitably designed experiment can resolve the n$\pi$ 
ambiguity inherent in polarization vector orientations. By 
obtaining observations with sufficient long and short spacings 
in \l2 this ambiguity can be resolved, and the correct RM determined. 

This paper completes the presentation of a rotation measure survey of 40 AGN 
suggested in \citet{tay00}. The first half of the observations appeared in 
\citet{zt03}. We present our observations and data reduction procedures in \S2. 
Results for individual sources are shown in \S3. In \S4 we consider the 
rotation measure properties of the sample as a whole, including a few additional 
sources in the literature. Conclusions appear in \S5.  

\section{Observations and Data Reduction}

The observations, performed on 2001 June 20 (2001.47), were carried out at 
seven widely separated frequencies between 8.1 and 15.2 GHz using the 10 
element VLBA. This 24 hour observation targeted the sources listed in 
Table~\ref{t1}. Due to an electrical short in the elevation system the 
Ft. Davis antenna was lost for the first 7.5 hours of the 24 hour run. Prior to 
self-calibration all processing was performed in the Astronomical Image 
Processing System \citep[AIPS;][]{van96}. AIPS procedures 
described in \citet{ugm} were employed and are indicated by eight letter 
capitalized words (e.g. VLBACPOL). Data collected at elevations less 
than 10\dg\ were flagged. Amplitude calibration was performed with the task 
APCAL. An opacity correction was employed at all frequencies as several 
antennas (Pie Town, Ft. Davis, Kitt Peak, and Hancock) reported rain 
during the observation. Plots of T$_{sys}$ versus airmass also indicated 
a variable opacity at North Liberty. The procedure VLBAPANG corrected 
the observations for varying parallactic angles of the alt-azimuth 
mounted VLBA antennas. VLBAMPCL was used on two minutes of data 
from 3C\,279 to remove errors due to clock and correlator model 
inaccuracies. A global fringe fit was run on all the data to remove 
the remaining delay and rate errors with the procedure VLBAFRNG. VLBAFRNG 
uses the AIPS task FRING, an implementation of the Schwab$-$Cotton 
algorithm \citep{schwab}. The delay offset between the right and left 
circularly polarized data  was removed using the procedure 
VLBACPOL \citep{cotton}. A bandpass correction table was made with 
BPASS using 1741$-$038 as a bandpass calibrator. The data were 
then averaged in frequency across the individual intermediate 
frequencies (IFs). 

Self-calibration was done using DIFMAP \citep{shep,spt} and AIPS in 
combination. Considerable radio frequency interference was present 
at 12 GHz on almost all baselines for 3C\,279 and 3C\,446, and was 
edited out. This resulted in the loss of 49\% of the visibilities 
for 3C\,279 and 23\% for 3C\,446 at 12 GHz. As the gain curves of 
the antennas used in the amplitude calibration are poorly known at 
12 GHz we compared the VLBA flux at 12 GHz with that from the VLA 
Polarization Monitoring web page\footnote{http://www.aoc.nrao.
edu/$\sim$smyers/calibration/} \citep{tmy00} and data from the 
University of Michigan Radio Astronomy Observatory (UMRAO; H. 
Aller, 2003, private communication). This comparison suggested a 
reduction of approximately 10\% in the gain solution was required 
at 12 GHz, and this was applied with the task SNCOR. Table~\ref{t1} 
lists the number of scans on each source, as well as the RMS and peak 
flux in the 15 GHz I map that this calibration produced. Scans were 
three and one half minutes long. If all ten antennas are present in 
7 scans the expected thermal noise at 15 GHz is approximately 
0.5 mJy beam$^{-1}$. 

Polarization leakage of the antennas (D-terms) were 
determined using the AIPS task LPCAL \citep{lpcal}. We chose 0552+398
as the D-term calibrator as it had a wide parallactic angle coverage, 
and a simple and nearly unresolved polarization structure. 
Plots of the real versus imaginary crosshand polarization data indicated 
that a satisfactory D-term solution was obtained. This was also 
verified in plots of the real and imaginary crosshand data 
versus (u,v) parallactic angle. After applying the D-term solution 
no variation was seen as a function of (u,v) parallactic angle.

\begin{figure}
\plotone{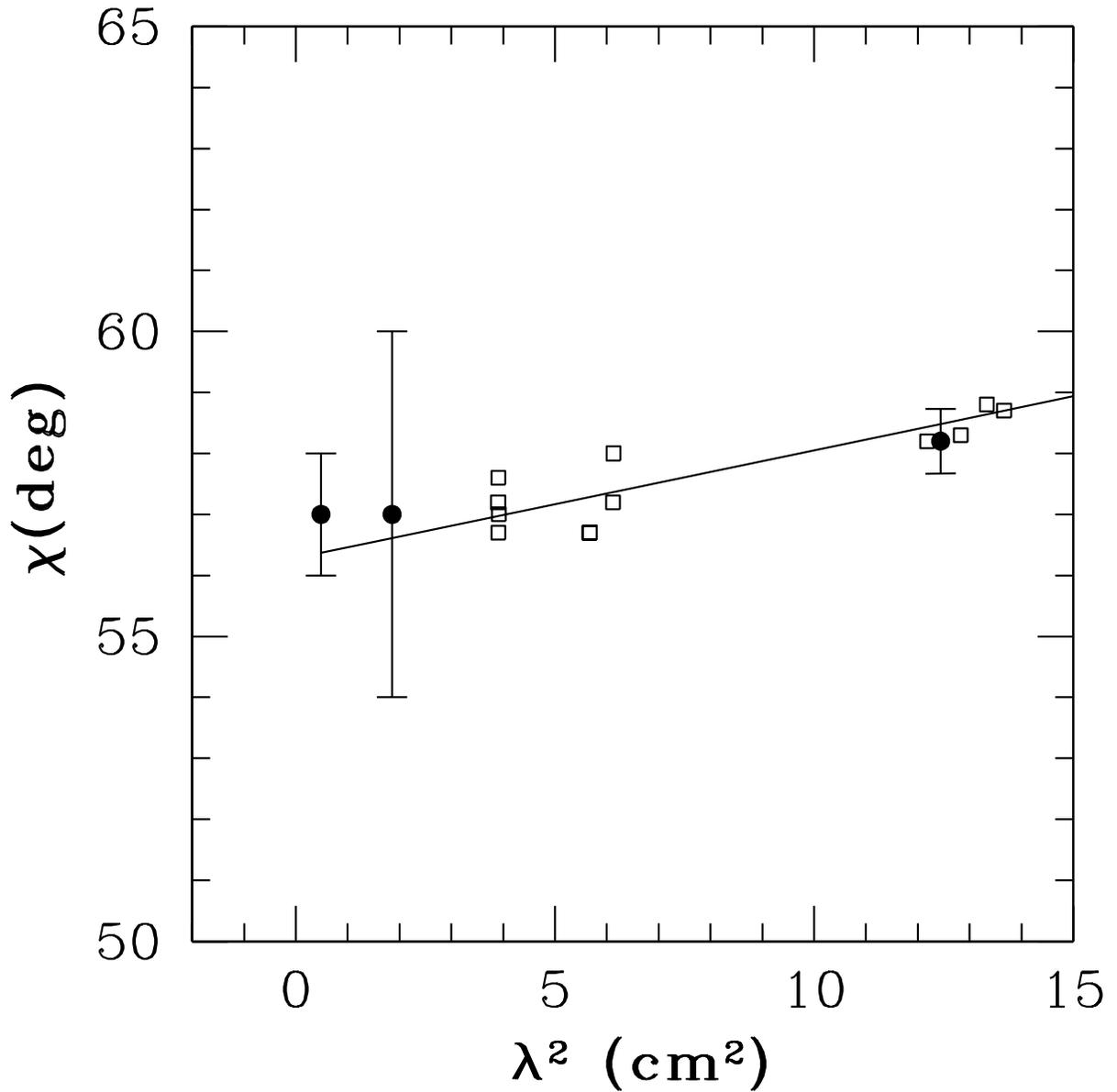}
\caption{EVPA calibration versus \l2 for 3C\,279. Filled circles 
are VLA polarization monitoring data, and open boxes are the VLBA 
EVPA's after the turns derived from Table~\ref{t3} were applied. The 
solid line represents a least-squares fit for a Faraday rotation 
\l2 law to the VLA data including the 5 GHz position angle 
(not shown). The fit represents an integrated RM of 31 $\pm$ 10 \radm.}
\label{evpa}
\end{figure}

Absolute electric vector position angle (EVPA)
calibration was determined by using the EVPA of 3C\,279 listed in the 
VLA Polarization Monitoring Program. We used the integrated Q and 
U fluxes from the VLBA data to derive a position angle, which we compared
to that listed on the polarization monitoring web page. This calibration 
scheme rests on the assumption that most of the polarized flux observed by 
the VLA is seen with the VLBA. To verify this the polarized flux 
observed by UMRAO, the VLA and the VLBA are listed in Table~\ref{t2} 
for 8 and 15 GHz, with their respective observation dates. The good 
agreement between these values for telescopes with very different 
resolutions makes us confident in our absolute EVPA calibration.   
The position angles for 3C\,279 were observed with the VLA in B array on 
2001 Jun 24. These position angles were in good agreement with nearly 
contemporaneous observations from the UMRAO data for 3C\,279. The EVPA 
calibration at 8 GHz was directly obtained from the polarization 
calibration website. Polarization monitoring observations at 8 and 22 
GHz were interpolated to produce position angles at 12 and 15 GHz, 
assuming the EVPA's obeyed a \l2\ Faraday rotation law. 
Fig.~\ref{evpa} shows the final calibrated VLBA EVPA's with the VLA 
EVPA's from the polarization monitoring webpage. 

The uncertainty in the EVPA calibration using the RM fit in Fig.~\ref{evpa} 
is approximately $\pm$1 degree. To this uncertainty we add in quadrature
the uncertainty derived from the individual Stokes Q and U maps. There
will be some additional uncertainty from the lack of simultaneous VLA
polarization observations which is difficult to quantify. Data
obtained with UMRAO and the VLA to establish the EVPA calibration were
taken within 1-5 days of the VLBA observation. If the lack of
simultaneous observations by the VLA and/or UMRAO were significant we
would expect all the fits to a \l2\ law to require a systematic
increase in their error budget. Although some sources do not show good
agreement to a \l2\ law many do, and thus we conclude that the
errors have been properly accounted for.

To perform the rotation measure analysis data cubes in \l2\ were 
constructed. The upper and lower pairs of 12 GHz IFs and all four 
15 GHz IFs were averaged to improve the signal to noise ratio 
and to obtain long and short spacings in \l2. Final frequencies 
used for the RM analysis are shown in Table~\ref{t3}. This provides 
adequate short and long spacings in \l2\ to properly recover 
RMs between $\pm$ 30000 \radm. The 12 and 15 GHz images used to produce 
the polarization angle maps were tapered to approximate the 8 GHz 
resolution, and a restoring beam matched to the 8 GHz beam was used. 
All images are naturally weighted. 

\section{Results}
Maps showing the rotation measure, RM corrected electric vectors and
spectral index between 8.5 and 12.1 GHz are presented. If the fits to
a $\lambda^2$ law do not appear satisfactory a reduced $\chi^2$ test
was performed. If the reduced $\chi^2$ indicates that a $\lambda^2$
law is ruled out at a 3$\sigma$ level or higher we remove the source
from consideration when examining the RM properties of the sample as a
whole (\S4).

\subsection{B0202+149}

\begin{figure}
\plotone{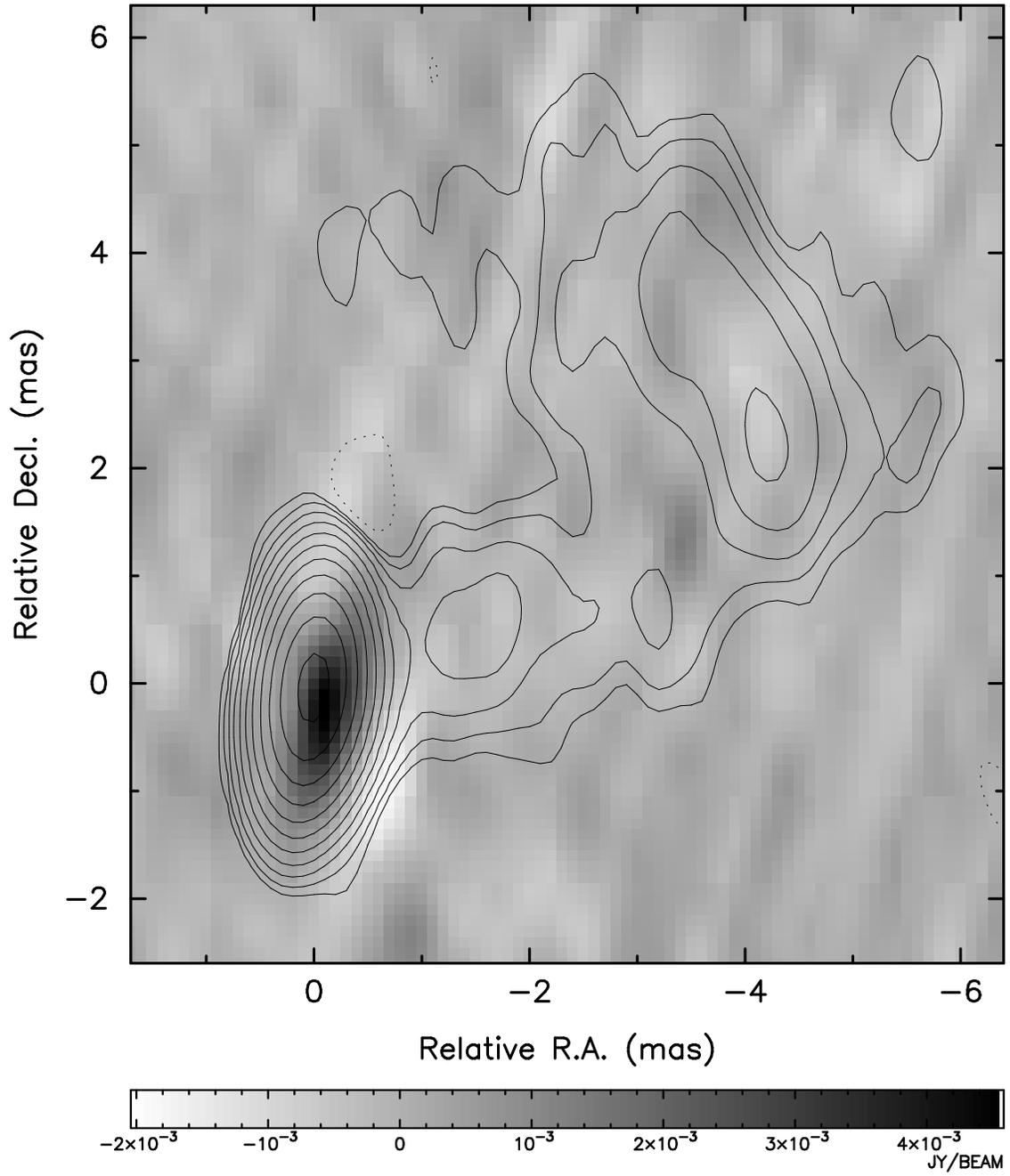}
\caption{Full resolution 15 GHz Stokes U flux in greyscale overlaid 
on Stokes I contours at 15 GHz for B0202+149. Contours start at 
1.2 mJy beam$^{-1}$ and increase by factors of two.}
\label{0202}
\end{figure}

This object was depolarized at 12 and 8 GHz, and thus no 
rotation measure image is provided. The object 
is classified as a blazar at a redshift of 0.405 \citep{perl}.
Observations of superluminal motion \citep{pyat}, a brightness
temperature in excess of of 10$^{12}$ K \citep{gm}, and the high 
probability of detection with EGRET \citep{mattox} all agree with a 
blazar identification for this source. This is surprising as unlike other 
blazars such as 3C\,279 and BL Lac \citep[and references therein]
{zt03} there is no detectable rotation measure in 0202+149. 
This is the case for another EGRET detected blazar, 0420$-$014
\citep{zt03}, whose depolarization seems to be explained by 
the superposition of components of differing position angles. 
For 0202+149 this may be the case. The source is 
only detected in polarization in Stokes U at 15 GHz and a 
full resolution image (Fig.~\ref{0202}) shows two components 
of Stokes U of opposite sign and but different magnitude. 
The negative U component is only weakly detected. Tapering 
and restoring with a beam matched to the 8 GHz resolution 
nearly eliminates the polarized components of 0202+149 
at 15 GHz.

\subsection{B0336$-$019}
\begin{figure}
\plottwo{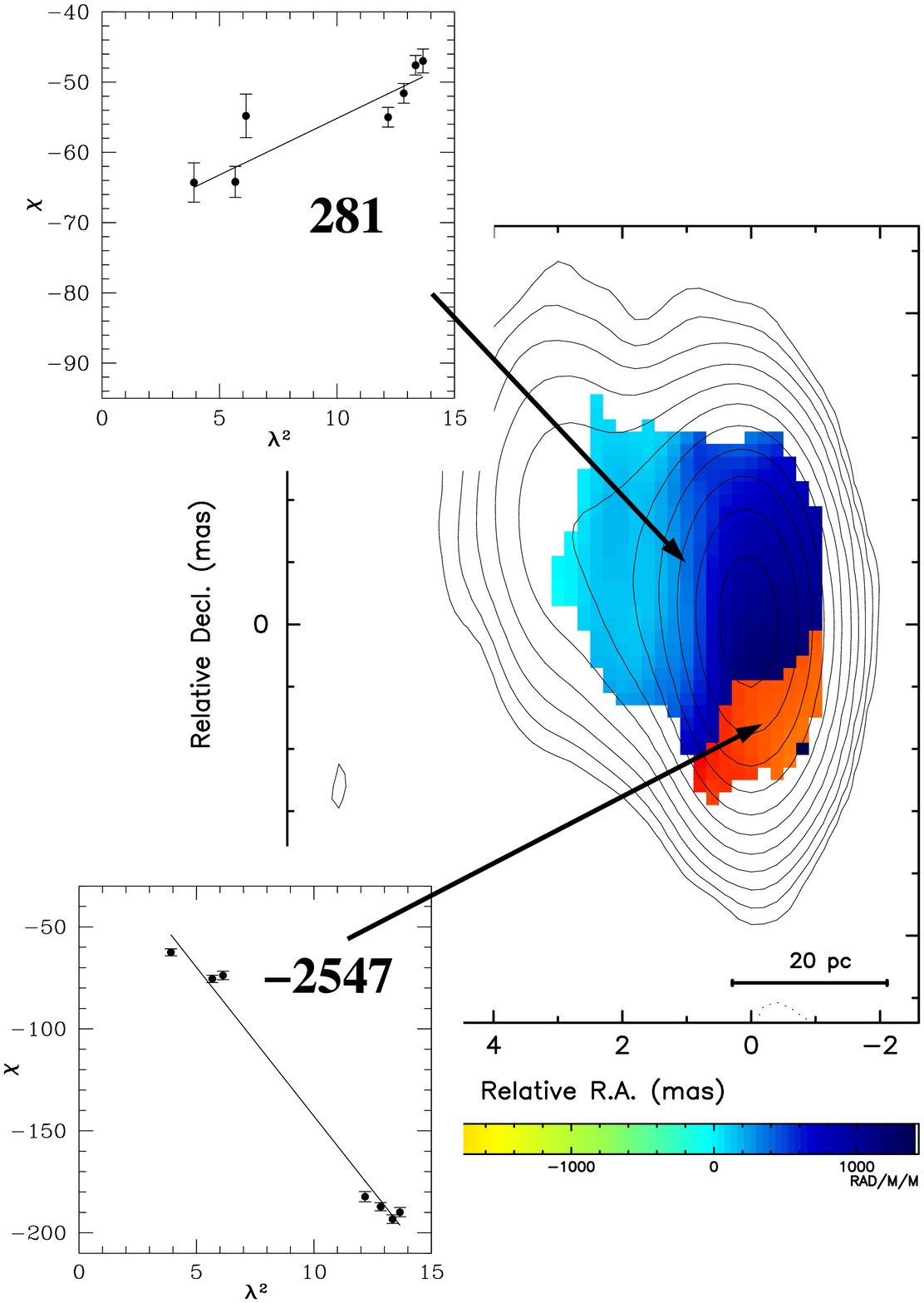}{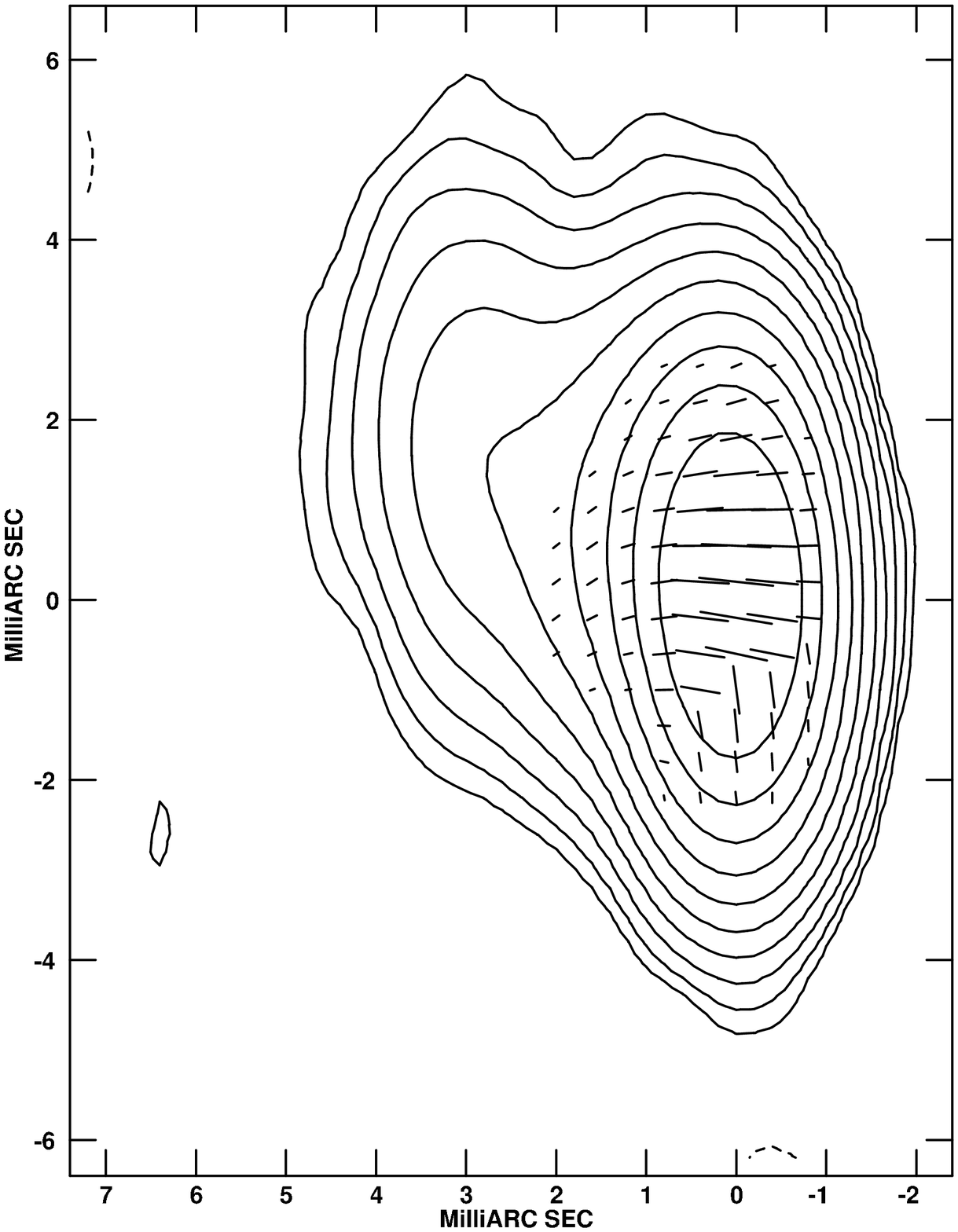}
\caption{(a) Rotation measure image (color) for 0336$-$019 overlaid on
Stokes I contours at 15 GHz. The inset is a plot of EVPA $\chi$
(deg) versus \l2 (cm$^2$). (b) Electric vectors (1 mas =
67 mJy beam$^{-1}$ polarized flux density) corrected for Faraday
Rotation overlaid on Stokes I contours. Contours start at 1.5 mJy
beam$^{-1}$ and increase by factors of two.}
\label{0336rm}
\end{figure}

\begin{figure}
\plotone{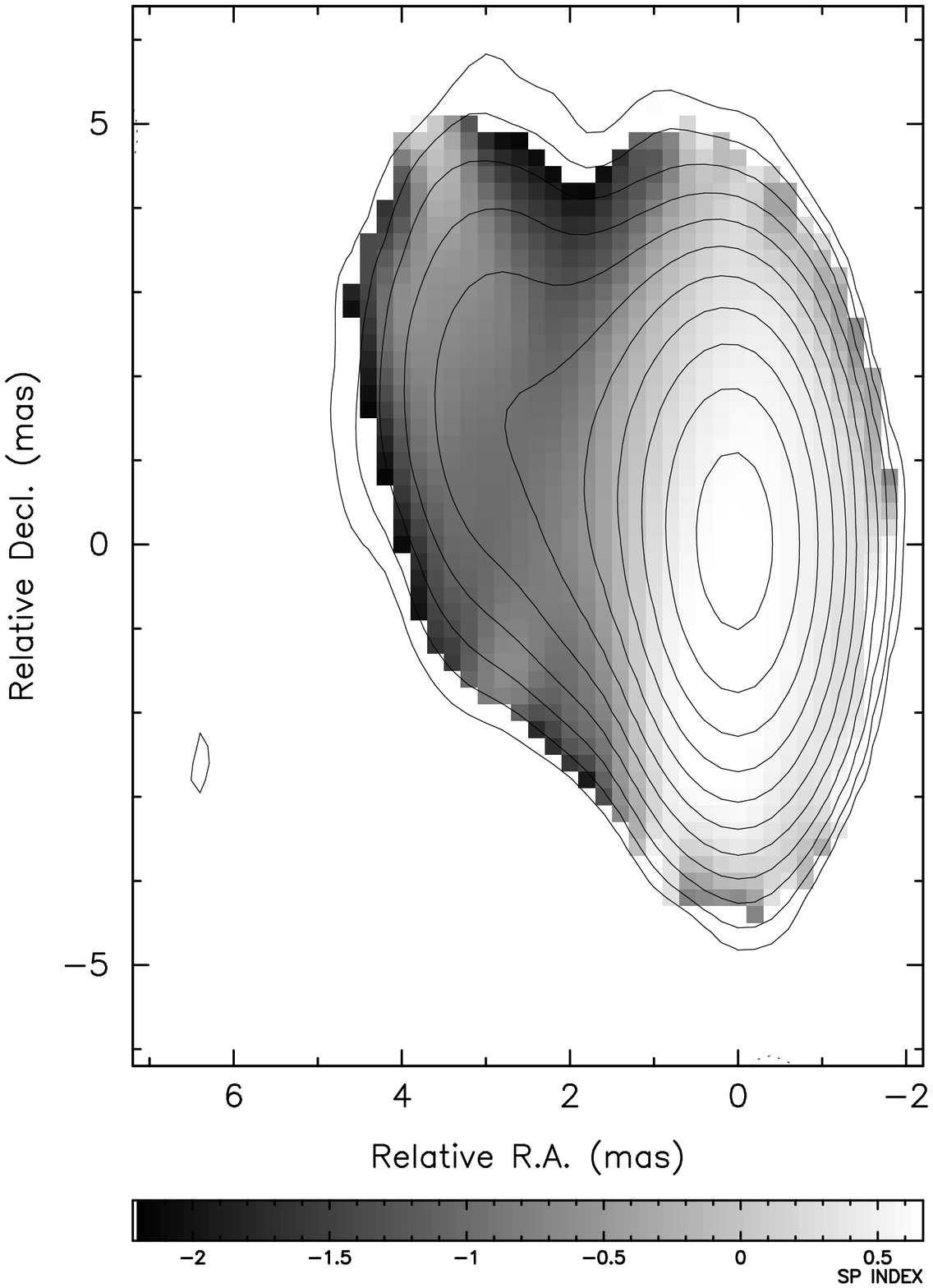}
\caption{Spectral index $\alpha_{12.1}^{8.1}$ plot for 0336$-$019 overlaid on
Stokes I contours at 15 GHz. Contours start at 1.5 mJy beam$^{-1}$ and
increase by factors of two.}
\label{0336si}
\end{figure}

This source has an RM of $-$2547 $\pm$ 33 \radm\, 
which decreases to 281 $\pm$ 37 \radm\ (Fig.~\ref{0336rm}a). A sharp 
border between the negative slope to the RM in the core and the 
positive slope in the jet coincides with a change in the intrinsic 
electric vector direction as shown in Fig.~\ref{0336rm}b. This 
change in the slope of the RM and electric vector orientation 
occurs as the spectral index changes from positive to negative 
(Fig.~\ref{0336si}).

The quality of the fits to a \l2\ law for this quasar appear suspect
(Fig.~\ref{0336rm}a). The reduced $\chi^2$ of the fits are 7.7 or larger. With 
5 degrees of freedom this implies that a \l2\ law can be ruled out 
with a confidence of more than 3$\sigma$.

\subsection{B0355+508}
\begin{figure}
\plottwo{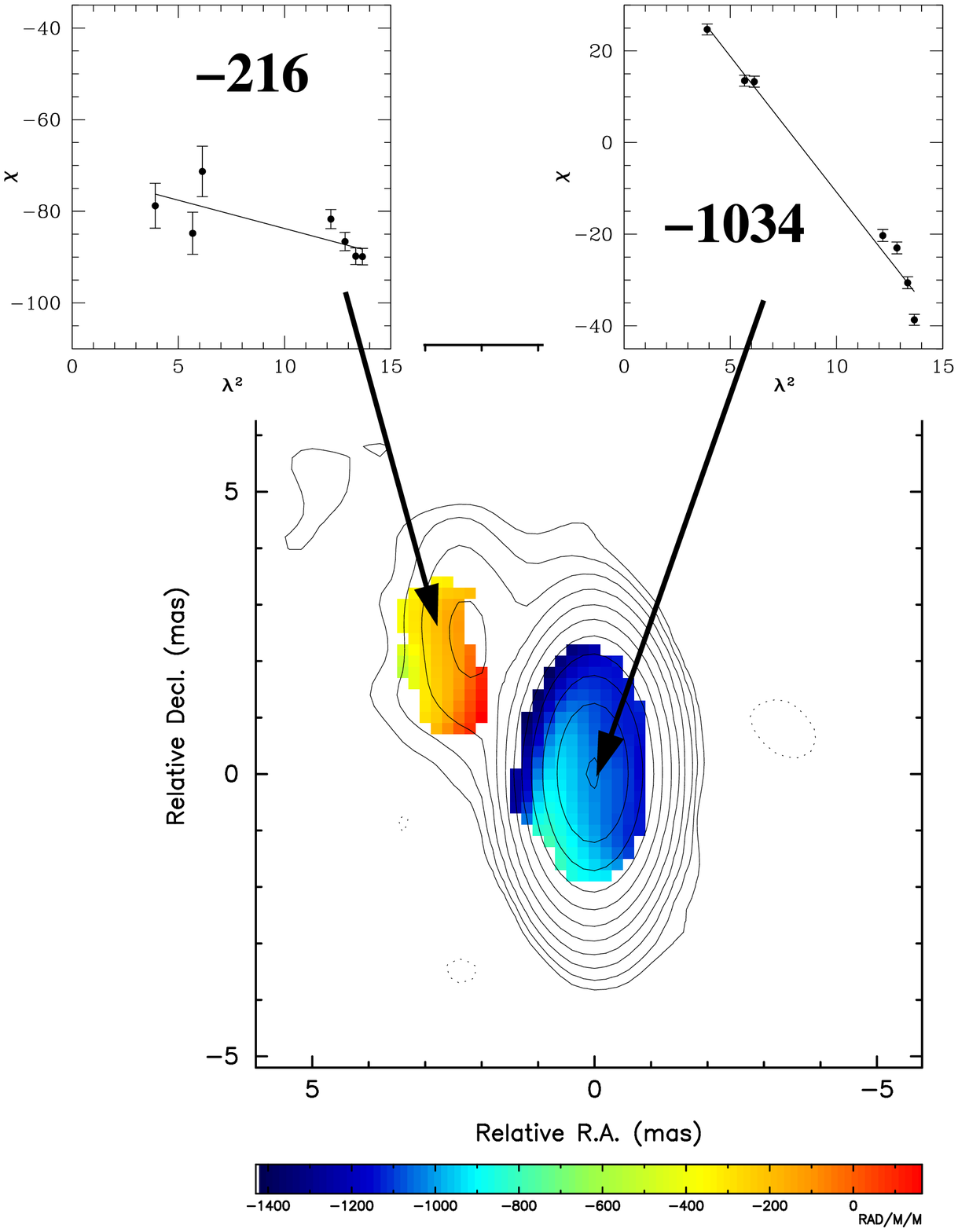}{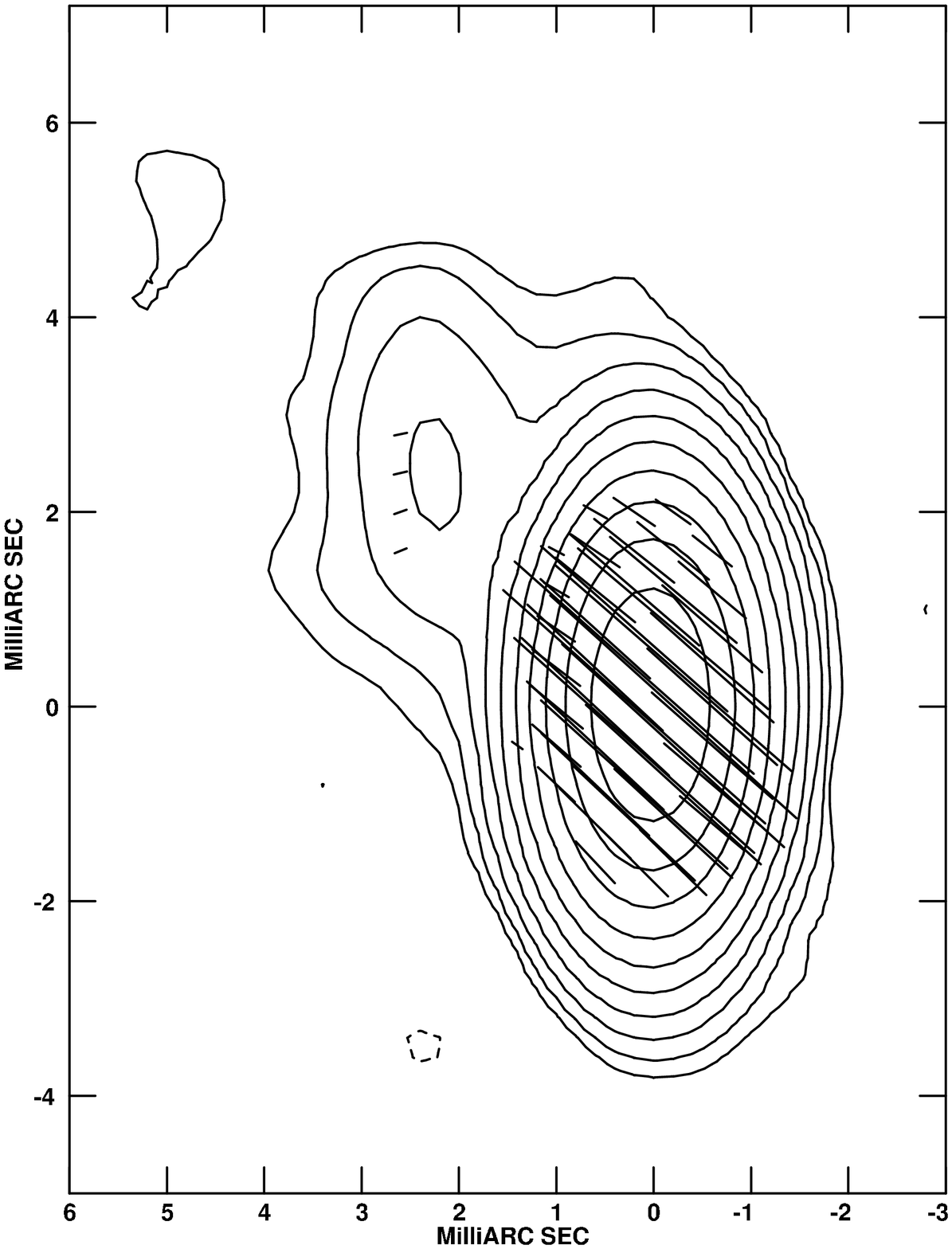}
\caption{(a) Rotation measure image (color) for 0355+508 overlaid on
Stokes I contours at 15 GHz. The inset is a plot of EVPA $\chi$
(deg) versus \l2 (cm$^2$). (b) Electric vectors (1 mas =
67 mJy beam$^{-1}$ polarized flux density) corrected for Faraday
Rotation overlaid on Stokes I contours. Contours start at 6.9 mJy
beam$^{-1}$ and increase by factors of two.}
\label{0355rm}
\end{figure}

\begin{figure}
\plotone{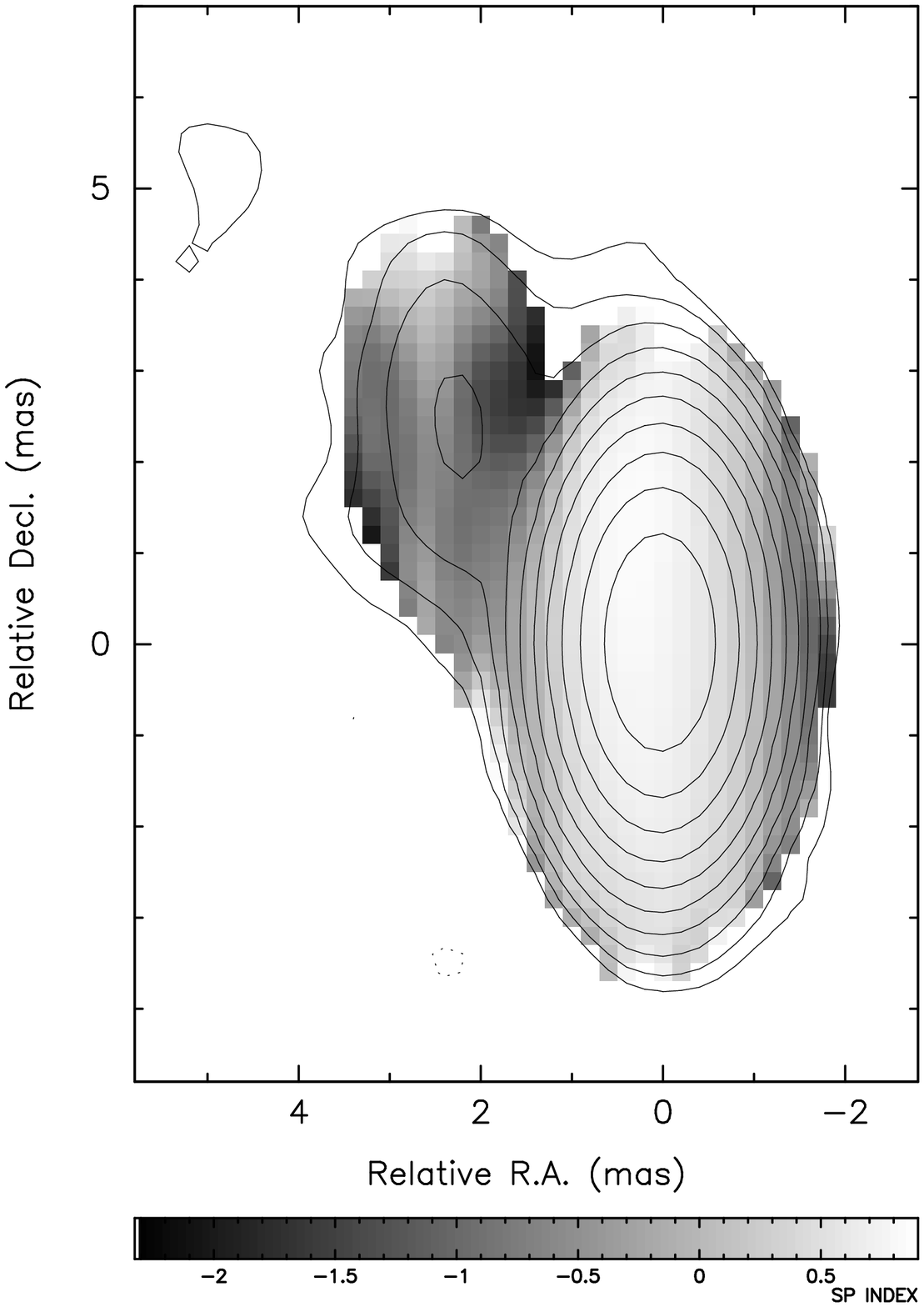}
\caption{Spectral index $\alpha_{12.1}^{8.1}$ plot for 0355+508 overlaid on
Stokes I contours at 15 GHz. Contours start at 6.9 mJy beam$^{-1}$ and
increase by factors of two.}
\label{0355si}
\end{figure}

Also known as NRAO 150, this source has no optical counterpart and 
thus  no redshift available. The RM in the core is 
$-1034 \pm 21$ \radm\ and this decreases by 
approximately a factor of five ($-216 \pm 65$ \radm) in the jet 
component (Fig.~\ref{0355rm}a). The core and jet have very different 
electric vector orientations (Fig.~\ref{0355rm}b), but it should be 
noted that the signal in the jet component is fairly weak as can be 
seen in the inset EVPA vs. \l2 plot in Fig.~\ref{0355rm}a. As 
Fig.~\ref{0355si} shows, the core is optically thick, while the jet 
component is optically thin. 

The 8 GHz core EVPA values suggest a non-linear variation of $\chi$
with $\lambda^2$. However, the reduced $\chi^2$ for the core cannot
rule out a Faraday rotation law at a level of $3\sigma$ or higher, and
we therefore conclude that the core of 0355+508 adheres to the
$\lambda^2$ law. The reduced $\chi^2$ for the RM fits in the jet can
rule out a \l2\ at a level of $3\sigma$ or more, and we conclude
that the data for the jet are not consistent with the $\lambda^2$
law.

\subsection{B0458$-$020}
\begin{figure}
\plottwo{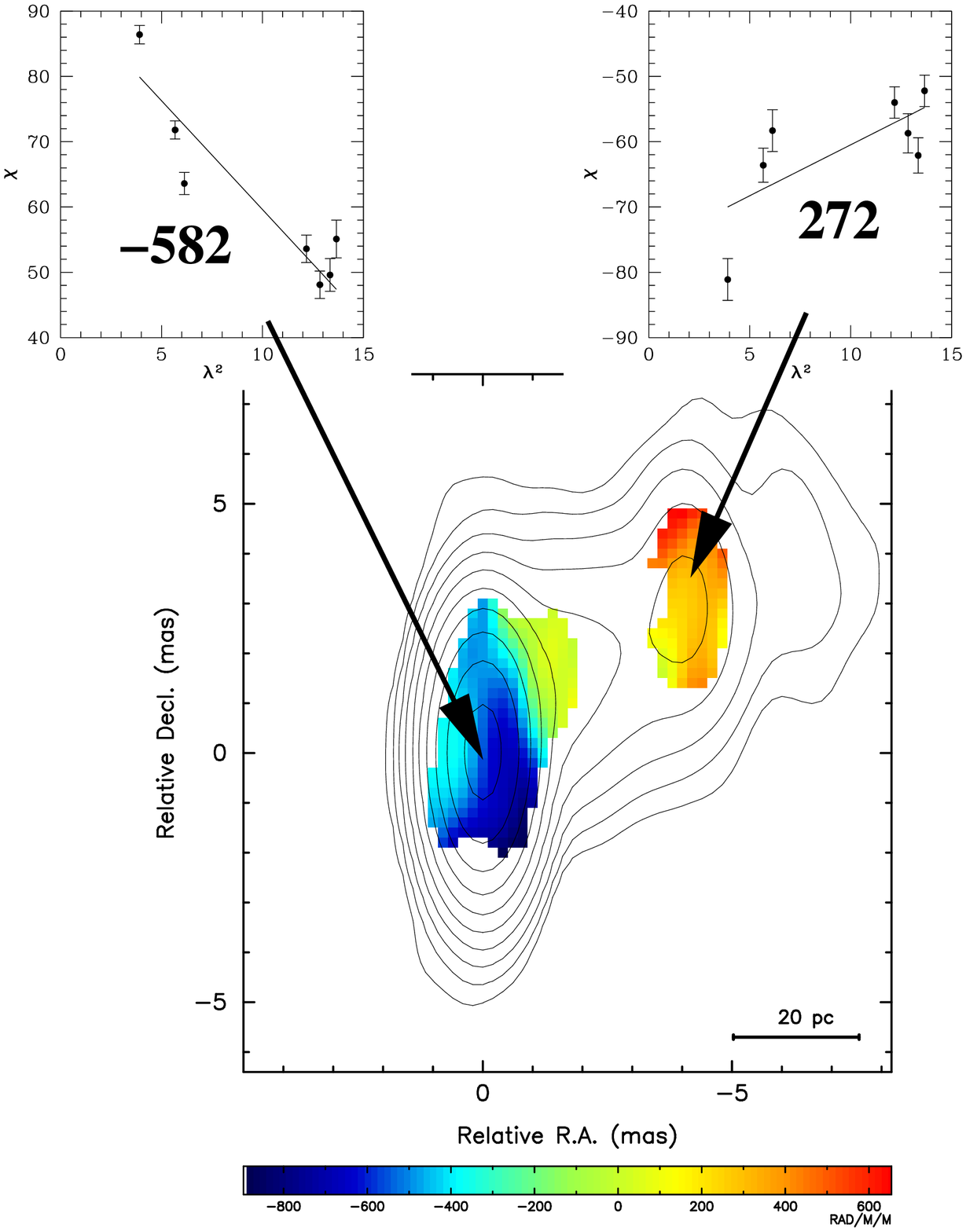}{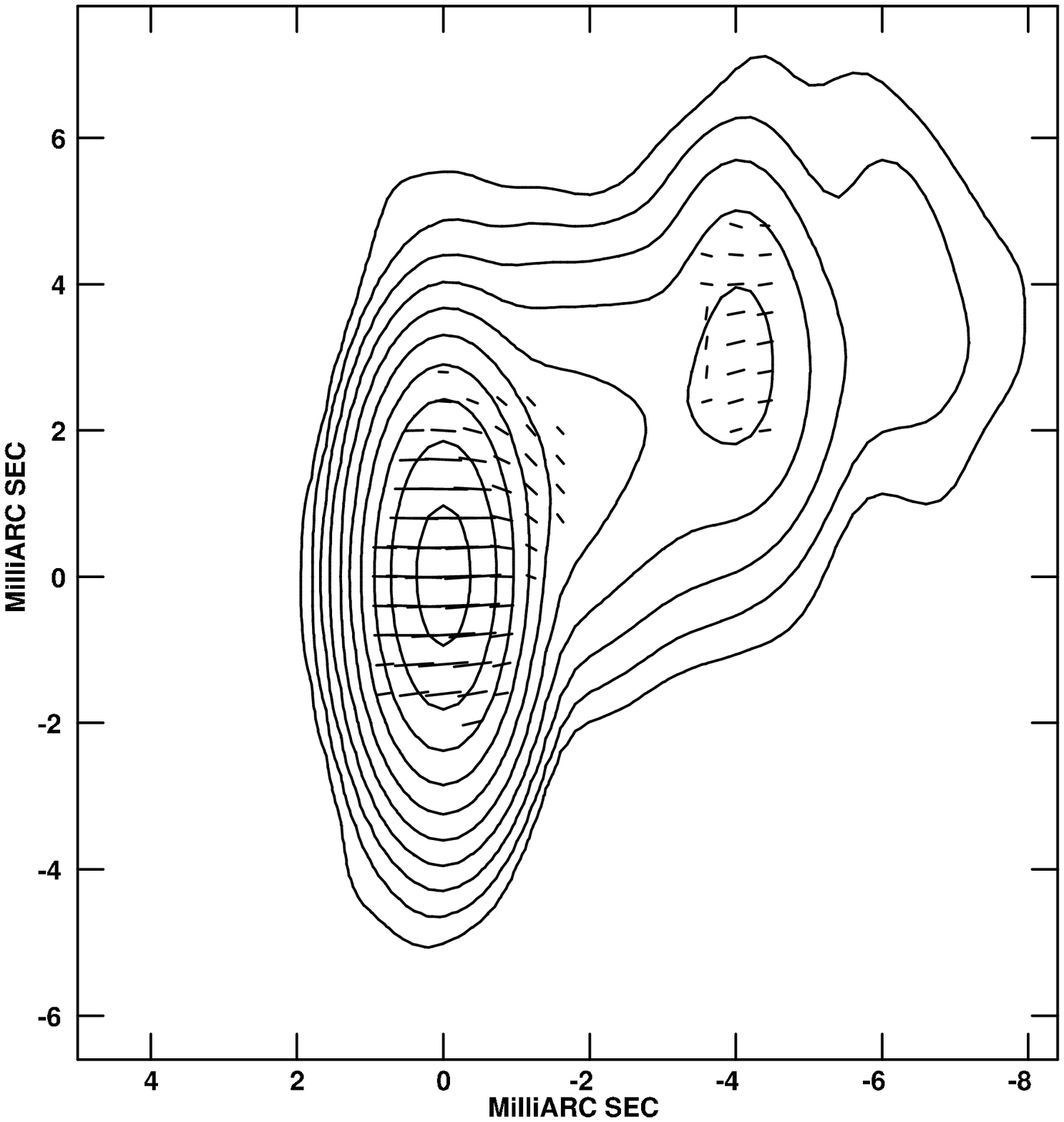}
\caption{(a) Rotation measure image (color) for 0458$-$020 overlaid on
Stokes I contours at 15 GHz. The inset is a plot of EVPA $\chi$
(deg) versus \l2 (cm$^2$). (b) Electric vectors (1 mas =
25 mJy beam$^{-1}$ polarized flux density) corrected for Faraday
Rotation overlaid on Stokes I contours. Contours start at 1.4 mJy
beam$^{-1}$ and increase by factors of two.}
\label{0458rm}
\end{figure}

\begin{figure}
\plotone{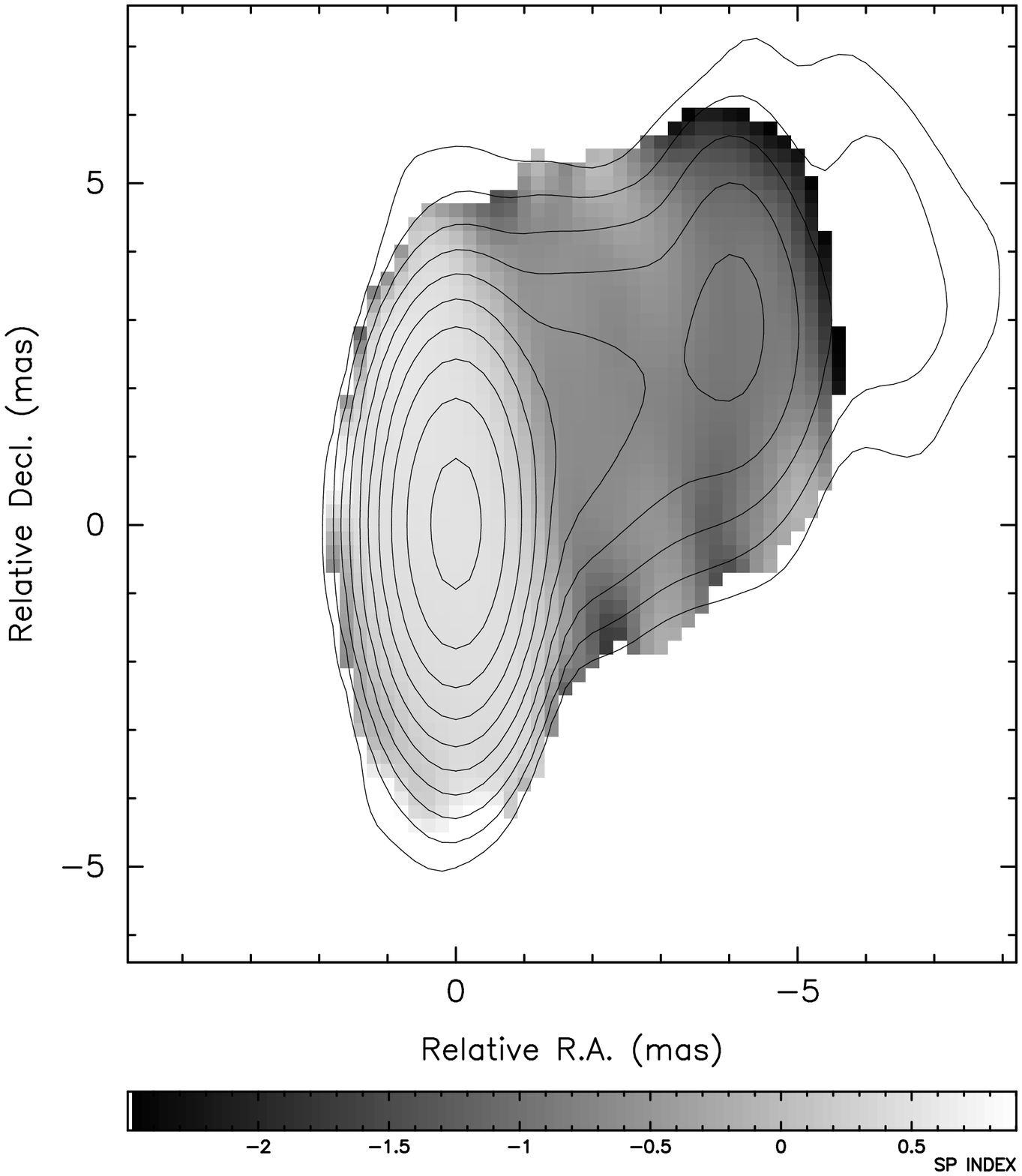}
\caption{Spectral index $\alpha_{12.1}^{8.1}$ plot for 0458$-$020 overlaid on
Stokes I contours at 15 GHz. Contours start at 1.4 mJy beam$^{-1}$ and
increase by factors of two.}
\label{0458si}
\end{figure}

Fig. 7a shows that the fits to a $\lambda^2$ law are not very
convincing for this source. The RM of $-582 \pm 32$ \radm\ has a
reduced $\chi^2$ of 12. The $3\sigma$ confidence level with 5 degrees
of freedom is approximately 3.1, and we reject the \l2\ law for
the core of 0458$-$020. Similarly, we reject the $\lambda^2$ law for
the jet as the reduced $\chi^2$ is 6. Assuming Faraday rotation does apply 
to 0458$-$020 we see in Fig.~\ref{0458rm}b that the jet and core have 
nearly the same electric vector alignments. 0458$-$020 has an optically 
thick core and optically thin jet (Fig.~\ref{0458si}). 

\subsection{B0552+398}
\citet{obs} classify 0552+398 as a gigahertz peaked spectrum (GPS) 
source. Infrared imaging suggests it is an interacting galaxy in a 
dense cluster \citep{hut}. \citet{wills} note that the redshift of 
0552+398 is uncertain because of a lack of firmly identified spectral 
lines. 

\begin{figure}
\plottwo{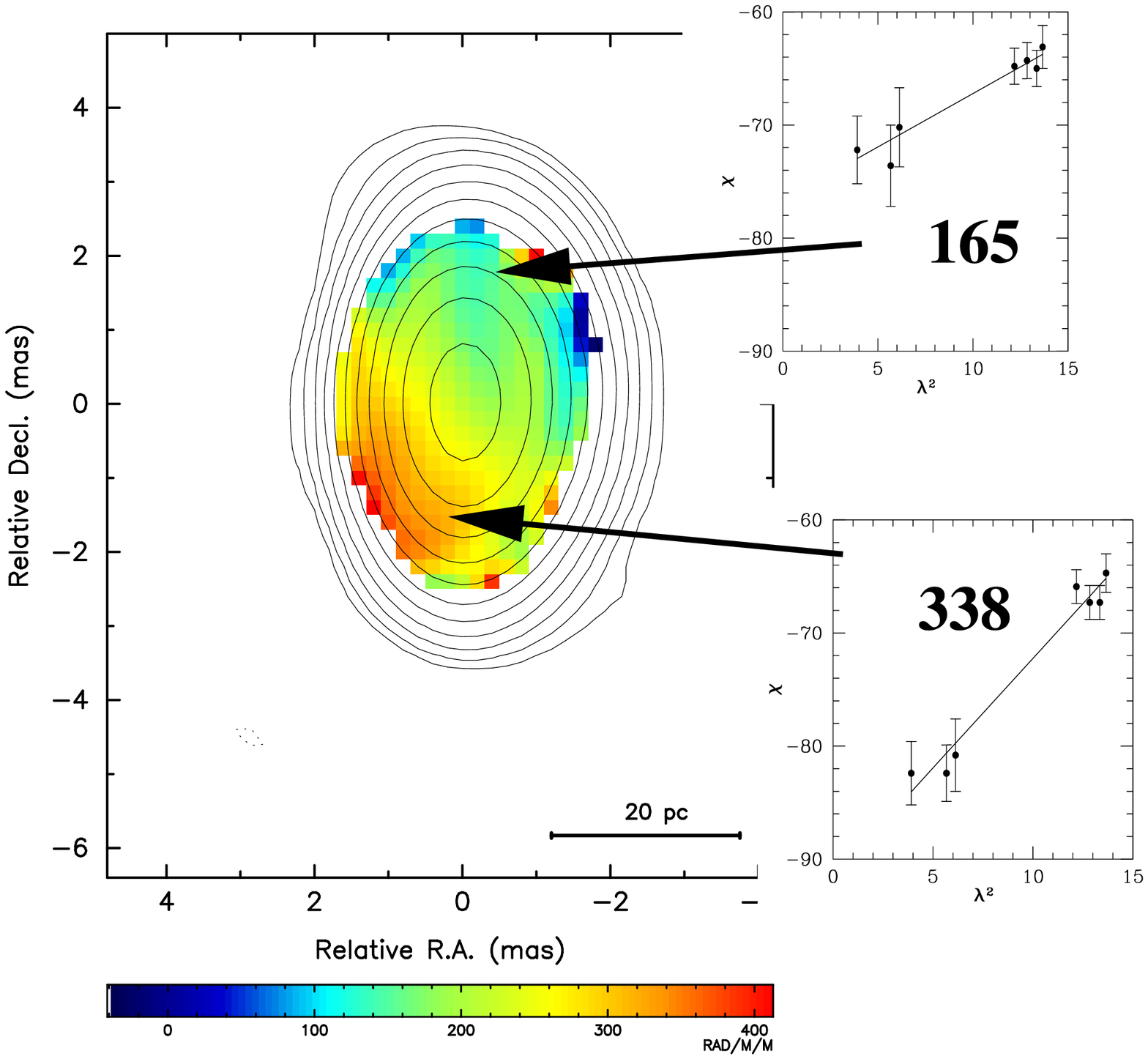}{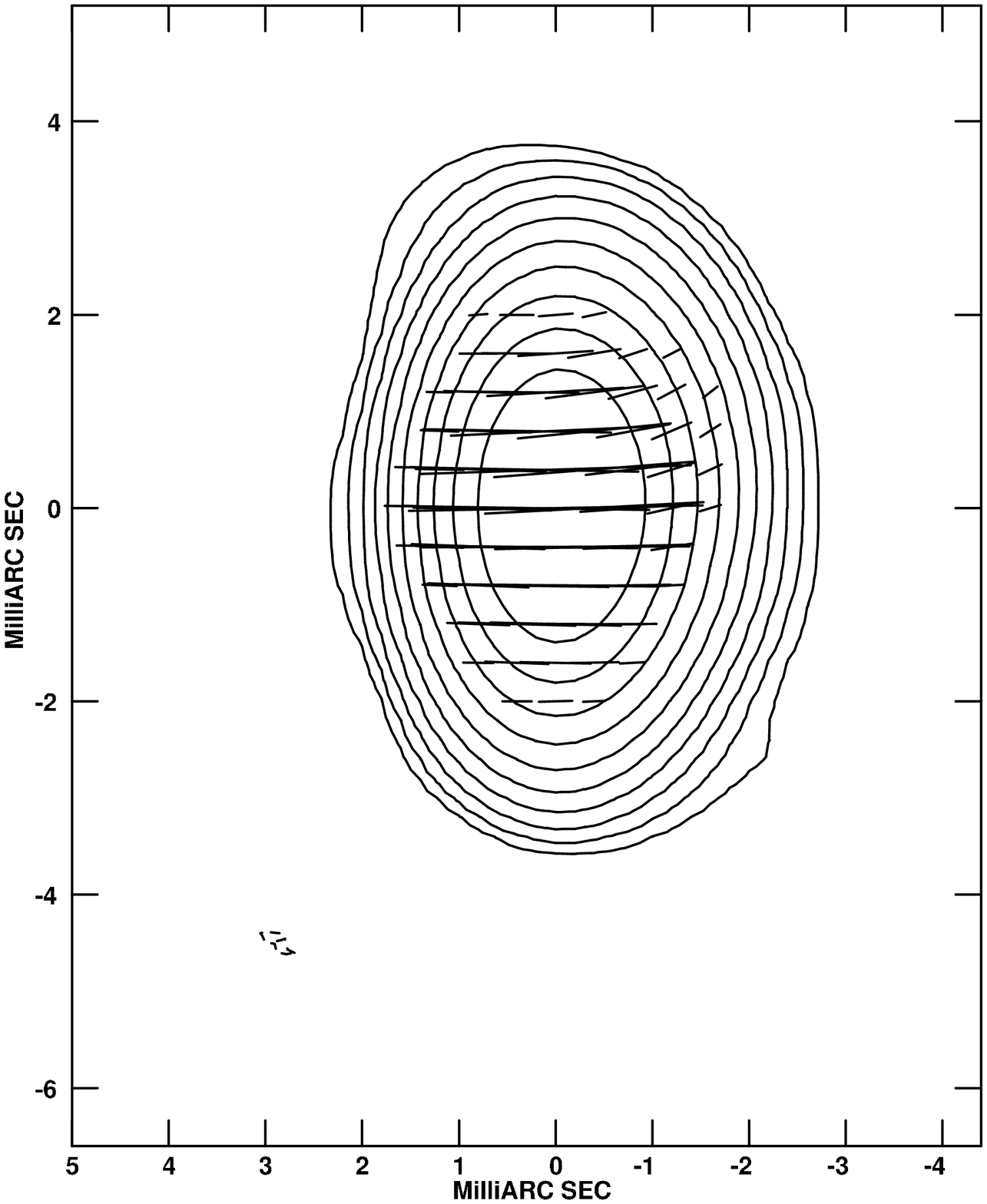}
\caption{(a) Rotation measure image (color) for 0552+398 overlaid on
Stokes I contours at 15 GHz. The inset is a plot of EVPA $\chi$
(deg) versus \l2 (cm$^2$). (b) Electric vectors (1 mas =
25 mJy beam$^{-1}$ polarized flux density) corrected for Faraday
Rotation overlaid on Stokes I contours. Contours start at 2.7 mJy
beam$^{-1}$ and increase by factors of two.}
\label{0552rm}
\end{figure}

\begin{figure}
\plotone{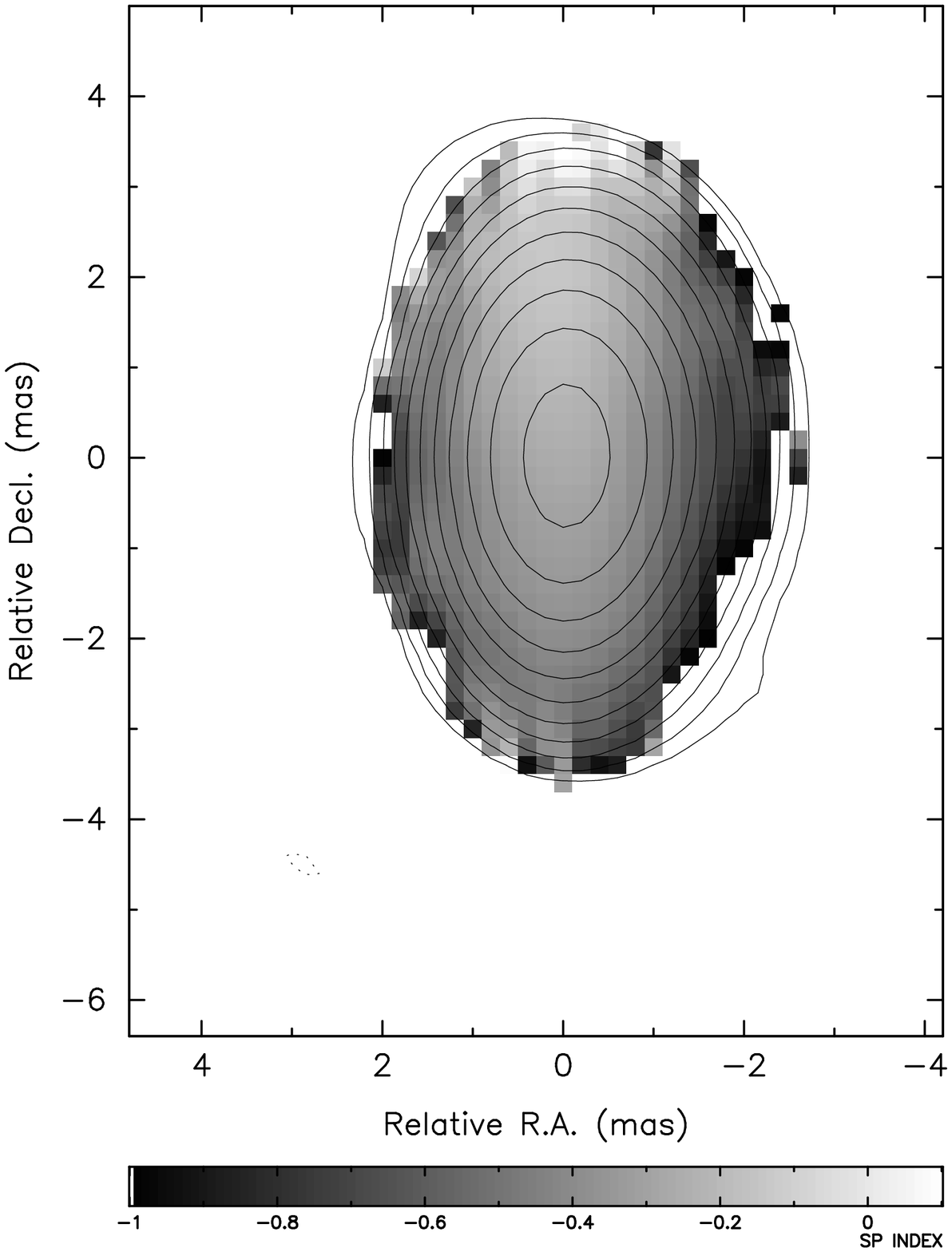}
\caption{Spectral index $\alpha_{12.1}^{8.1}$ plot for 0552+398 overlaid on
Stokes I contours at 15 GHz. Contours start at 2.7 mJy beam$^{-1}$ and
increase by factors of two.}
\label{0552si}
\end{figure}

We are just able to resolve a rotation measure gradient across this 
source, as shown in Fig.~\ref{0552rm}a. The RM changes from 338 $\pm$ 
39 \radm\ to 165 $\pm$ 45 \radm. This gradient is across a 
projected distance of 20 pc. This distance would incorporate the 
high RM core and lower RM jet in 3C\,273 shown in \citet{zt01}. The 
8$-$15 GHz RM image of 3C\,273 in \citet{zt01} showed lower RMs than 
the the higher resolution 15$-$43 GHz RM images. Therefore a much 
higher RM may be hidden under the coarse spatial resolution of our 
image. RM corrected electric vectors are aligned East-West 
(Fig.~\ref{0552rm}b). The spectral index changes from approximately 0
in the north to $-$0.5 or less in the south (Fig.~\ref{0552si}).

\subsection{B0605$-$085}
\begin{figure}
\plottwo{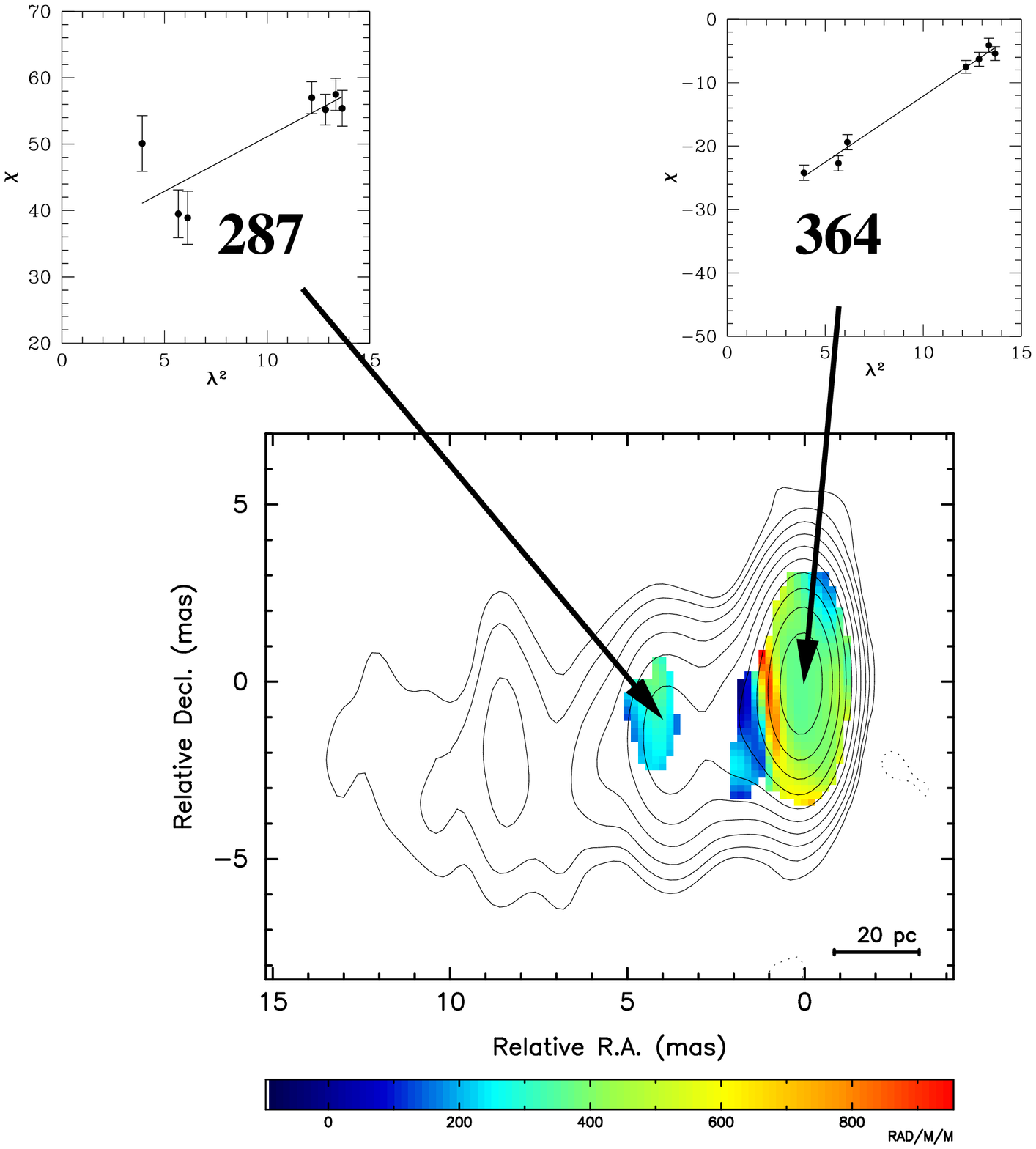}{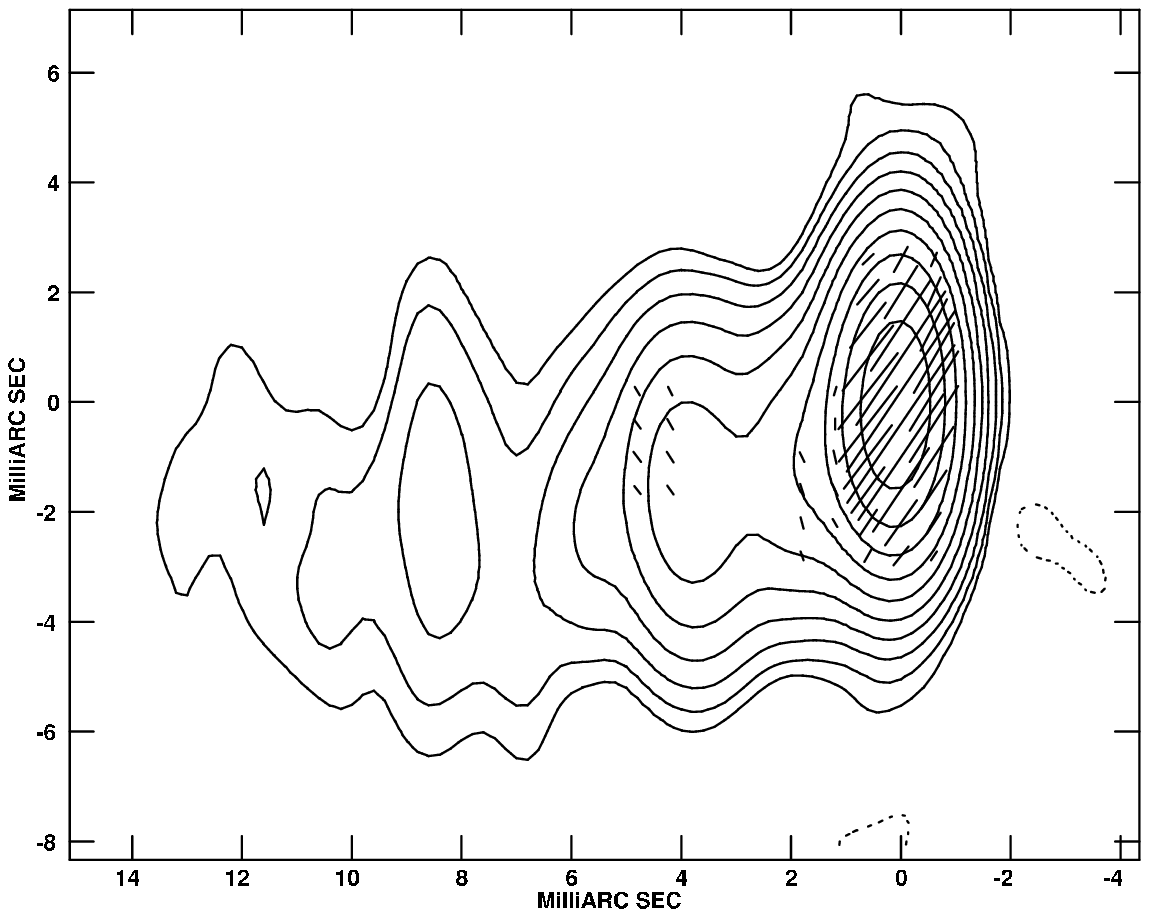}
\caption{(a) Rotation measure image (color) for 0605$-$085 overlaid on
Stokes I contours at 15 GHz. The inset is a plot of EVPA $\chi$
(deg) versus \l2 (cm$^2$). (b) Electric vectors (1 mas =
25 mJy beam$^{-1}$ polarized flux density) corrected for Faraday
Rotation overlaid on Stokes I contours. Contours start at 1.5 mJy
beam$^{-1}$ and increase by factors of two.}
\label{0605rm}
\end{figure}

\begin{figure}
\plotone{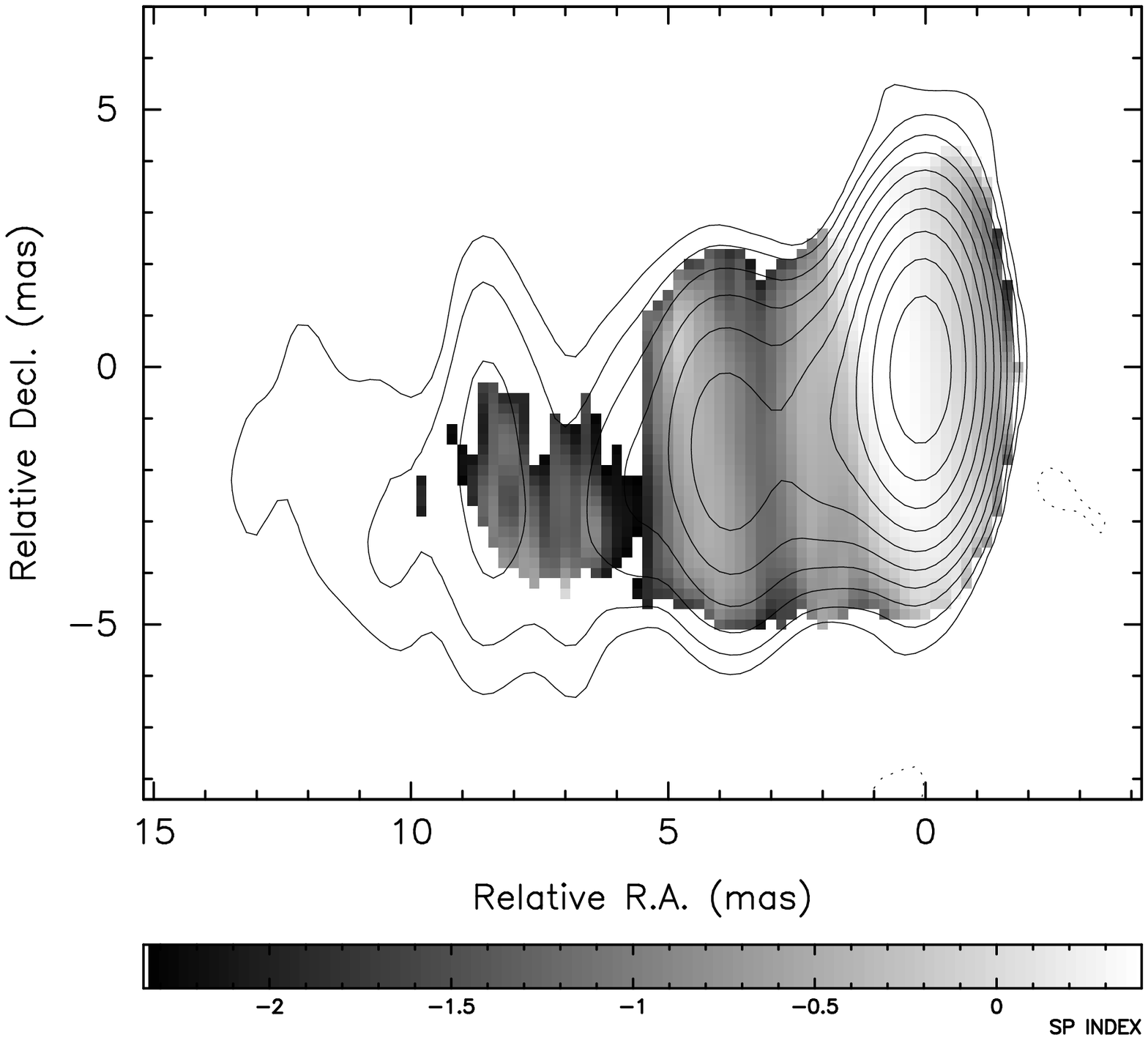}
\caption{Spectral index $\alpha_{12.1}^{8.1}$ plot for 0605$-$085 overlaid on
Stokes I contours at 15 GHz. Contours start at 1.5 mJy beam$^{-1}$ and
increase by factors of two.}
\label{0605si}
\end{figure}

The core and jet component 4 mas east of the core show similar RMs, 
although there is a lower SNR at the higher frequencies for the jet 
component (Fig.~\ref{0605rm}a). The inset plots in Fig.~\ref{0605rm}a have 
RMs of 364 $\pm$ 20 \radm\ and 287 $\pm$ 57 \radm. About 2 mas SE 
of the core there seems to be a flattening of the RM slope, and this 
is coincident with a change in the RM corrected EVPA (Fig.~\ref{0605rm}b). 
Fig.~\ref{0605si} shows that this region 2 mas from the core marks 
the transition to a negative spectral index. 

As the SNR for the RM fit in the jet appears rather low we examined
the reduced $\chi^2$ for both the jet and core RM fits. The core RM
fits all have reduced $\chi^2$ consistent with a $\lambda^2$ law with
values less than the 3$\sigma$ level. Even the apparently poor RM fit
in the jet has a reduced $\chi^2$ of 2.1, and thus consistent with a
$\lambda^2$ law interpretation.

\subsection{B0736+017}
\begin{figure}
\plottwo{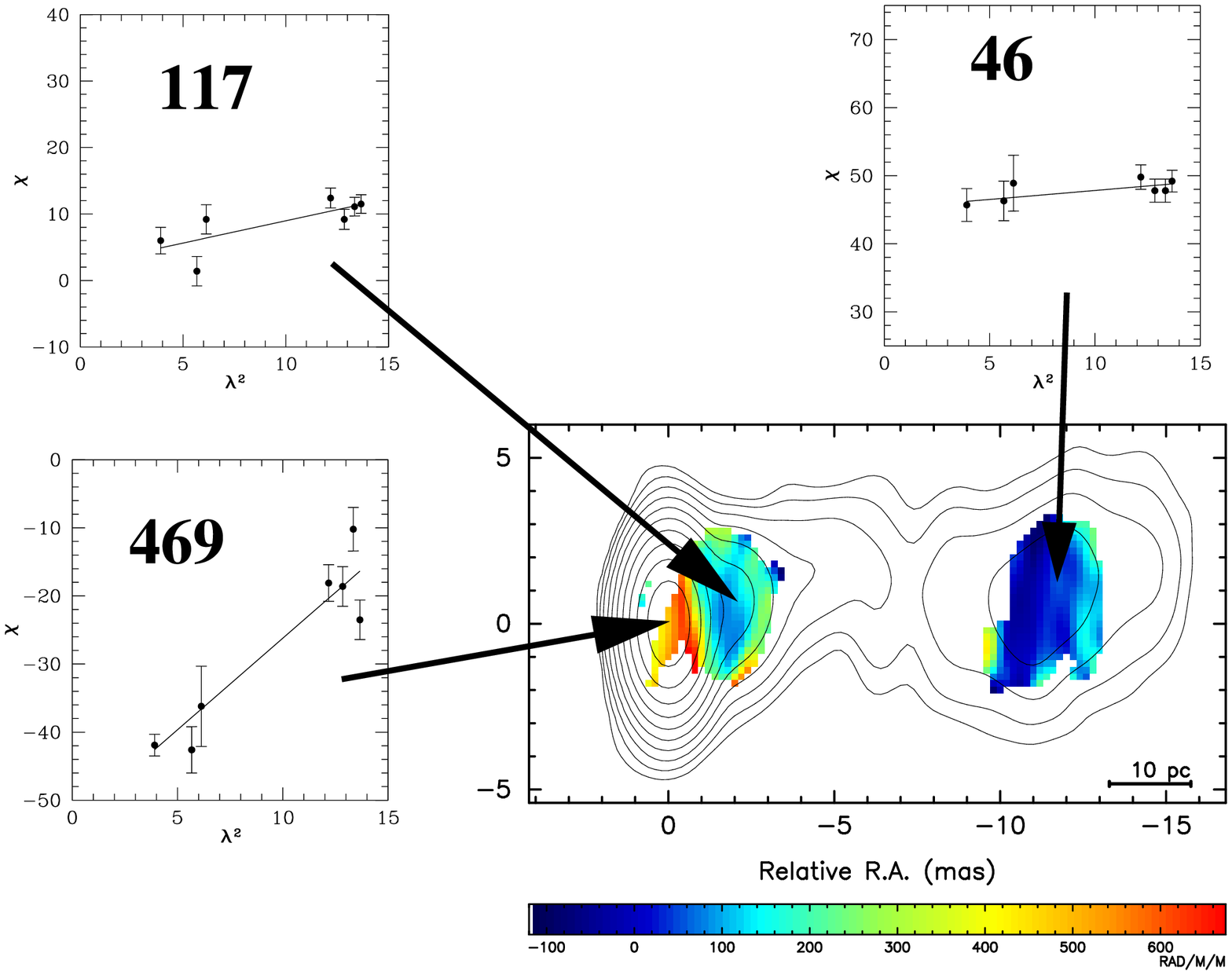}{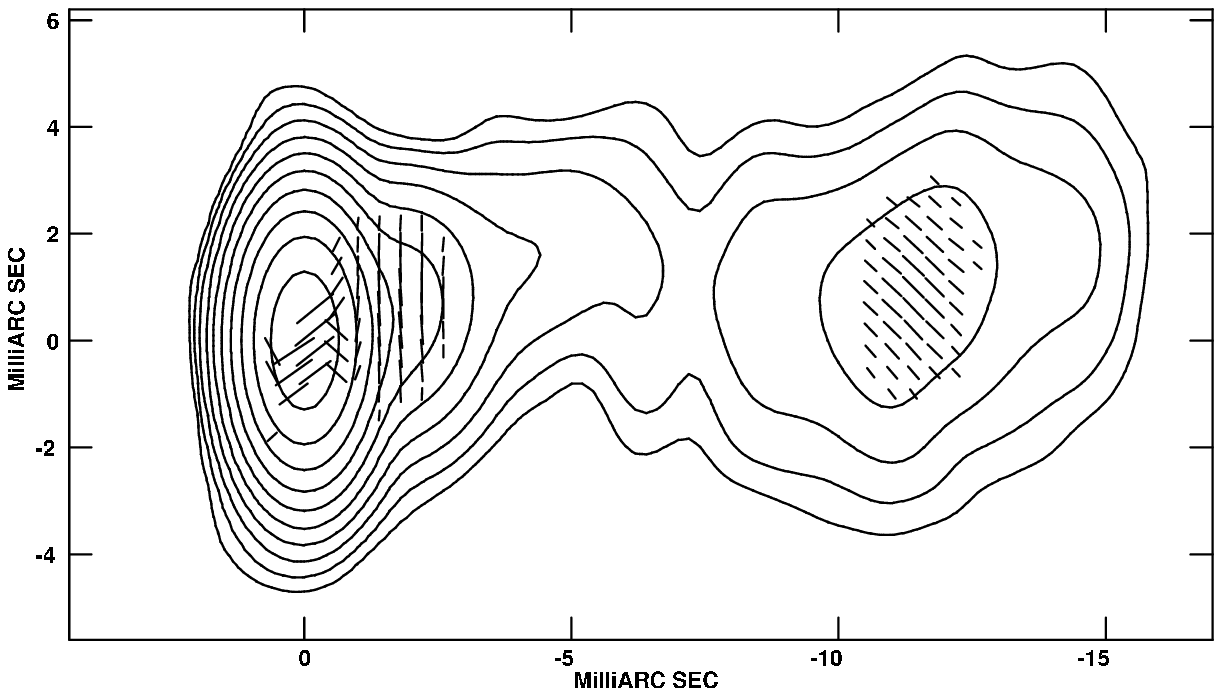}
\caption{(a) Rotation measure image (color) for 0736+017 overlaid on
Stokes I contours at 15 GHz. The inset is a plot of EVPA $\chi$
(deg) versus \l2 (cm$^2$). (b) Electric vectors (1 mas =
17 mJy beam$^{-1}$ polarized flux density) corrected for Faraday
Rotation overlaid on Stokes I contours. Contours start at 1.5 mJy
beam$^{-1}$ and increase by factors of two.}
\label{0736rm}
\end{figure}

\begin{figure}
\plotone{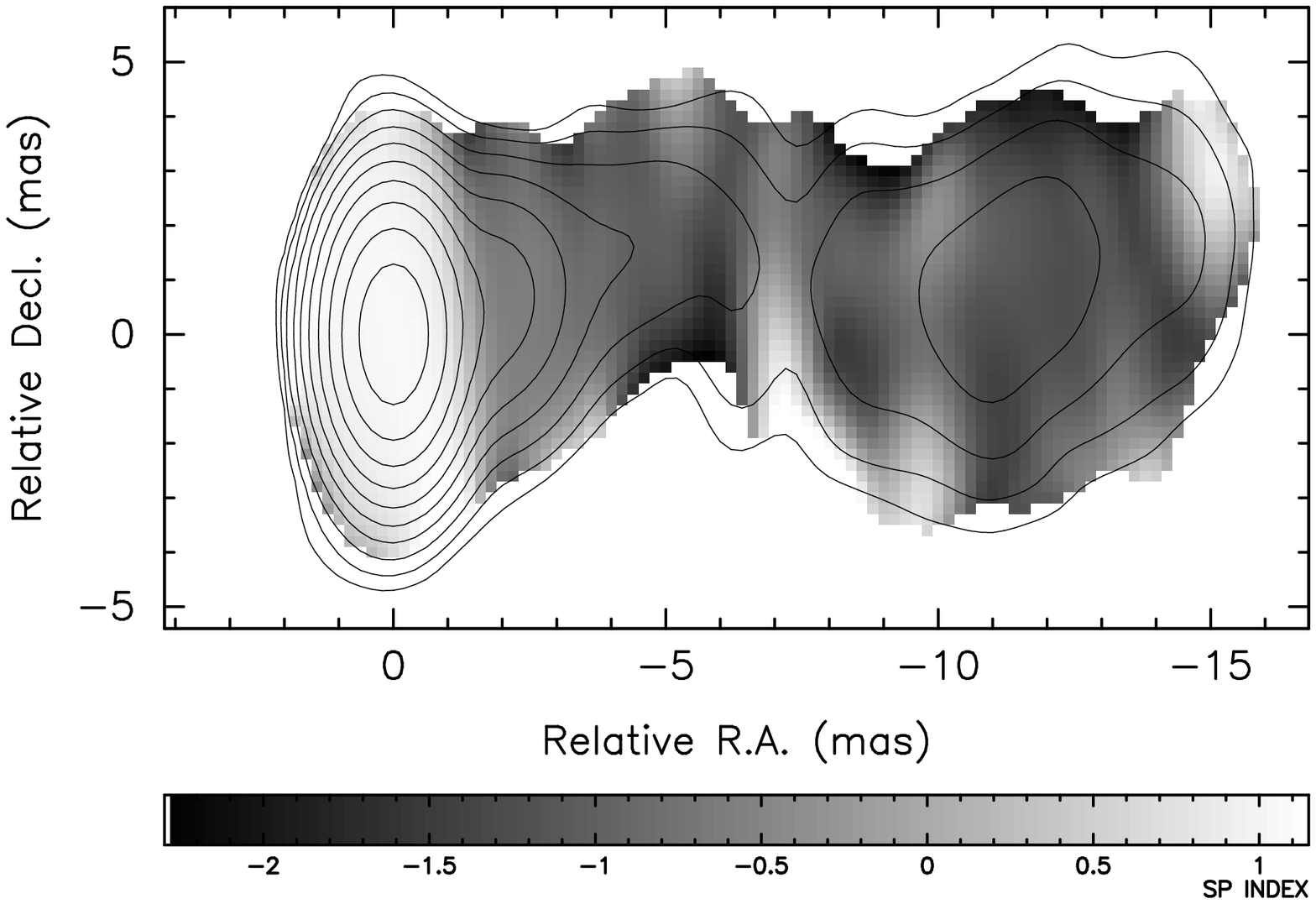}
\caption{Spectral index $\alpha_{12.1}^{8.1}$ plot for 0736+017 overlaid on
Stokes I contours at 15 GHz. Contours start at 1.5 mJy beam$^{-1}$ and
increase by factors of two.}
\label{0736si}
\end{figure}

This quasar has recently been shown to exhibit a dramatic optical 
flare, and shows evidence for microvariability \citep{cjt}.
The weakly polarized core (0.6\%) has an RM of $469 \pm 40$ \radm\ 
(Fig.~\ref{0736rm}a), but approximately 50\% of the pixels within a beam area
centered on the core have a reduced $\chi^2$ which rules out a
$\lambda^2$ law at a level greater than $3\sigma$. We thus reject the
optically thick (Fig.~\ref{0736si}) core as a region where the Faraday rotation
law applies. Beyond a beamwidth ($\approx$1mas west) of the core the reduced
$\chi^2$ values are consistent with a $\lambda^2$ law, and the spectrum
changes to optically thin 2-3mas west of the core (Fig.~\ref{0736si}).

\subsection{B0748+126}
\begin{figure}
\plottwo{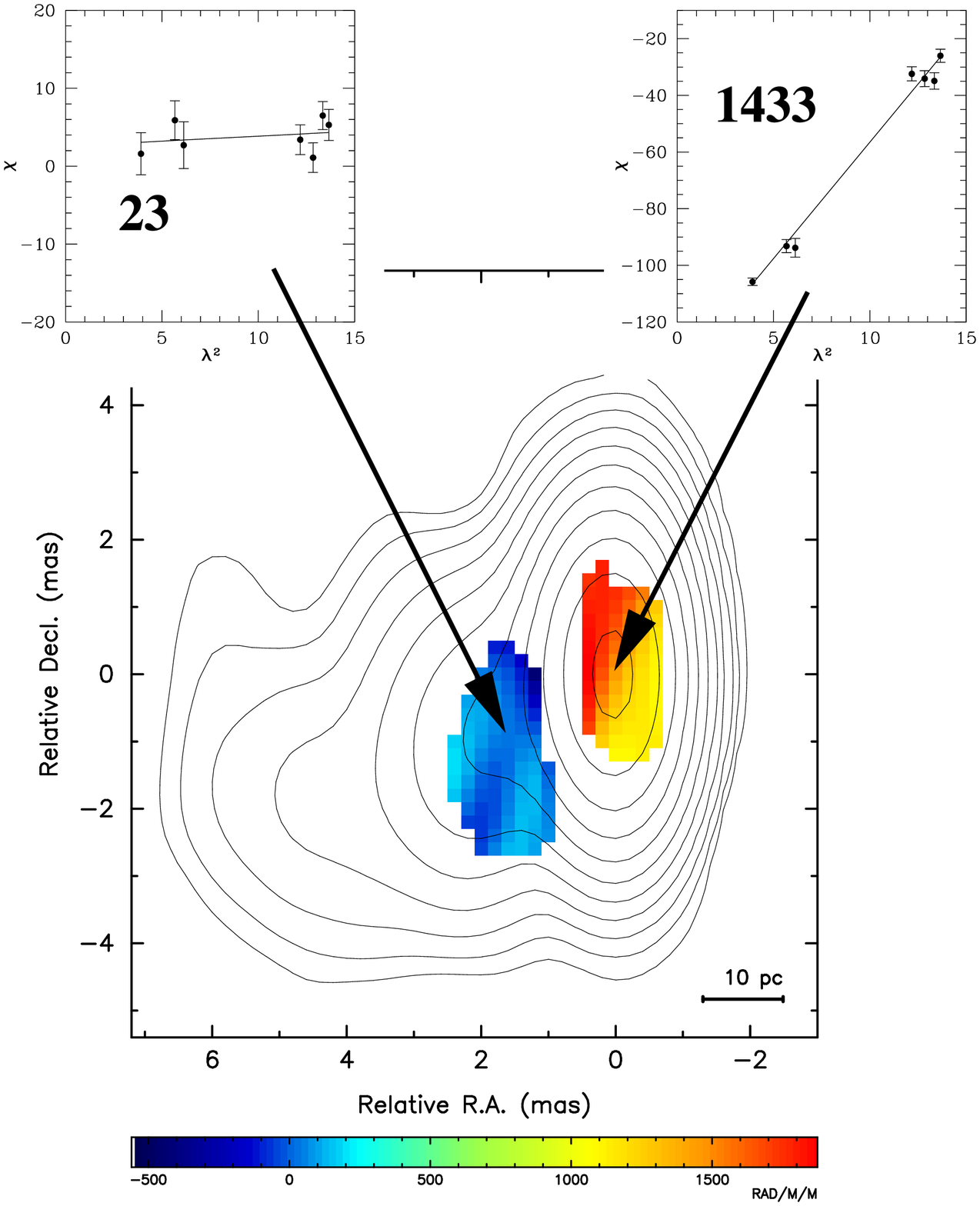}{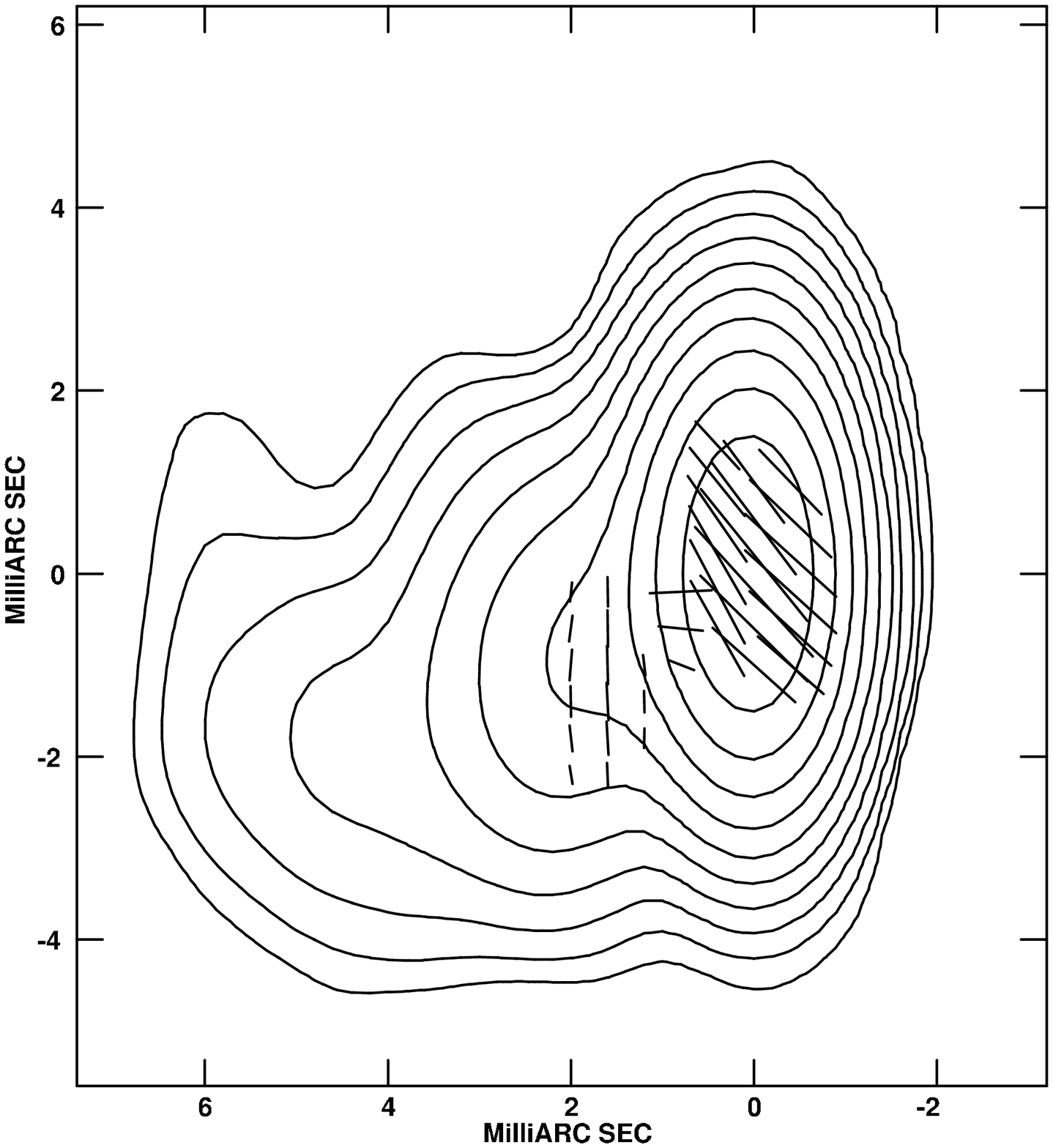}
\caption{(a) Rotation measure image (color) for 0748+126 overlaid on
Stokes I contours at 15 GHz. The inset is a plot of EVPA $\chi$
(deg) versus \l2 (cm$^2$). (b) Electric vectors (1 mas =
17 mJy beam$^{-1}$ polarized flux density) corrected for Faraday
Rotation overlaid on Stokes I contours. Contours start at 1.2 mJy
beam$^{-1}$ and increase by factors of two.}
\label{0748rm}
\end{figure}

\begin{figure}
\plotone{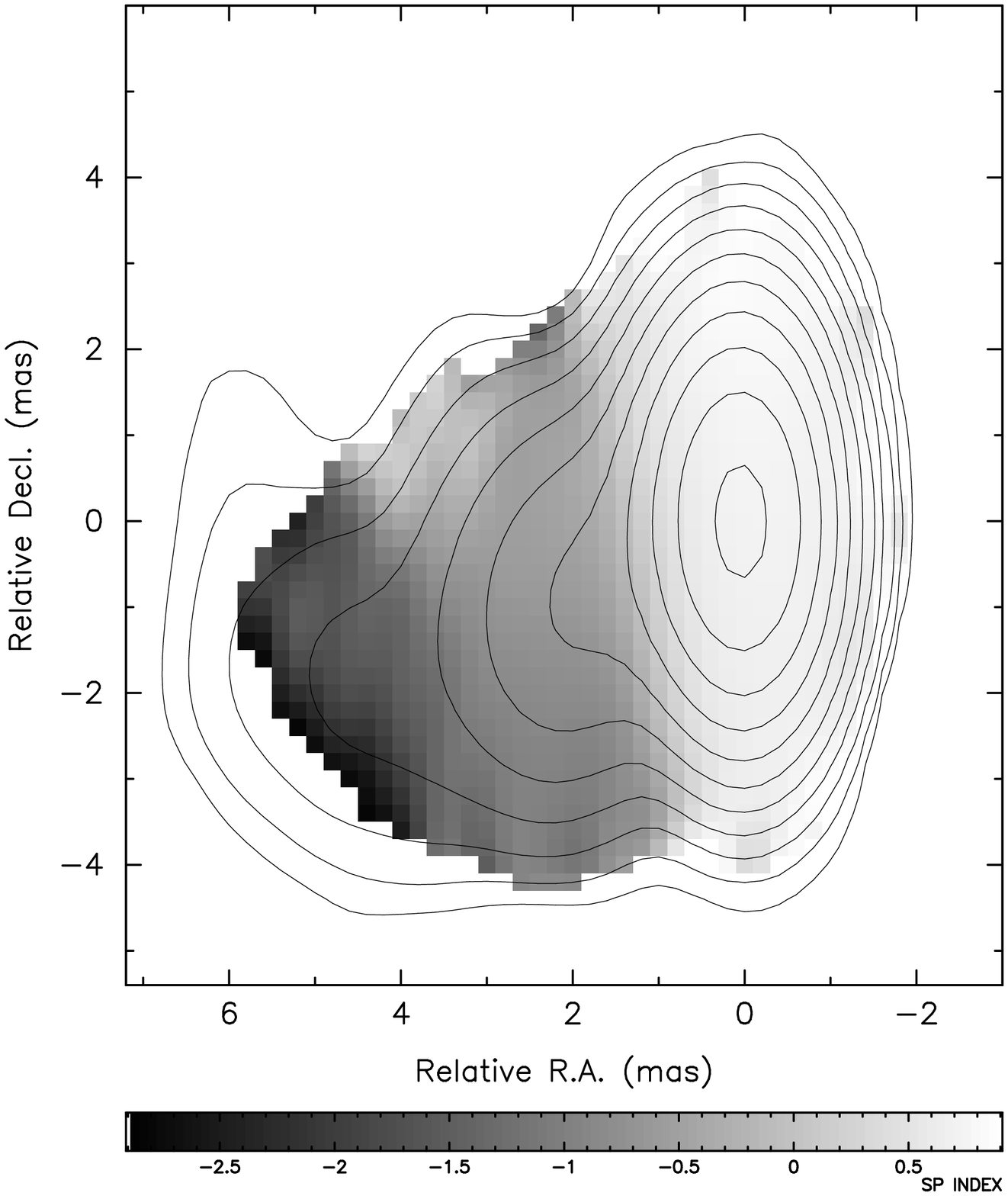}
\caption{Spectral index $\alpha_{12.1}^{8.1}$ plot for 0748+126 overlaid on
Stokes I contours at 15 GHz. Contours start at 1.2 mJy beam$^{-1}$ and
increase by factors of two.}
\label{0748si}
\end{figure}

A typical quasar core RM of 1433 $\pm$ 34 \radm\ and jet RM consistent 
with 0 (23 $\pm$ 40 \radm) are shown in Fig.~\ref{0748rm}a. These two 
RM regions have EVPAs which differ by $\approx$ 45\dg\ 
(Fig.~\ref{0748rm}b). Fig.~\ref{0748si} shows the typical flat 
spectrum core and steep spectrum jet. \citet{wills} report an 
uncertainty in the published redshift.

\subsection{B1055+018}
\begin{figure}
\plottwo{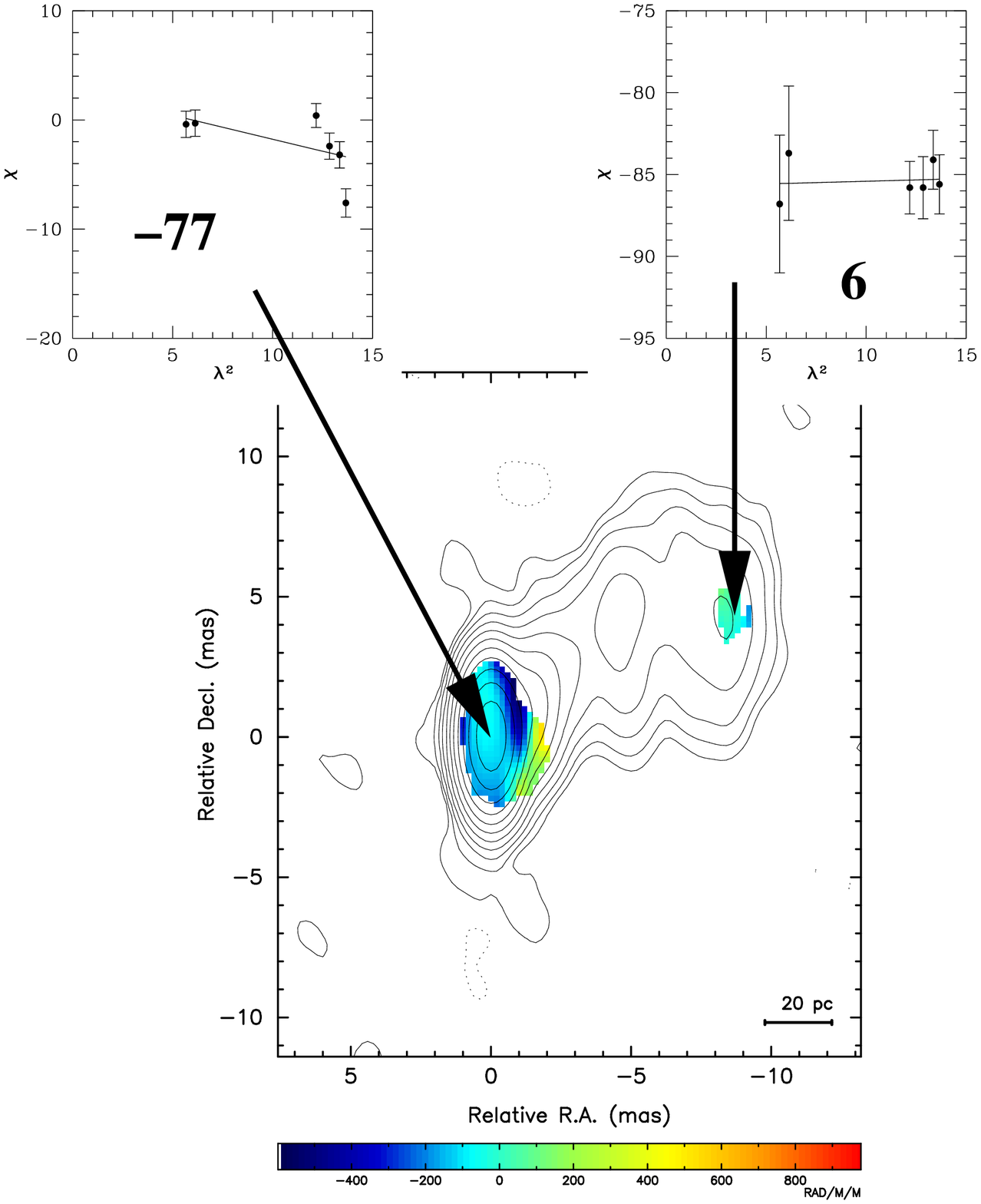}{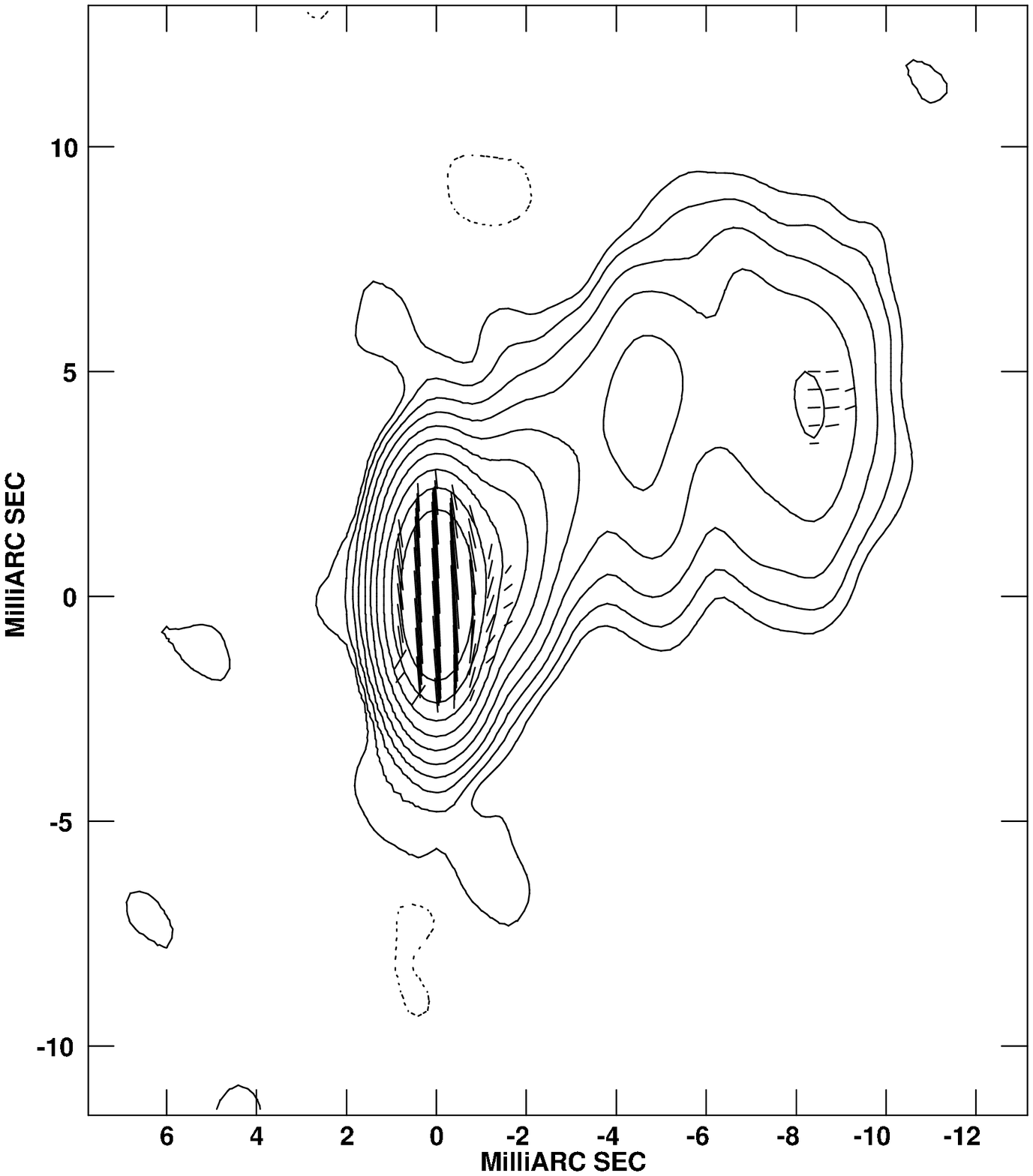}
\caption{(a) Rotation measure image (color) for 1055+018 overlaid on
Stokes I contours at 12.5 GHz. The inset is a plot of EVPA $\chi$
(deg) versus \l2 (cm$^2$). (b) Electric vectors (1 mas =
25 mJy beam$^{-1}$ polarized flux density) corrected for Faraday
Rotation overlaid on Stokes I contours. Contours start at 2.4 mJy
beam$^{-1}$ and increase by factors of two.}
\label{1055rm}
\end{figure}

\begin{figure}
\plotone{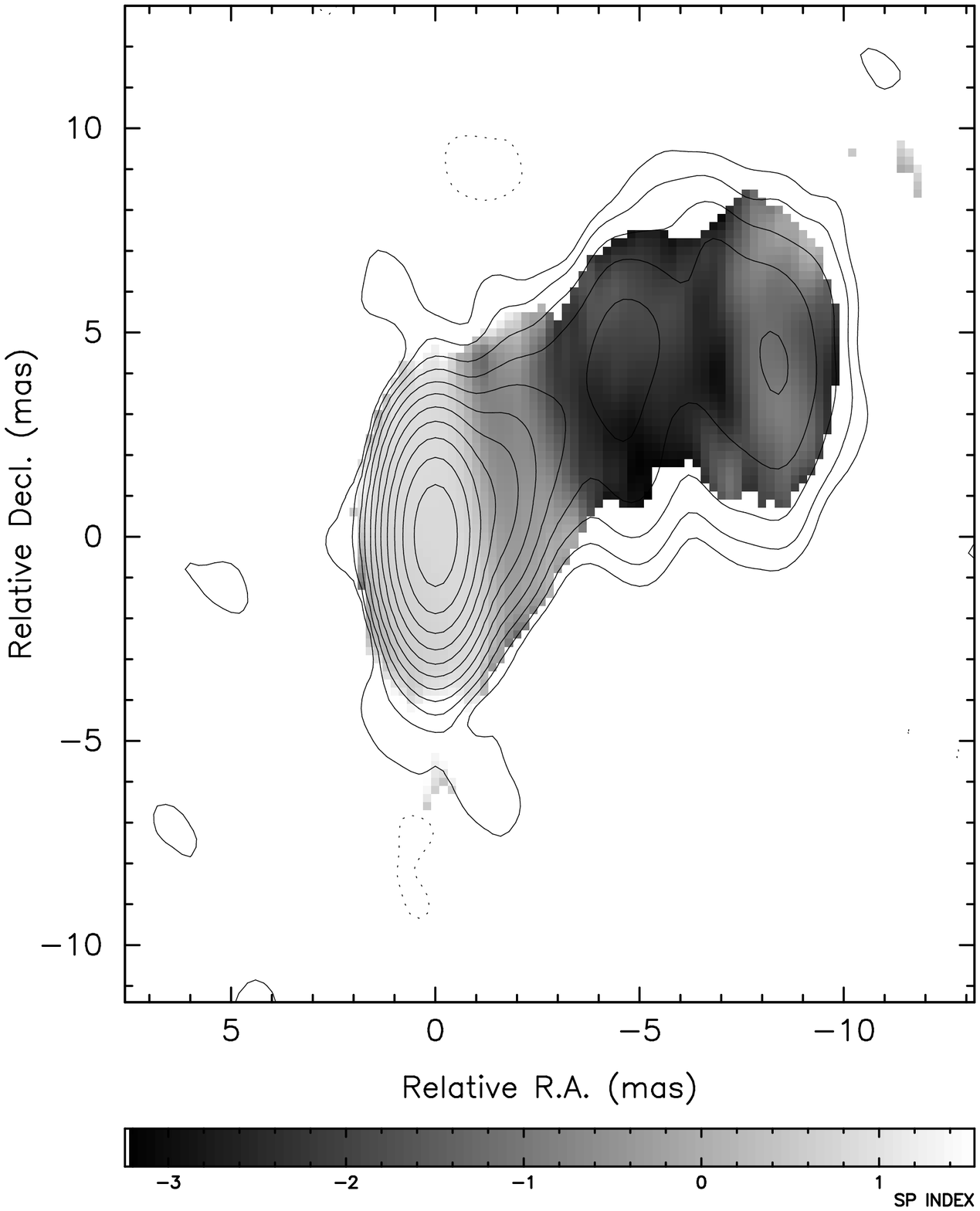}
\caption{Spectral index $\alpha_{12.1}^{8.1}$ plot for 1055+018 overlaid on
Stokes I contours at 12.5 GHz. Contours start at 2.4 mJy beam$^{-1}$ and
increase by factors of two.}
\label{1055si}
\end{figure}

An error in the observing schedule caused the loss of the 15 GHz data for 
1055+018, so the total intensity contours in Figs~\ref{1055rm}a \& b, 
and Fig.~\ref{1055si} are for 12.5 GHz. Table~\ref{big} shows the core 
of this BL Lacertae object is relatively weakly polarized at 8.1 GHz, 
and the core and jet RMs ($-77 \pm 25$ and $6 \pm 73$ \radm) are 
consistent with zero. Approximately 50\% of pixels have a
reduced $\chi^2$ which is not consistent with a $\lambda^2$ law. For
the same reason we reject the \l2\ law for the jet of 1055+018. The core is 
optically thick and the jet component 9 mas NW of the core 
is optically thin. The jet does not exhibit the interesting ``spine \& sheath'' 
polarization structure found by \citet{att} at 5 GHz. Our observations 
probably lack the sensitivity to reveal the sheath structure which Attridge, 
Roberts \& Wardle observed. 

\subsection{3C\,279}
\begin{figure}
\plottwo{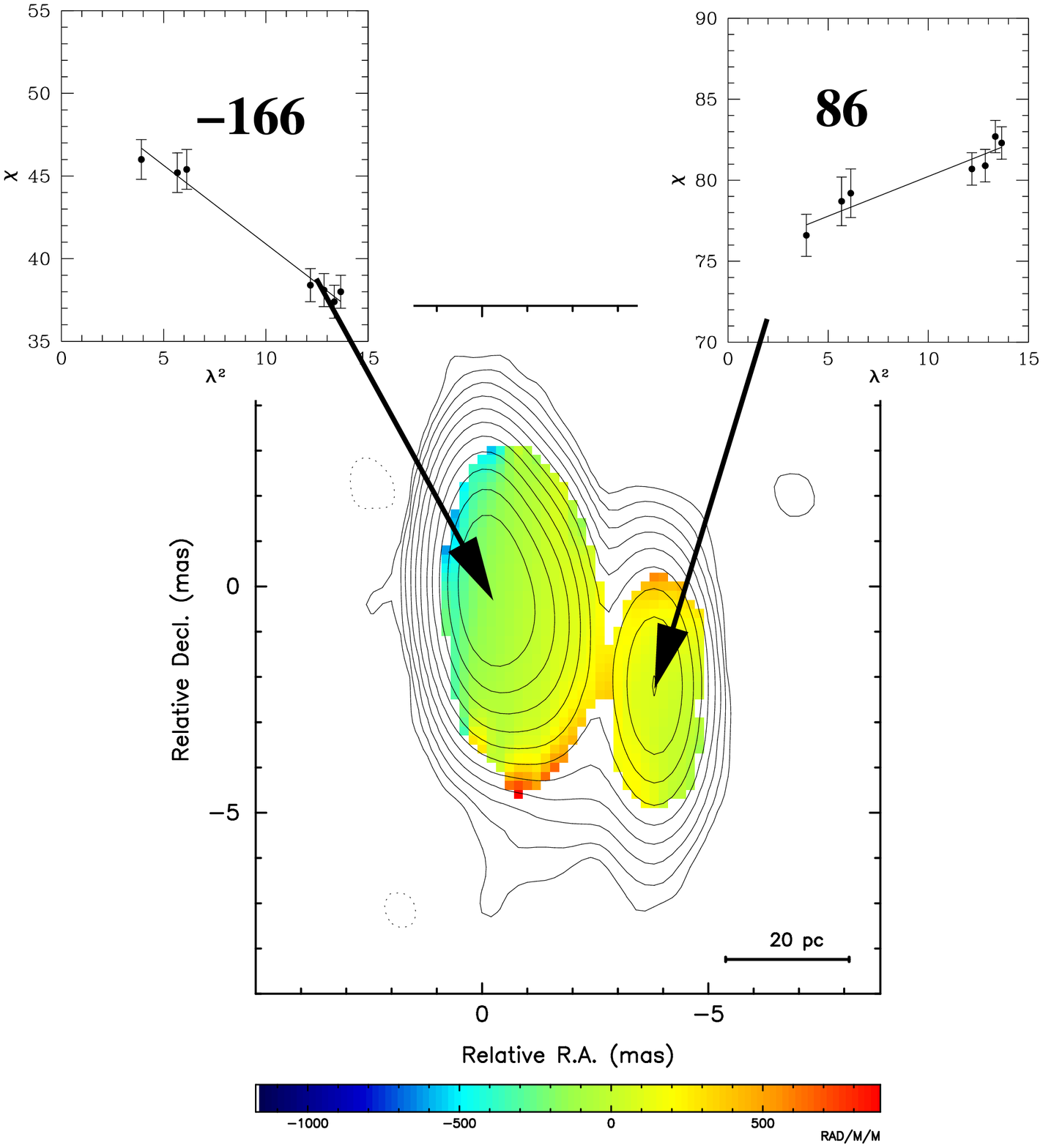}{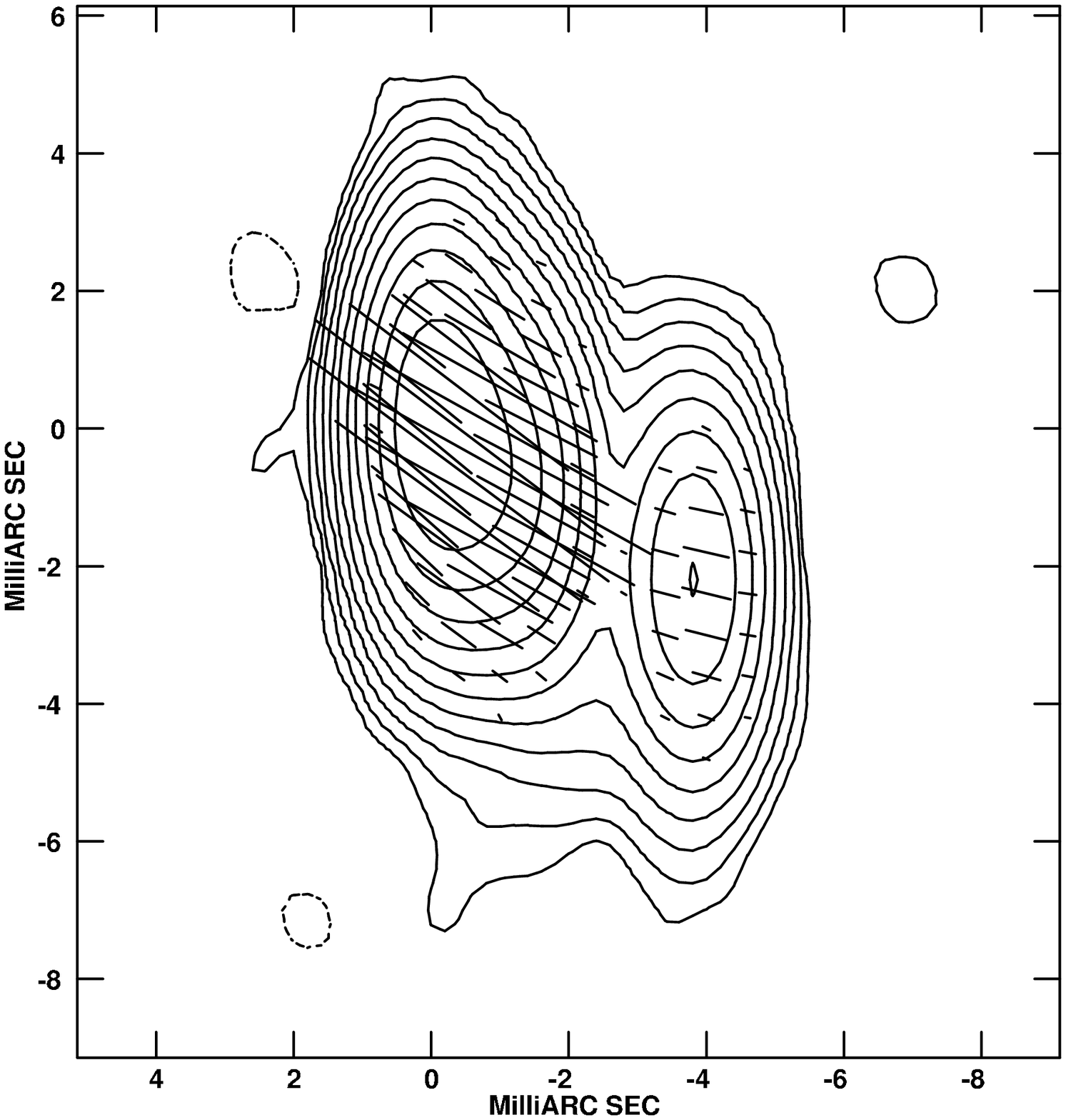}
\caption{(a) Rotation measure image (color) for 3C\,279 overlaid on
Stokes I contours at 15 GHz. The inset is a plot of EVPA $\chi$
(deg) versus \l2 (cm$^2$). (b) Electric vectors (1 mas =
250 mJy beam$^{-1}$ polarized flux density) corrected for Faraday
Rotation overlaid on Stokes I contours. Contours start at 5.3 mJy
beam$^{-1}$ and increase by factors of two.}
\label{279rm}
\end{figure}

\begin{figure}
\plotone{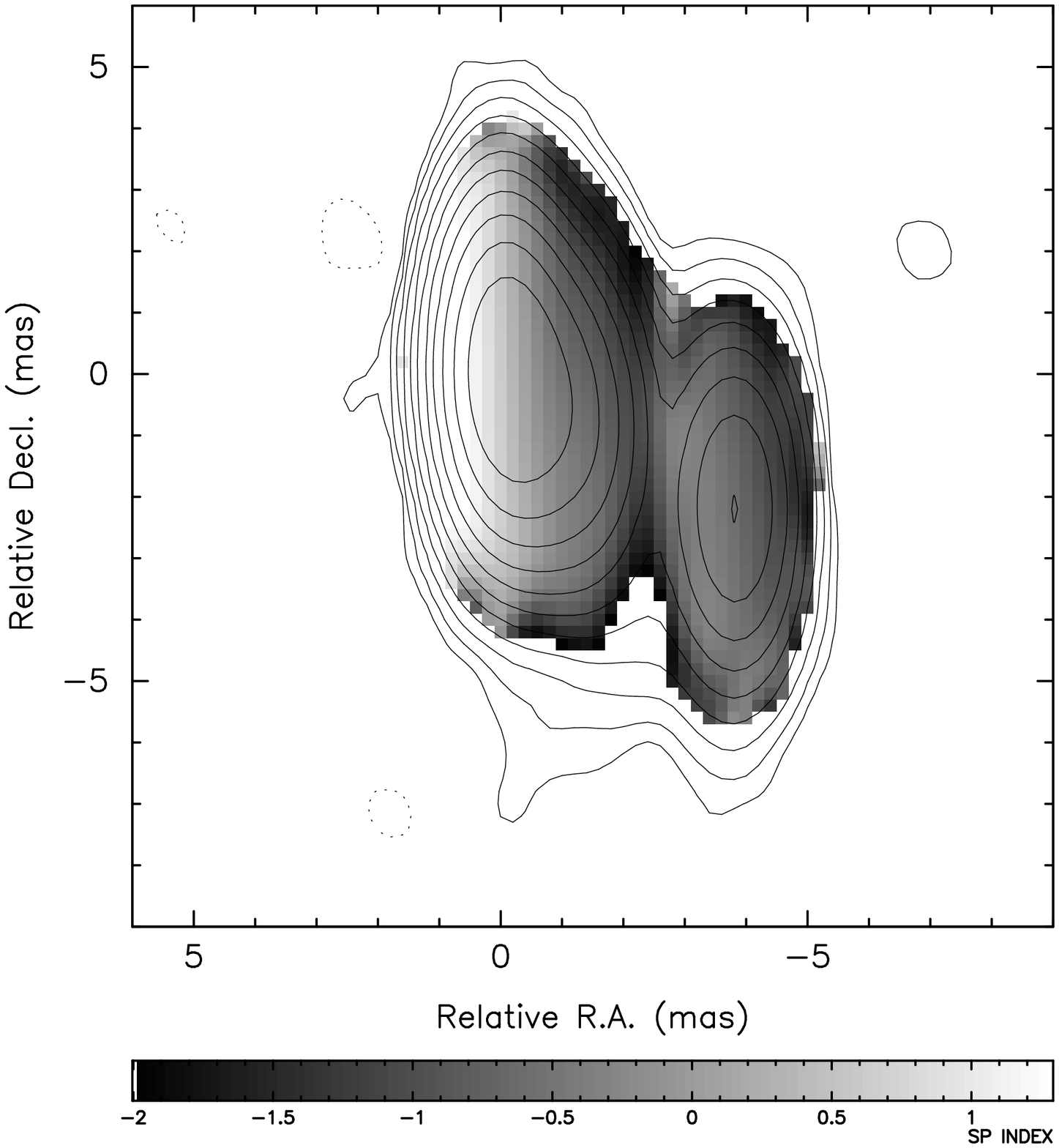}
\caption{Spectral index $\alpha_{12.1}^{8.1}$ plot for 3C\,279 overlaid on
Stokes I contours at 15 GHz. Contours start at 5.3 mJy beam$^{-1}$ and
increase by factors of two.}
\label{279si}
\end{figure}

This paper presents the fifth epoch of RM monitoring for the quasar 
3C\,279. Previous epochs were presented in \citet{tay98,tay00} 
and \citet{zt01,zt03}. The core RM (Fig.~\ref{279rm}a) is $-$166
$\pm$ 19 \radm\ and component C4 (4 mas west of the core) has an RM 86 
$\pm$ 21 \radm. The RM corrected EVPA (Fig.~\ref{279rm}b) of the core 
is 50\dg\ and for C4 is 76\dg. Within a milliarcsecond of the core the 
spectral index becomes negative (Fig.~\ref{279si}).

\subsection{B1546+027}
\begin{figure}
\plottwo{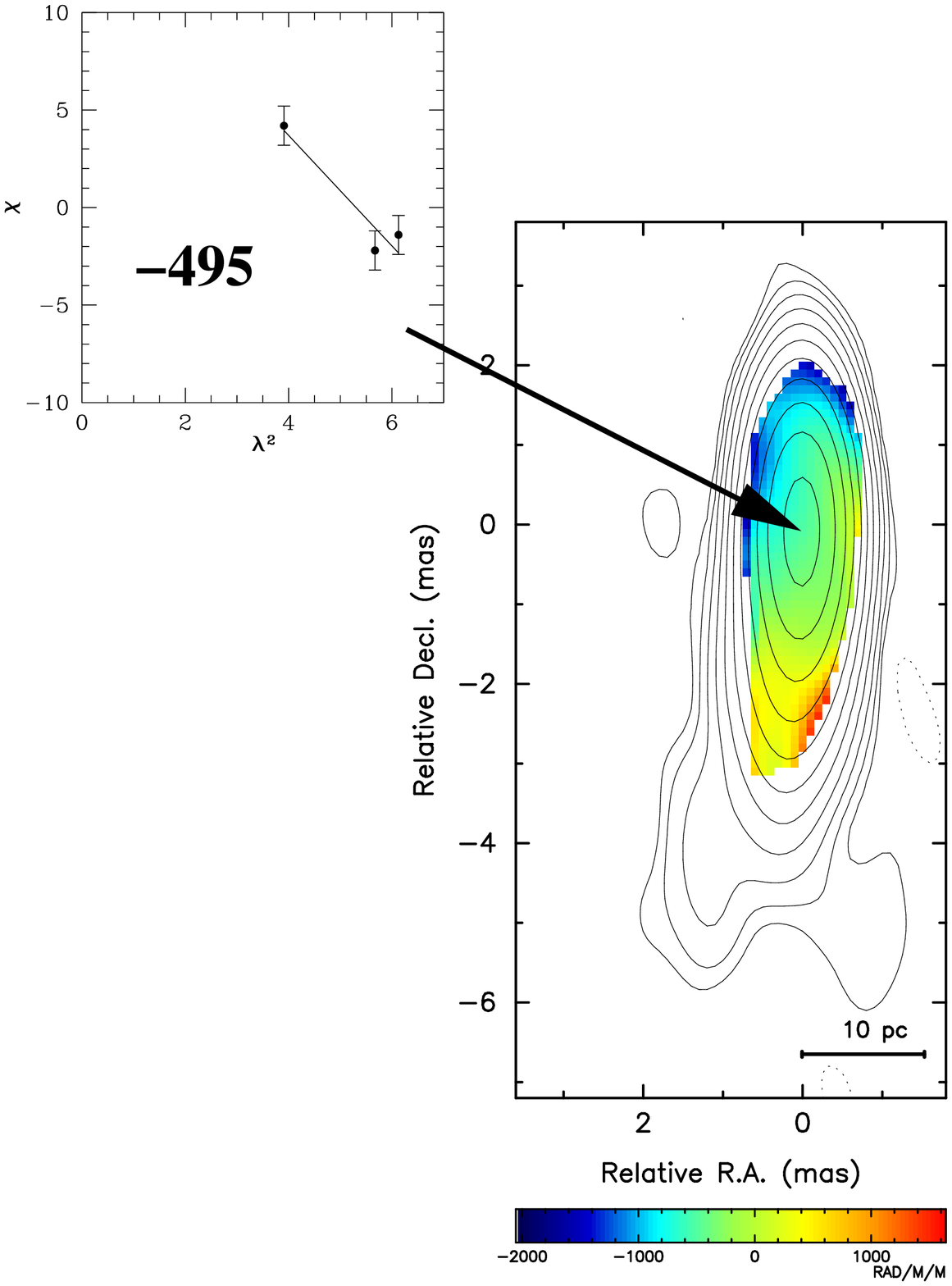}{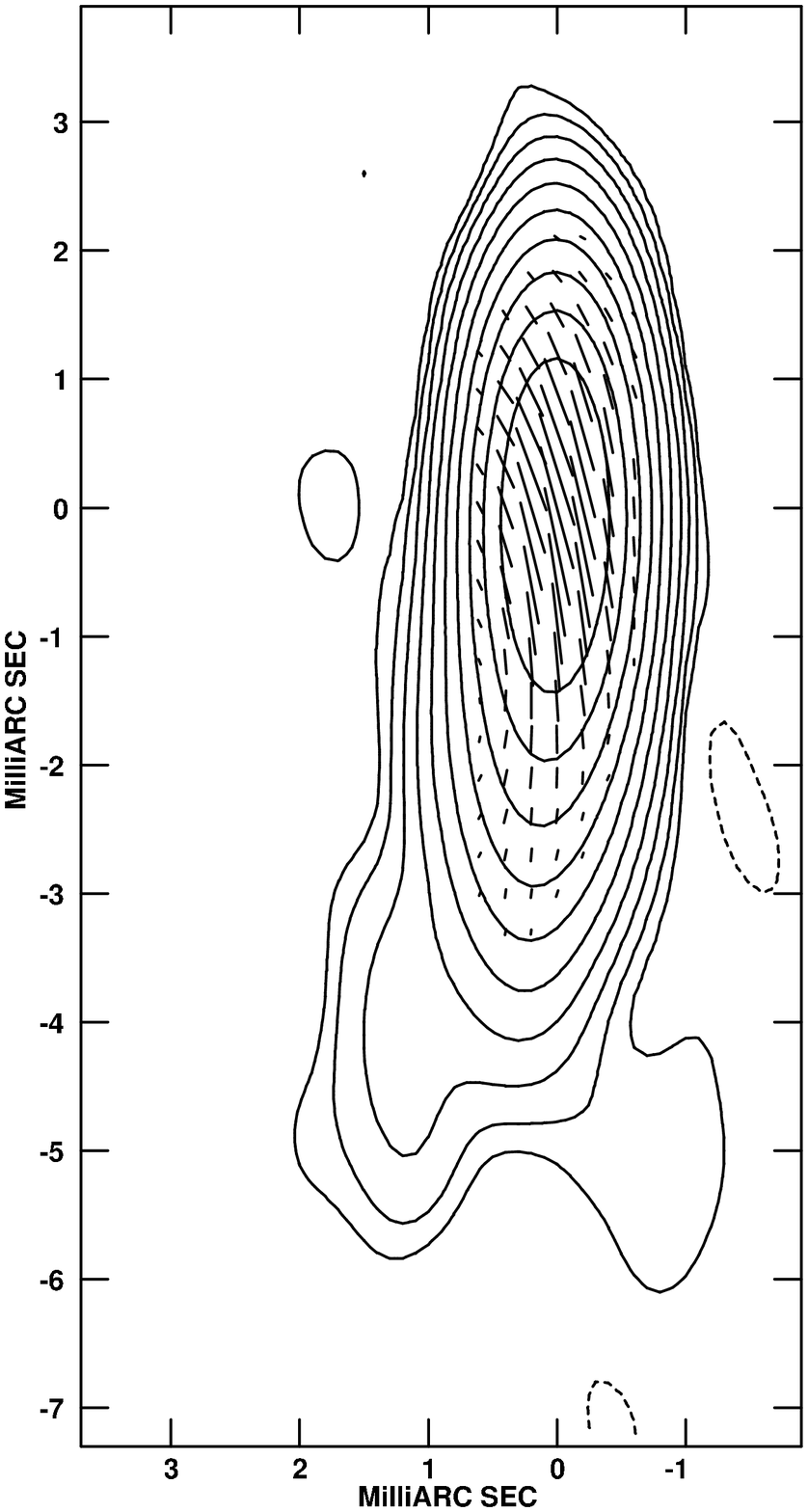}
\caption{(a) Rotation measure image (color) for 1546+027 overlaid on
Stokes I contours at 15 GHz. The inset is a plot of EVPA $\chi$
(deg) versus \l2 (cm$^2$). (b) Electric vectors (1 mas =
67 mJy beam$^{-1}$ polarized flux density) corrected for Faraday
Rotation overlaid on Stokes I contours. Contours start at 1.2 mJy
beam$^{-1}$ and increase by factors of two.}
\label{1546rm}
\end{figure}

\begin{figure}
\plotone{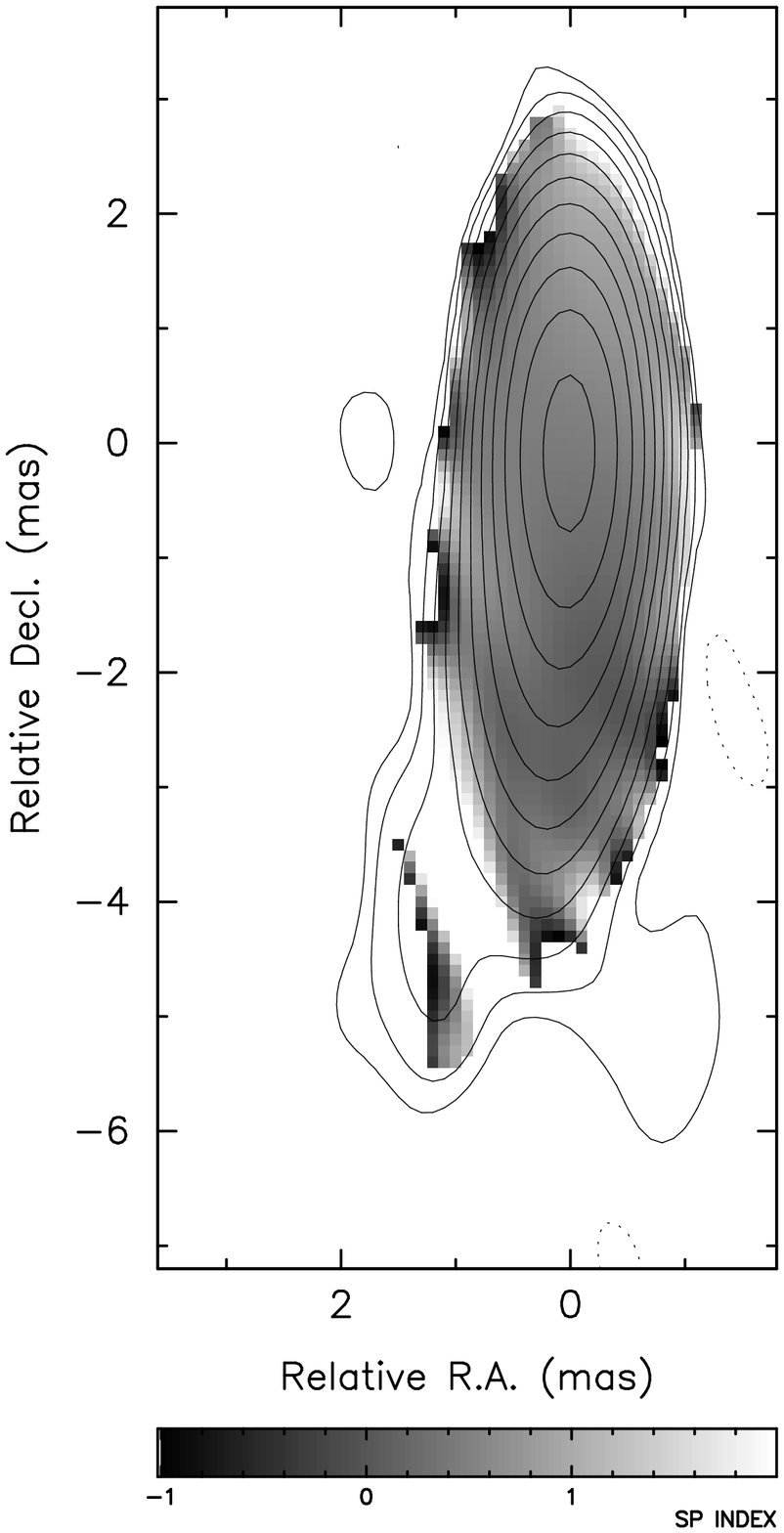}
\caption{Spectral index $\alpha_{15.1}^{12.5}$ plot for 1546+027 overlaid on
Stokes I contours at 15 GHz. Contours start at 1.2 mJy beam$^{-1}$ and
increase by factors of two.}
\label{1546si}
\end{figure}

A error in the observation caused the loss of the 8 GHz data 
for 1546+027. The RM in the core (Fig.~\ref{1546rm}a) of $-$495 
$\pm$ 105 \radm\ is over 15 and 12 GHz only. There is a change in 
the RM corrected EVPA from N$-$S as one proceeds along the jet 
(Fig.~\ref{1546rm}b). Fig.~\ref{1546si}, the spectral index map, 
shows that 1546+027 has an inverted spectrum up to 15 GHz.

\subsection{B1548+056}
\begin{figure}
\plottwo{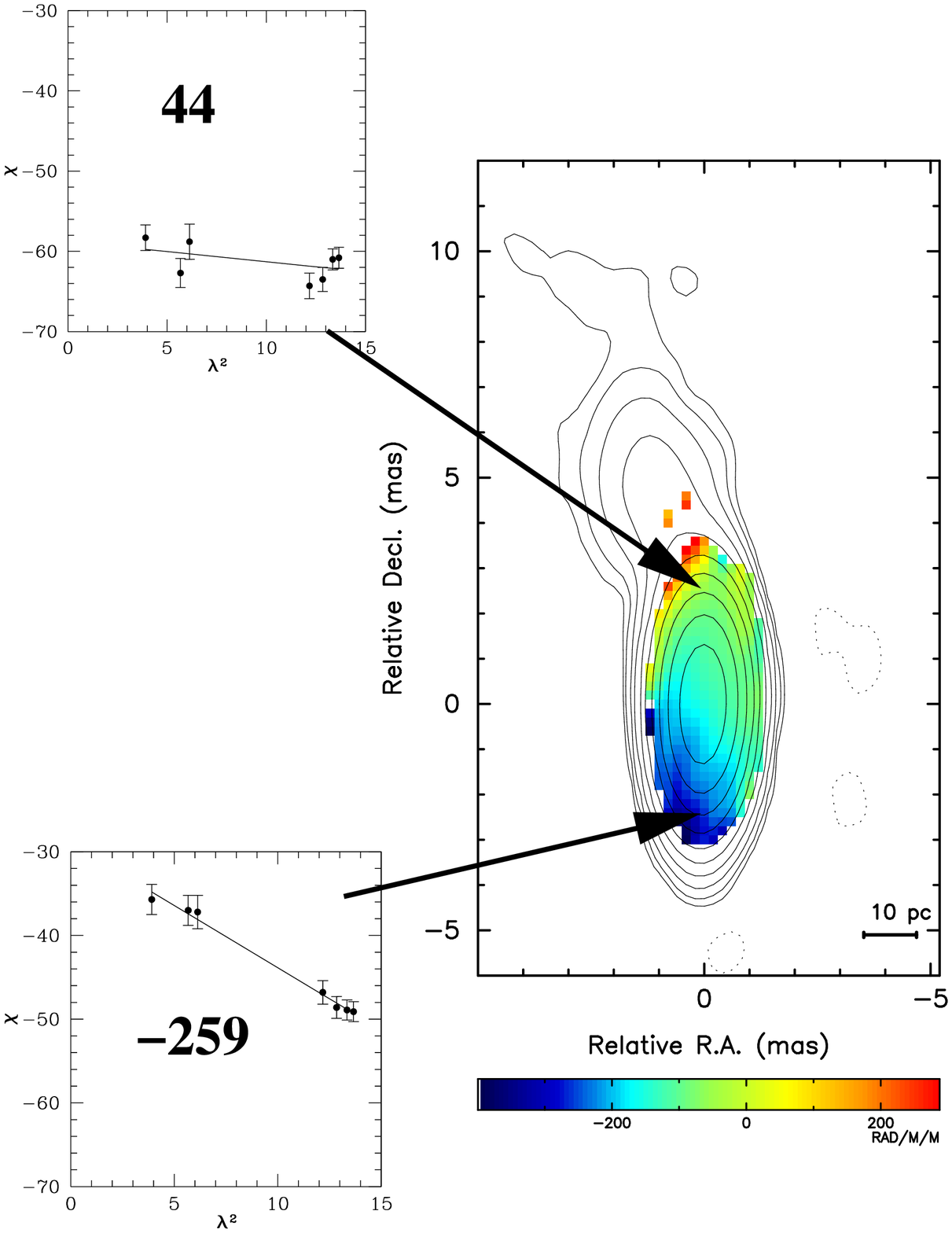}{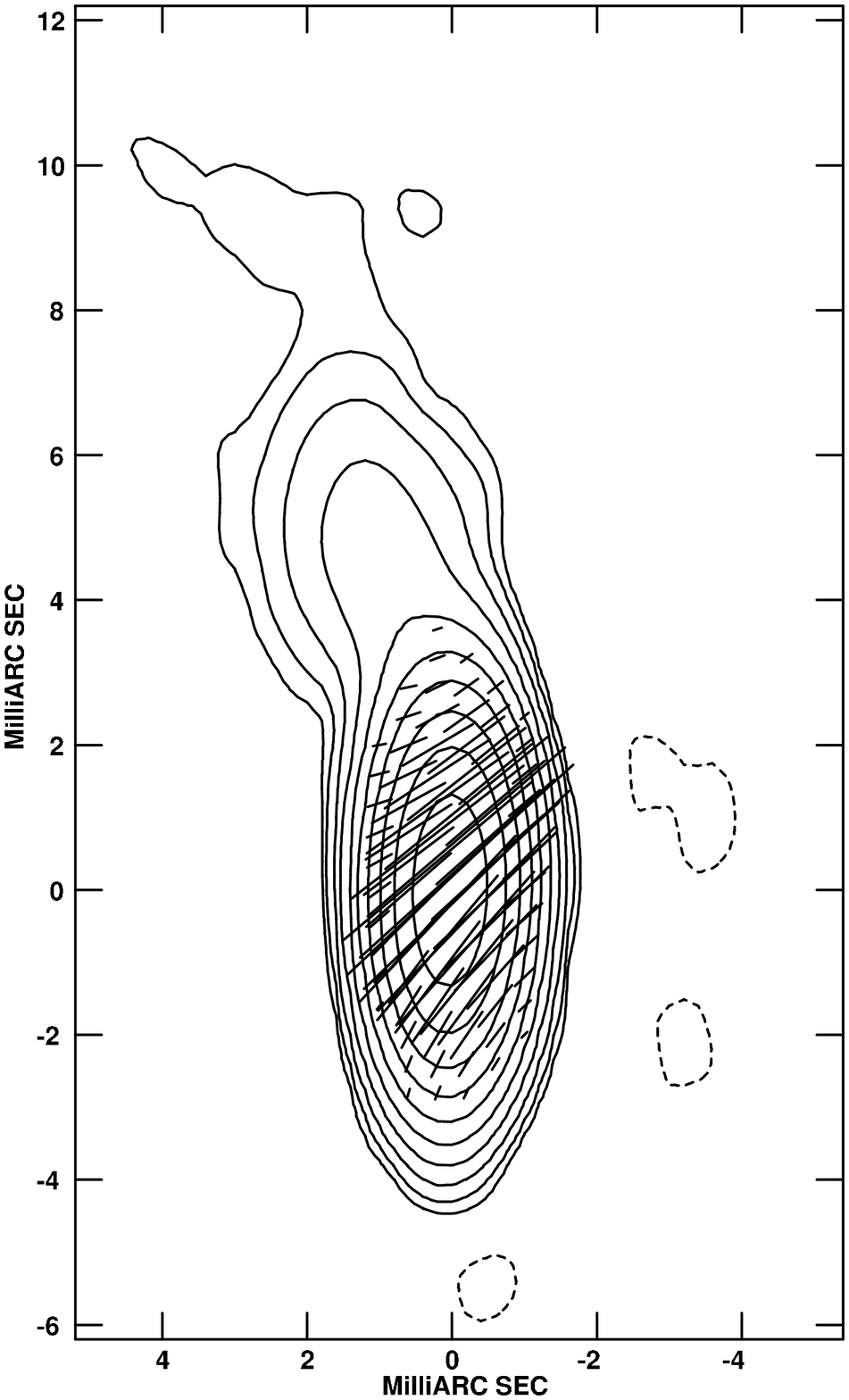}
\caption{(a) Rotation measure image (color) for 1548+056 overlaid on
Stokes I contours at 15 GHz. The inset is a plot of EVPA $\chi$
(deg) versus \l2 (cm$^2$). (b) Electric vectors (1 mas =
50 mJy beam$^{-1}$ polarized flux density) corrected for Faraday
Rotation overlaid on Stokes I contours. Contours start at 2.4 mJy
beam$^{-1}$ and increase by factors of two.}
\label{1548rm}
\end{figure}

\begin{figure}
\plotone{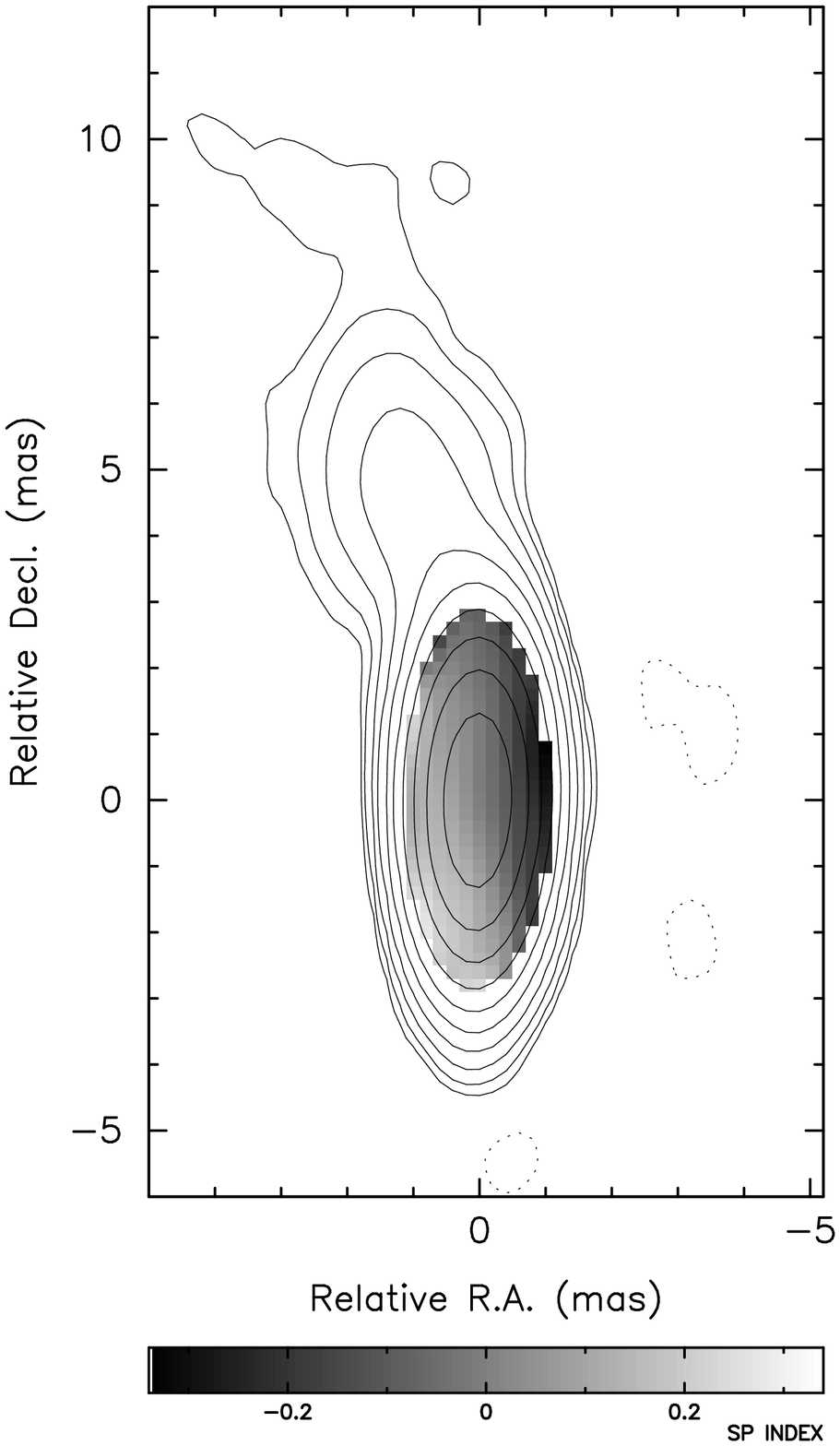}
\caption{Spectral index $\alpha_{12.1}^{8.1}$ plot for 1548+056 overlaid on
Stokes I contours at 15 GHz. Contours start at 2.4 mJy beam$^{-1}$ and
increase by factors of two.}
\label{1548si}
\end{figure}

A gradient in rotation measure is visible across 1548+056 from 
south$-$north in Fig.~\ref{1548rm}a. Three mas south of the peak 
the RM is $-$259 $\pm$ 27 \radm, while 3 mas north of the peak the 
RM has declined to 44 $\pm$ 59 \radm. This occurs over a projected 
distance of less than 60 pc. The RM corrected electric vectors 
maintain a roughly constant orientation across the source 
(Fig.~\ref{1548rm}b). There is a flat spectral index along the RM 
gradient as shown in Fig.~\ref{1548si}. 

\subsection{B1741$-$038}
\begin{figure}
\plottwo{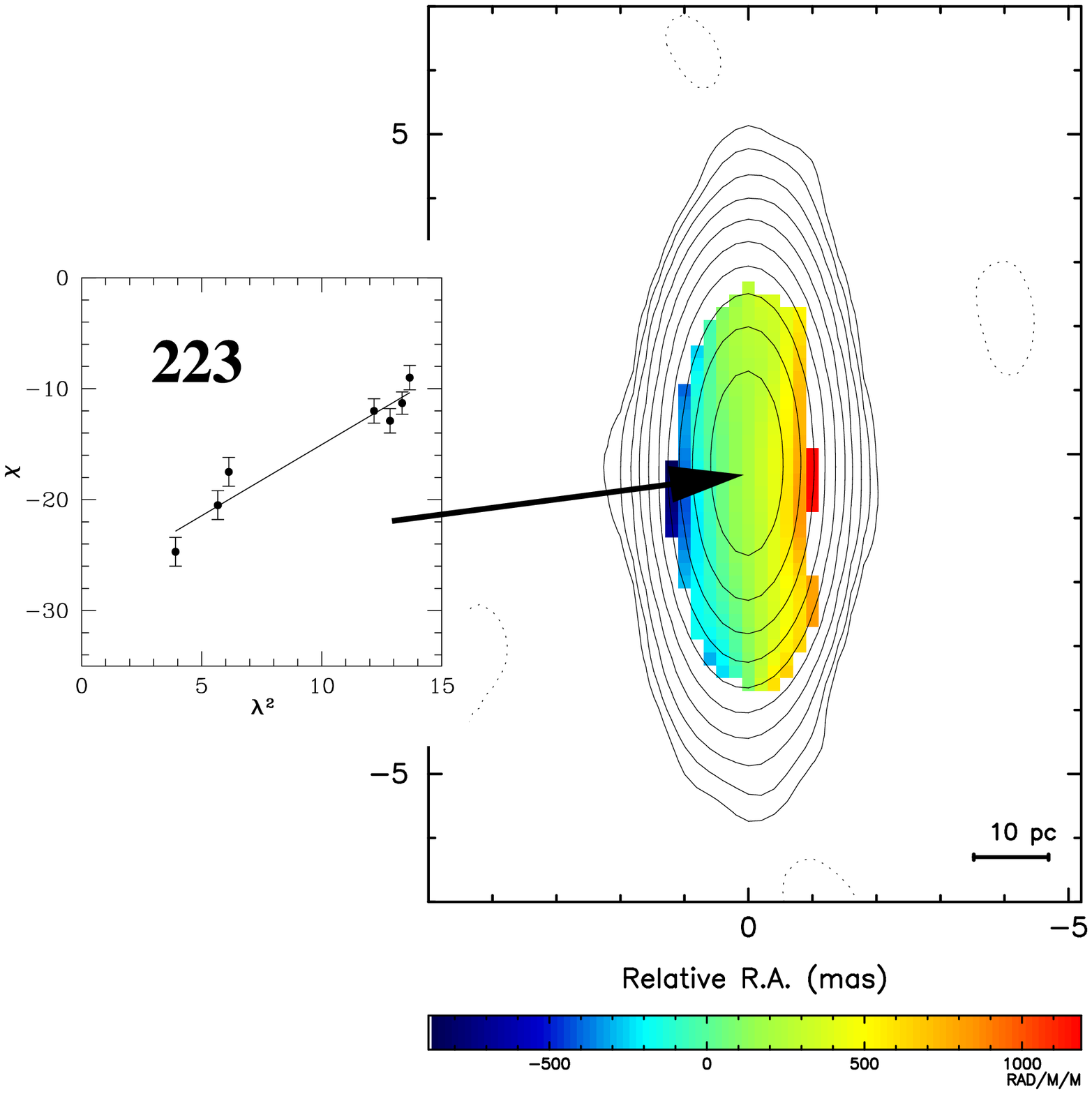}{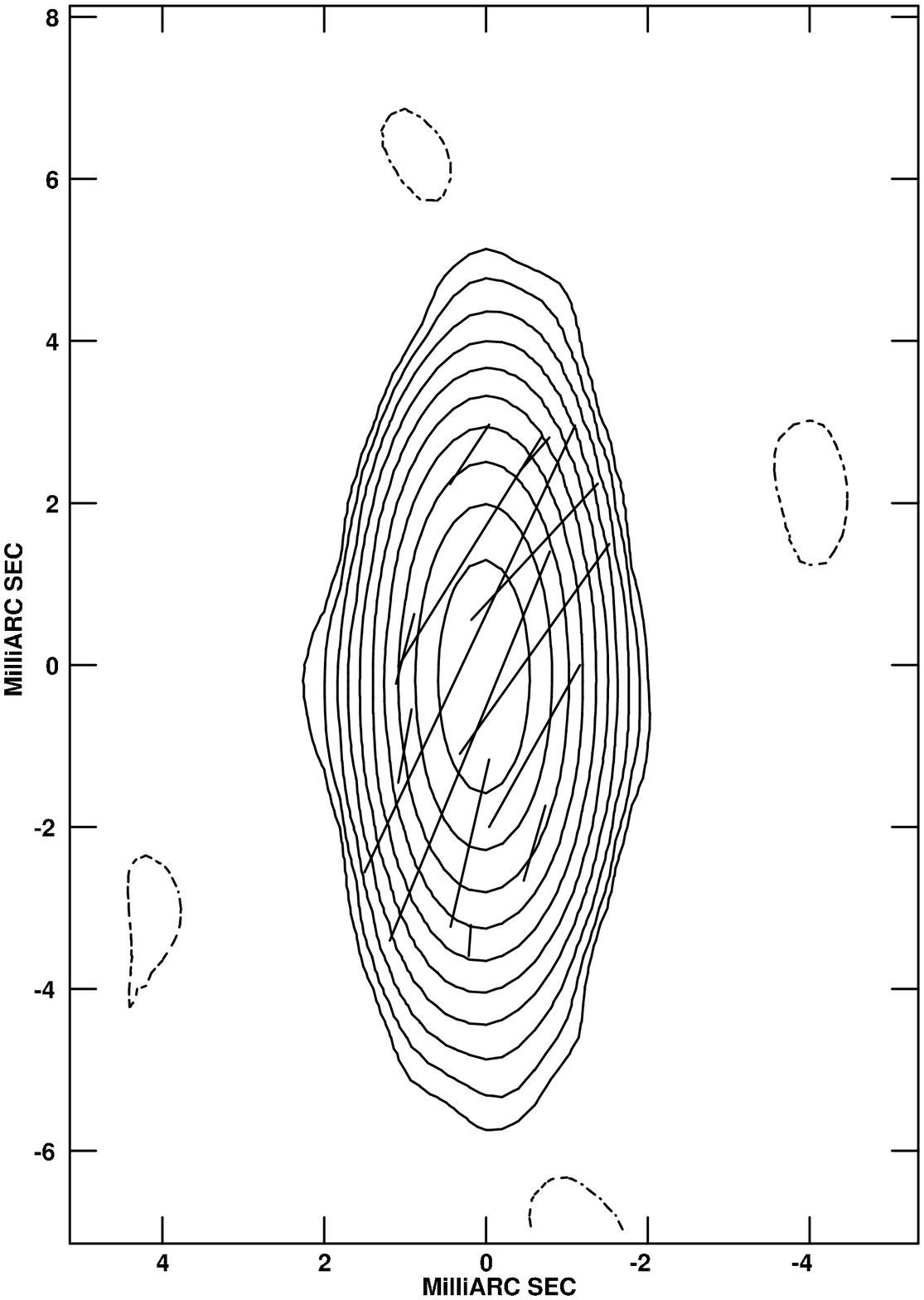}
\caption{(a) Rotation measure image (color) for 1741$-$038 overlaid on
Stokes I contours at 15 GHz. The inset is a plot of EVPA $\chi$
(deg) versus \l2 (cm$^2$). (b) Electric vectors (1 mas =
100 mJy beam$^{-1}$ polarized flux density) corrected for Faraday
Rotation overlaid on Stokes I contours. Contours start at 5.4 mJy
beam$^{-1}$ and increase by factors of two.}
\label{1741rm}
\end{figure}

\begin{figure}
\plotone{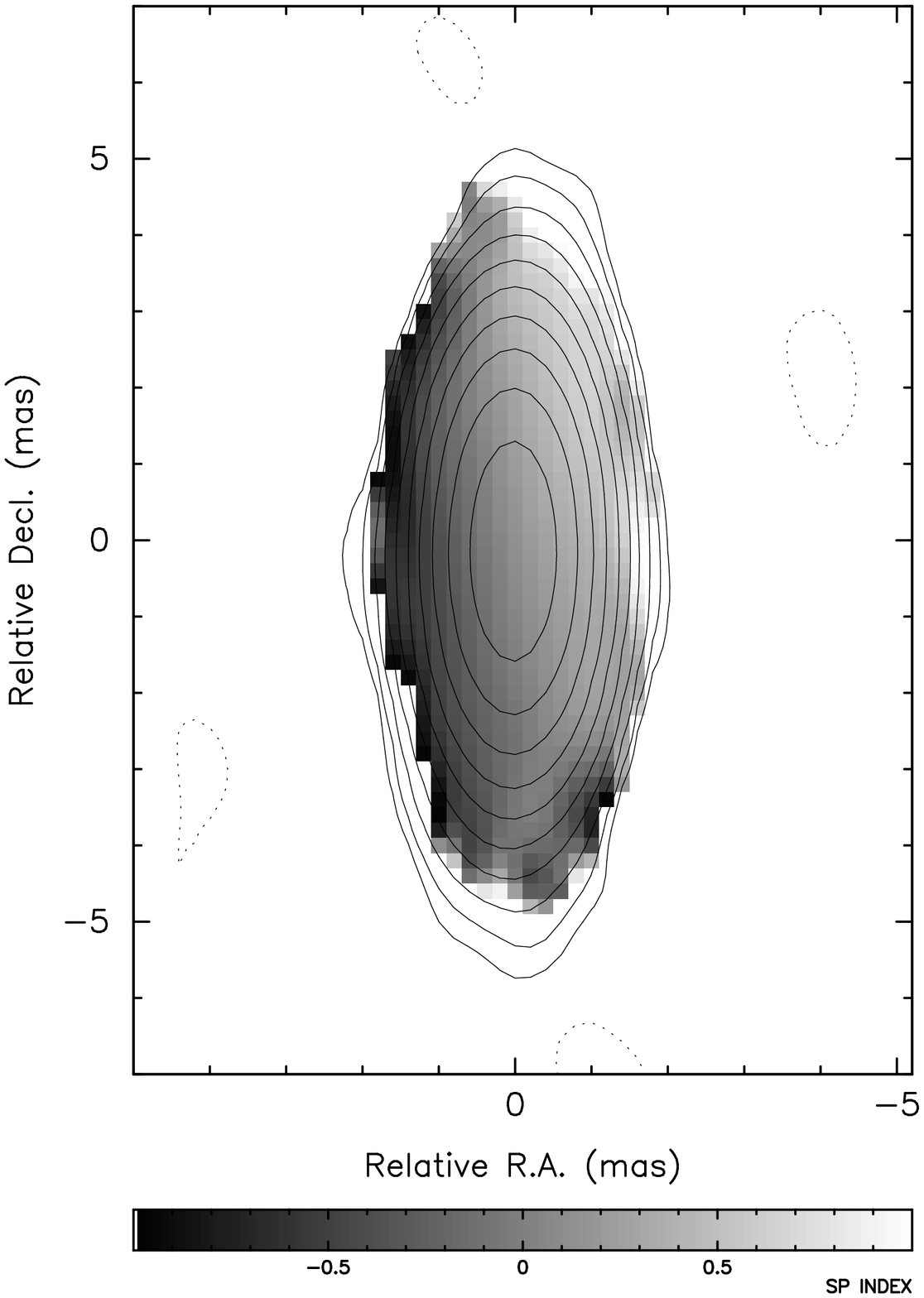}
\caption{Spectral index $\alpha_{12.1}^{8.1}$ plot for 1741$-$038 overlaid on
Stokes I contours at 15 GHz. Contours start at 5.4 mJy beam$^{-1}$ and
increase by factors of two.}
\label{1741si}
\end{figure}

1741$-$038 was one of the first three sources detected with a Space 
VLBI experiment \citep{levy}. This quasar is essentially unresolved 
(Fig.~\ref{1741rm}a), and has a core RM of 223 $\pm$ 20 \radm. 
The RM corrected electric vectors are oriented along a SE-NW axis 
(Fig.~\ref{1741rm}b). The spectrum steepens from S$-$N as shown in 
Fig.~\ref{1741si}.

\subsection{B1749+096}
\begin{figure}
\plottwo{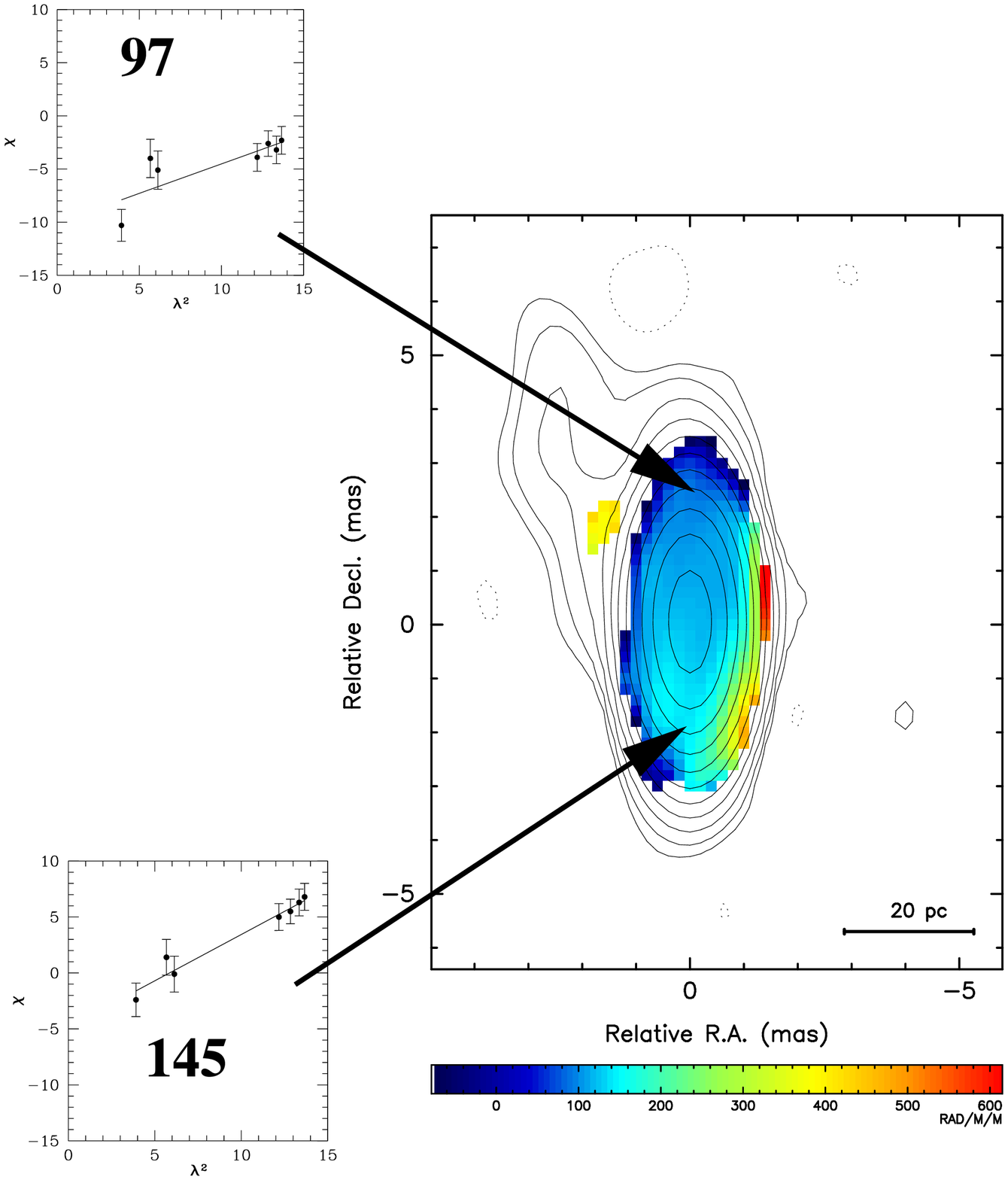}{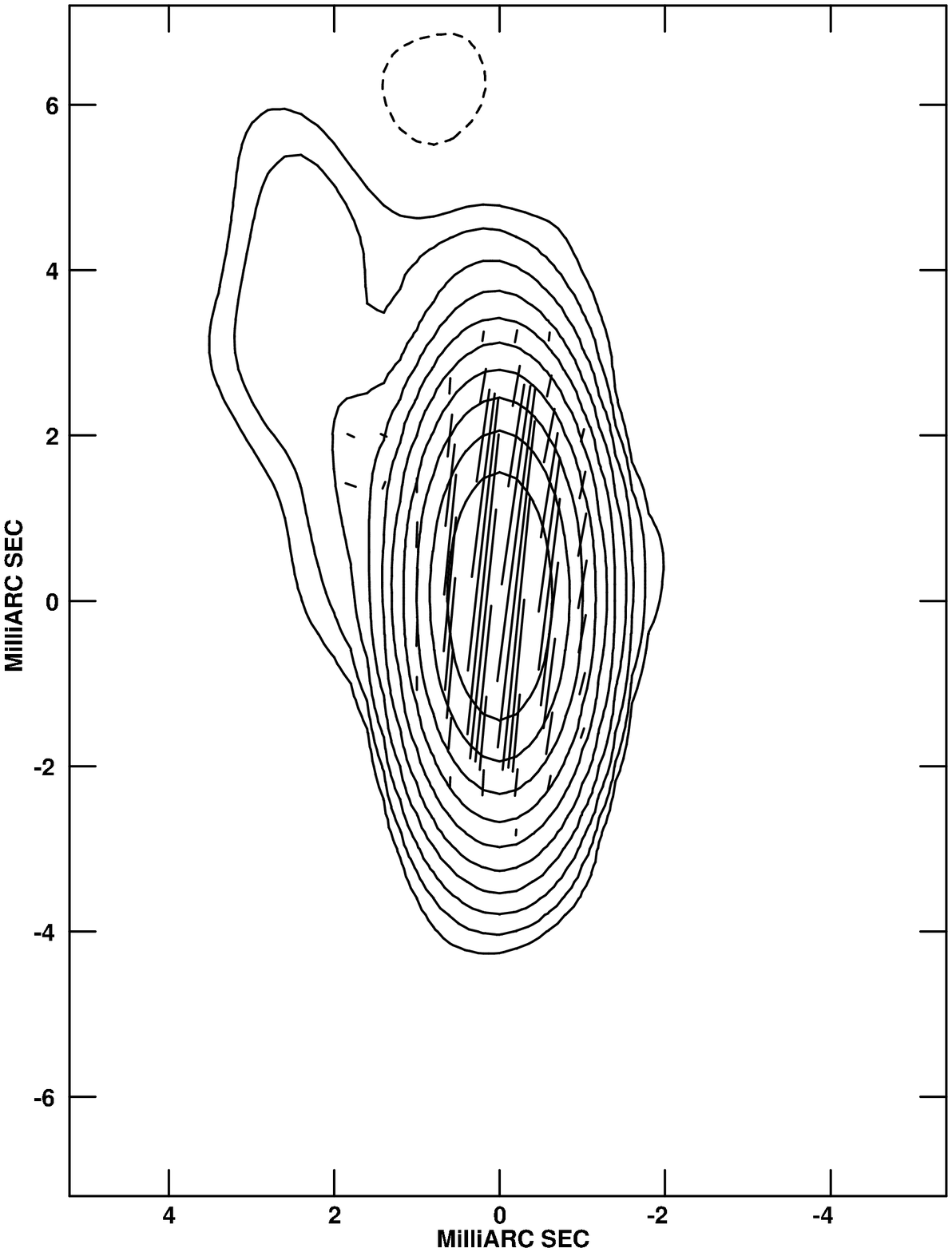}
\caption{(a) Rotation measure image (color) for 1749+096 overlaid on
Stokes I contours at 15 GHz. The inset is a plot of EVPA $\chi$
(deg) versus \l2 (cm$^2$). (b) Electric vectors (1 mas =
67 mJy beam$^{-1}$ polarized flux density) corrected for Faraday
Rotation overlaid on Stokes I contours. Contours start at 2.1 mJy
beam$^{-1}$ and increase by factors of two.}
\label{1749rm}
\end{figure}

\begin{figure}
\plotone{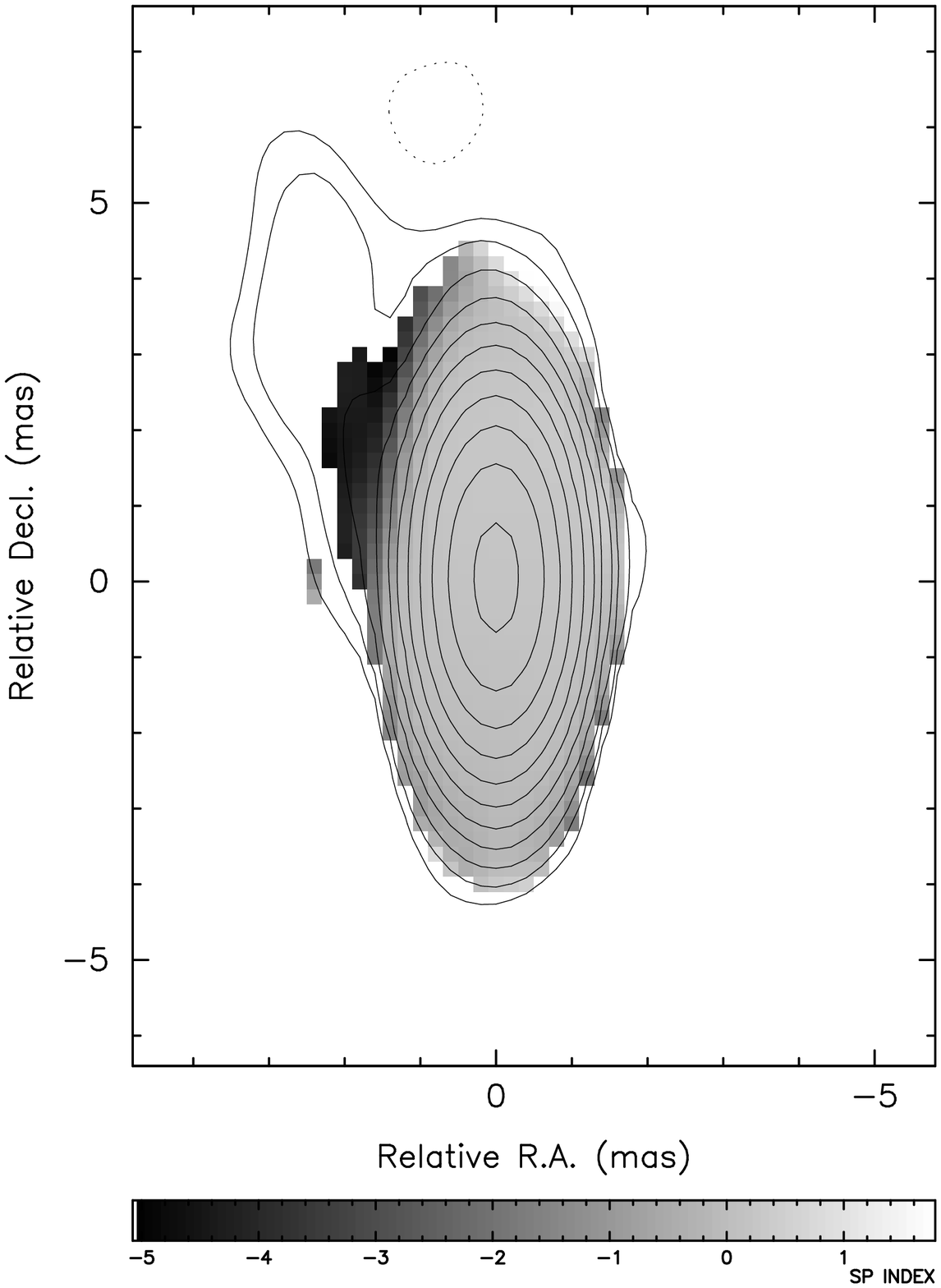}
\caption{Spectral index $\alpha_{12.1}^{8.1}$ plot for 1749+096 overlaid on
Stokes I contours at 15 GHz. Contours start at 2.1 mJy beam$^{-1}$ and
increase by factors of two.}
\label{1749si}
\end{figure}

The BL Lac object 1749+096 has a fairly uniform RM distribution. 
The fits in the inset plots of Fig.~\ref{1749rm}a have RMs of 145 
$\pm$ 24 \radm\ and 97 $\pm$ 25 \radm, which are essentially the same 
within the errors. The RM corrected electric vectors appear roughly 
perpendicular to the projected direction of the jet 
(Fig.~\ref{1749rm}b). 1749+096 is dominated by a 
flat spectrum core (Fig.~\ref{1749si}), thus the magnetic vectors 
are parallel to the electric vectors \citep{hugh} in Fig.~\ref{1749rm}b.

\subsection{B2021+317}
\begin{figure}
\plottwo{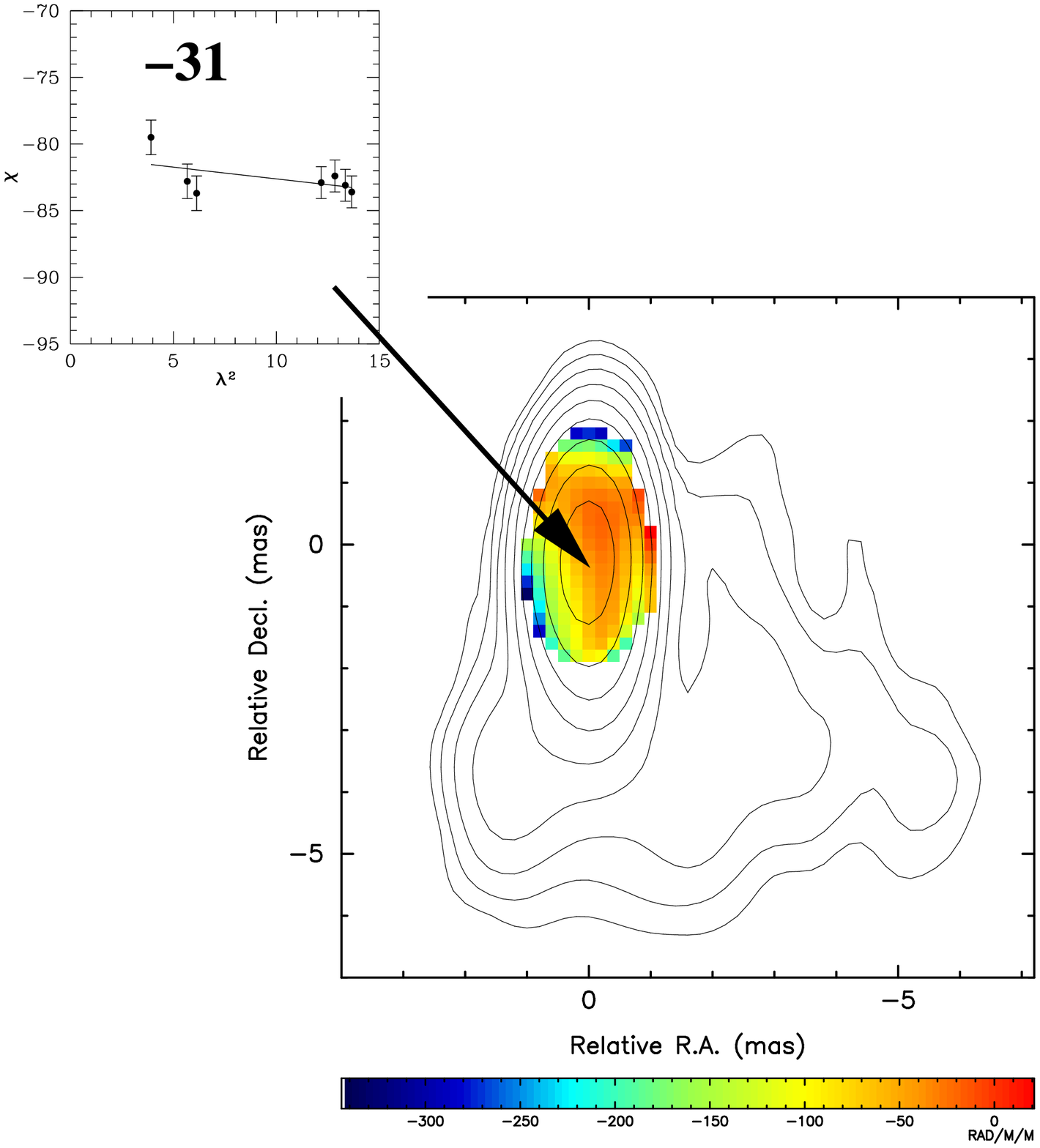}{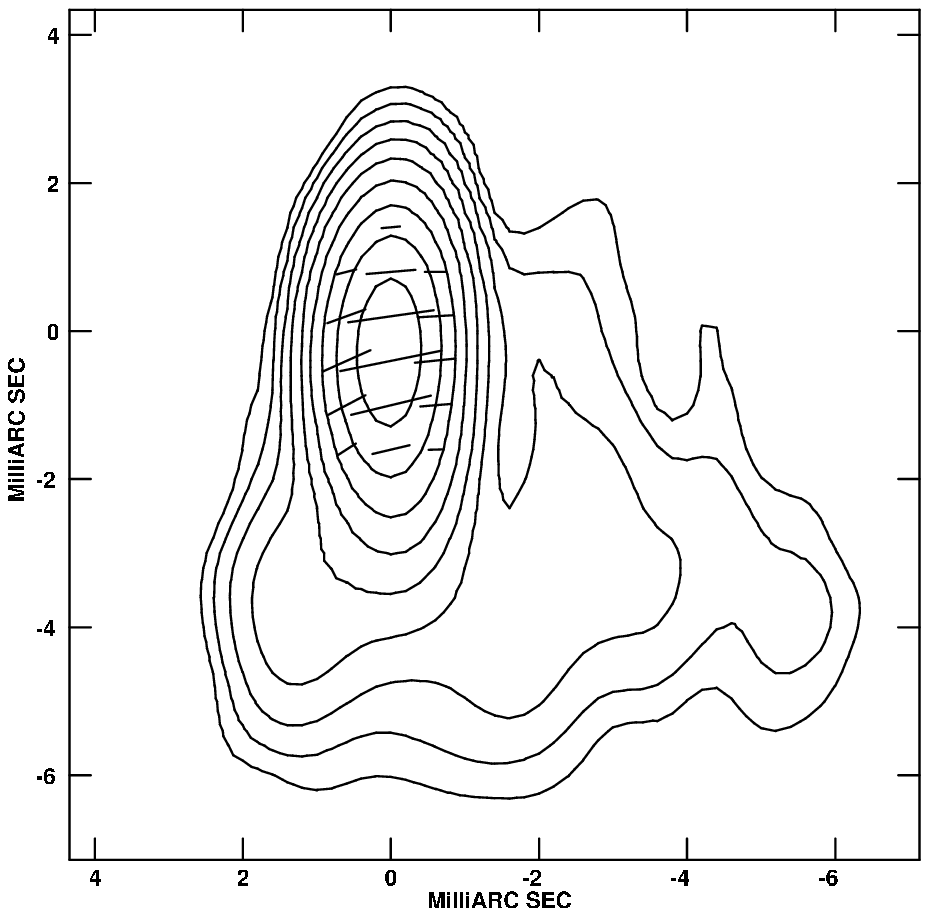}
\caption{(a) Rotation measure image (color) for 2021+317 overlaid on
Stokes I contours at 15 GHz. The inset is a plot of EVPA $\chi$
(deg) versus \l2 (cm$^2$). (b) Electric vectors (1 mas =
12.5 mJy beam$^{-1}$ polarized flux density) corrected for Faraday
Rotation overlaid on Stokes I contours. Contours start at 1.2 mJy
beam$^{-1}$ and increase by factors of two.}
\label{2021rm}
\end{figure}

\begin{figure}
\plotone{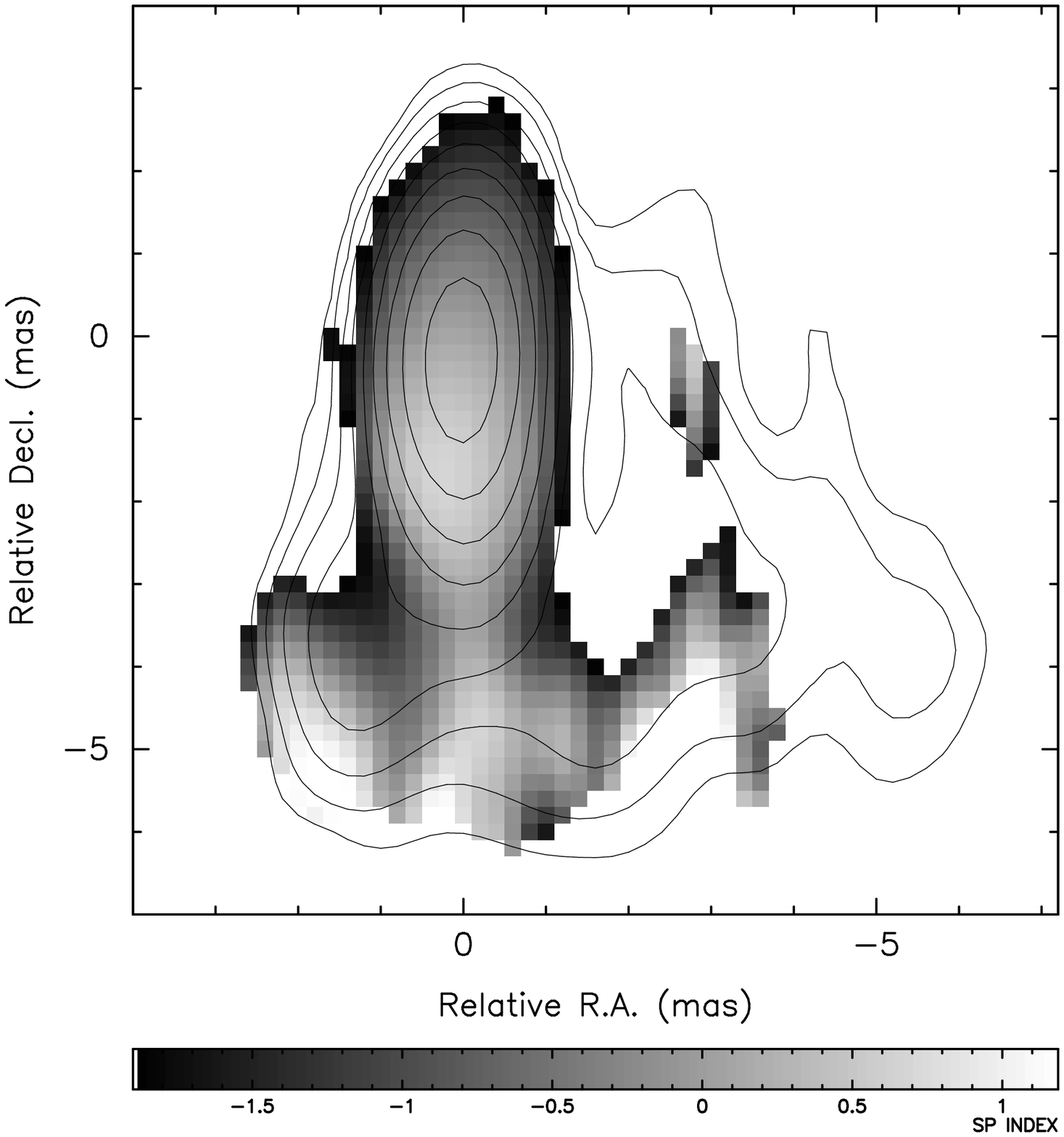}
\caption{Spectral index $\alpha_{12.1}^{8.1}$ plot for 2021+317 overlaid on
Stokes I contours at 15 GHz. Contours start at 1.2 mJy beam$^{-1}$ and
increase by factors of two.}
\label{2021si}
\end{figure}

This source lacks an optical counterpart, and has an RM consistent 
with zero ($-$31 $\pm$ 21 \radm) in Fig~\ref{2021rm}a. 
The NRAO VLA Sky Survey \citep[NVSS;][]{nvss} image shows a jet extending 
2 arcminutes to the NE, an extreme misalignment with the 
structure seen on parsec scales. The jet has a very diffuse and poorly 
ordered structure. The RM corrected electric vectors are oriented E-W 
(Fig~\ref{2021rm}b). The core is optically thick 
(Fig.~\ref{2021si}), and the magnetic vectors are therefore parallel to 
the electric vectors in Fig~\ref{2021rm}b.

\subsection{B2201+315}
\begin{figure}
\plottwo{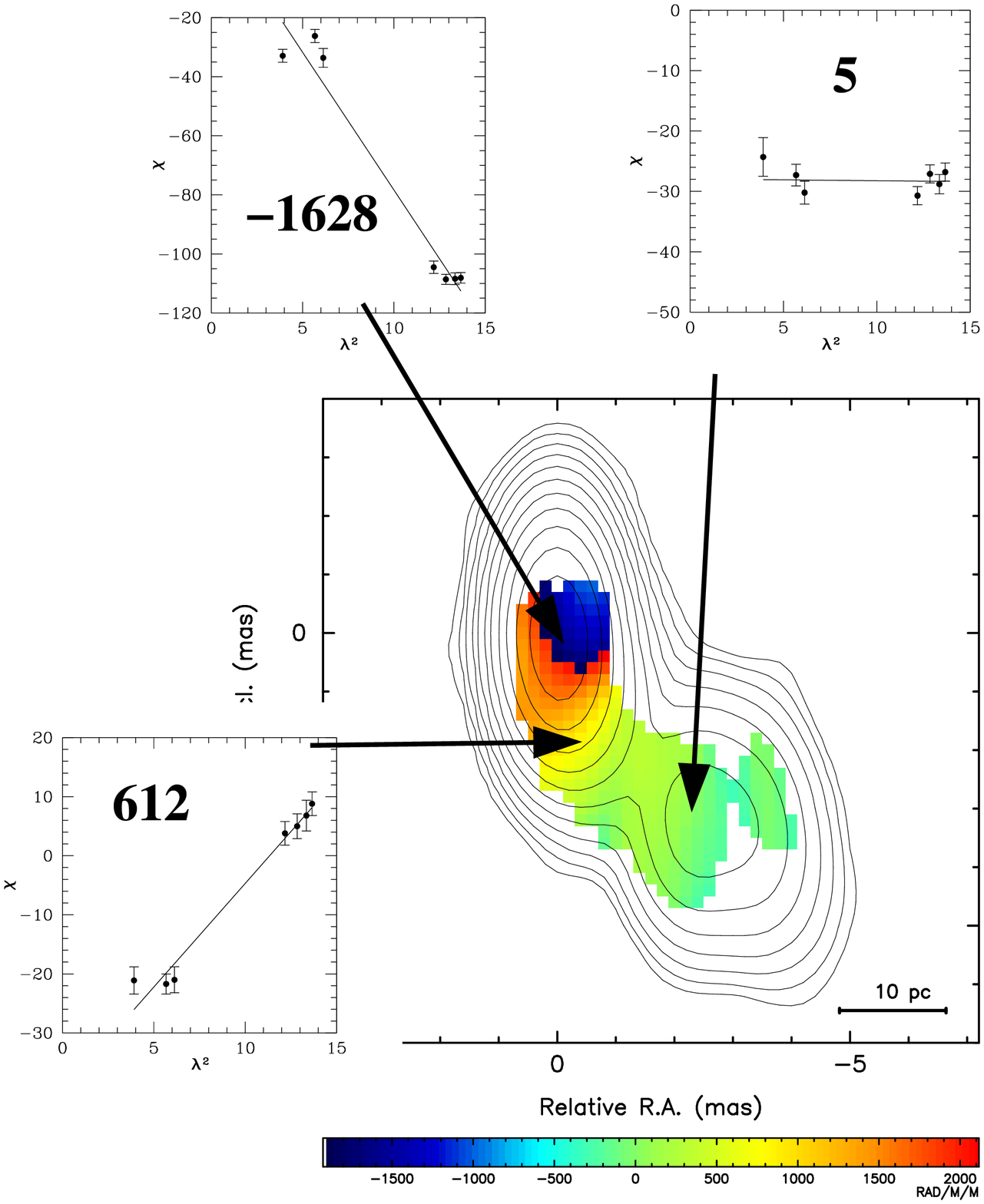}{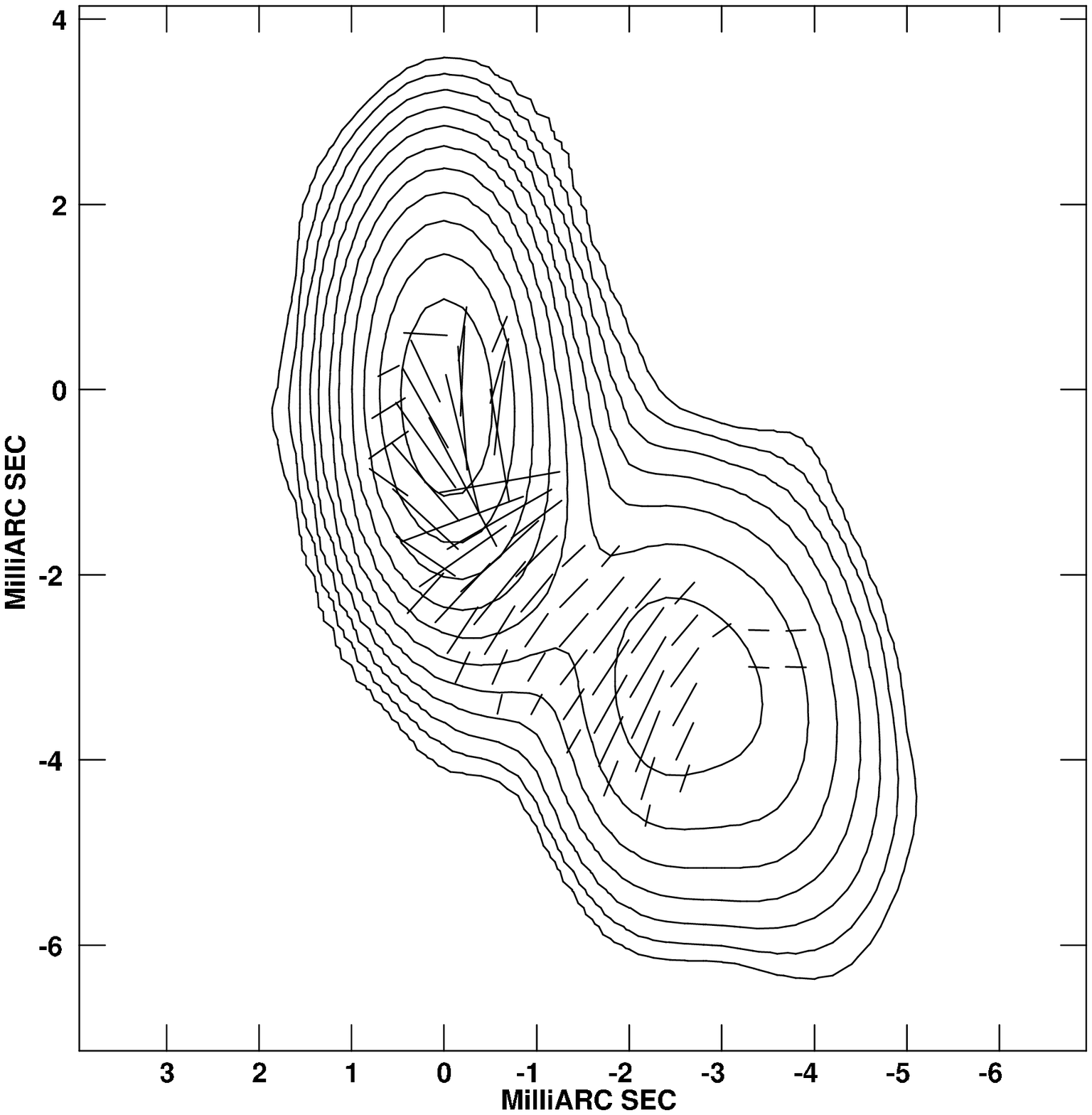}
\caption{(a) Rotation measure image (color) for 2201+315 overlaid on
Stokes I contours at 15 GHz. The inset is a plot of EVPA $\chi$
(deg) versus \l2 (cm$^2$). (b) Electric vectors (1 mas =
10 mJy beam$^{-1}$ polarized flux density) corrected for Faraday
Rotation overlaid on Stokes I contours. Contours start at 1.2 mJy
beam$^{-1}$ and increase by factors of two.}
\label{2201rm}
\end{figure}

\begin{figure}
\plotone{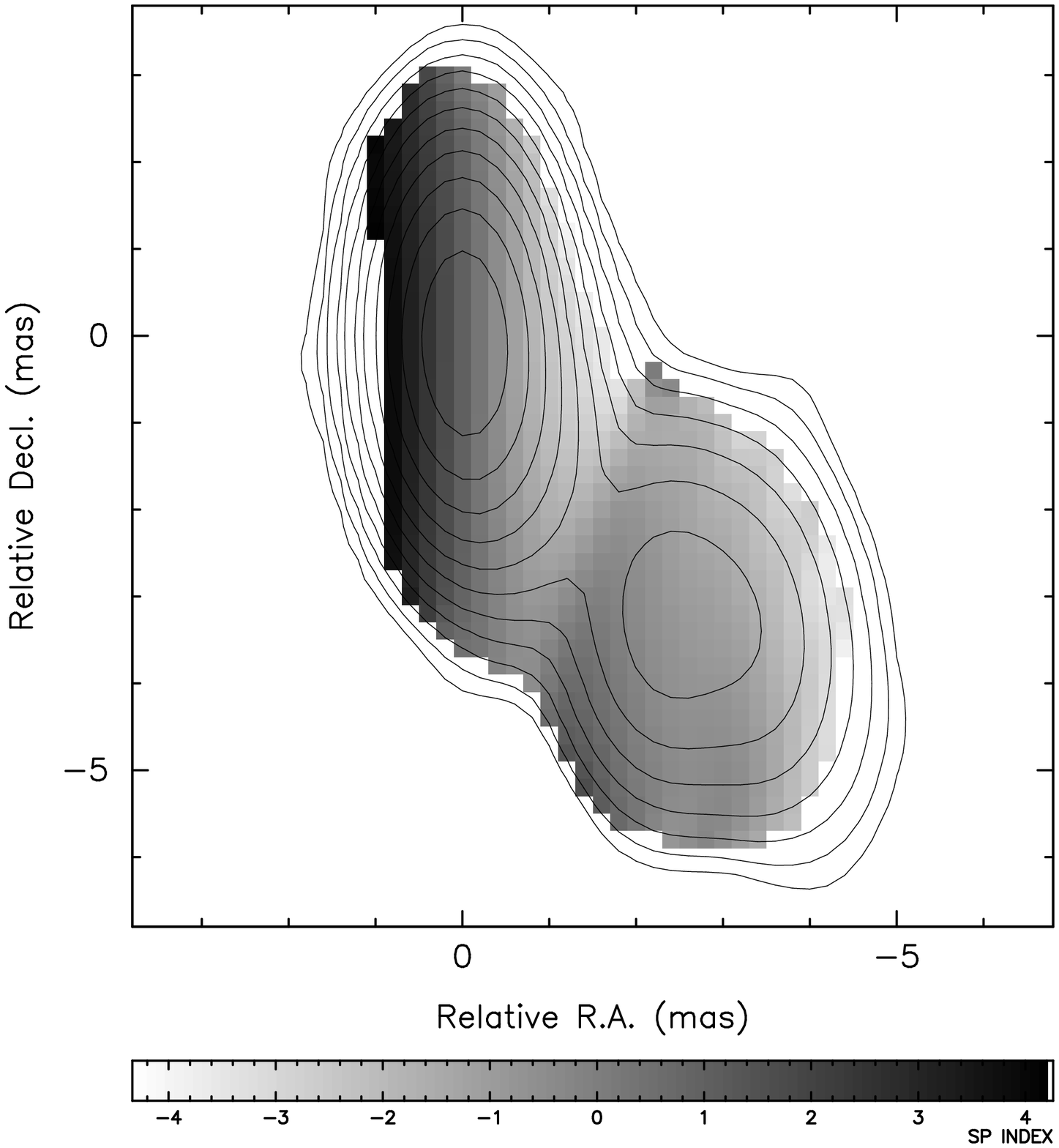}
\caption{Spectral index $\alpha_{12.1}^{8.1}$ plot for 2201+315 overlaid on
Stokes I contours at 15 GHz. Contours start at 1.2 mJy beam$^{-1}$ and
increase by factors of two.}
\label{2201si}
\end{figure}

There is a sign change in the slope of the RM across the core 
of this quasar from $-$1628 $\pm$ 36 to 612 $\pm$ 36 \radm\ (Fig.~\ref{2201rm}a). 
RM corrected electric vectors appear in  Fig.~\ref{2201rm}b.
The 12 and 15 GHz position angles in the core do not appear to follow the 
slope set by the 8 GHz position angles. This may result from optical 
depth effects as the core is optically thick (Fig.~\ref{2201si}). 
Nearly half of the pixels
of the core have a reduced $\chi^2$ greater than the $3\sigma$
level. Beyond 2 mas SW of the core the RM fit $\chi^2$ do become
consistent with a $\lambda^2$ law. Four mas southwest of the core 
the RM has decreased to 5 $\pm$ 33 \radm, or consistent with zero. 

\subsection{3C\,446}
\begin{figure}
\plottwo{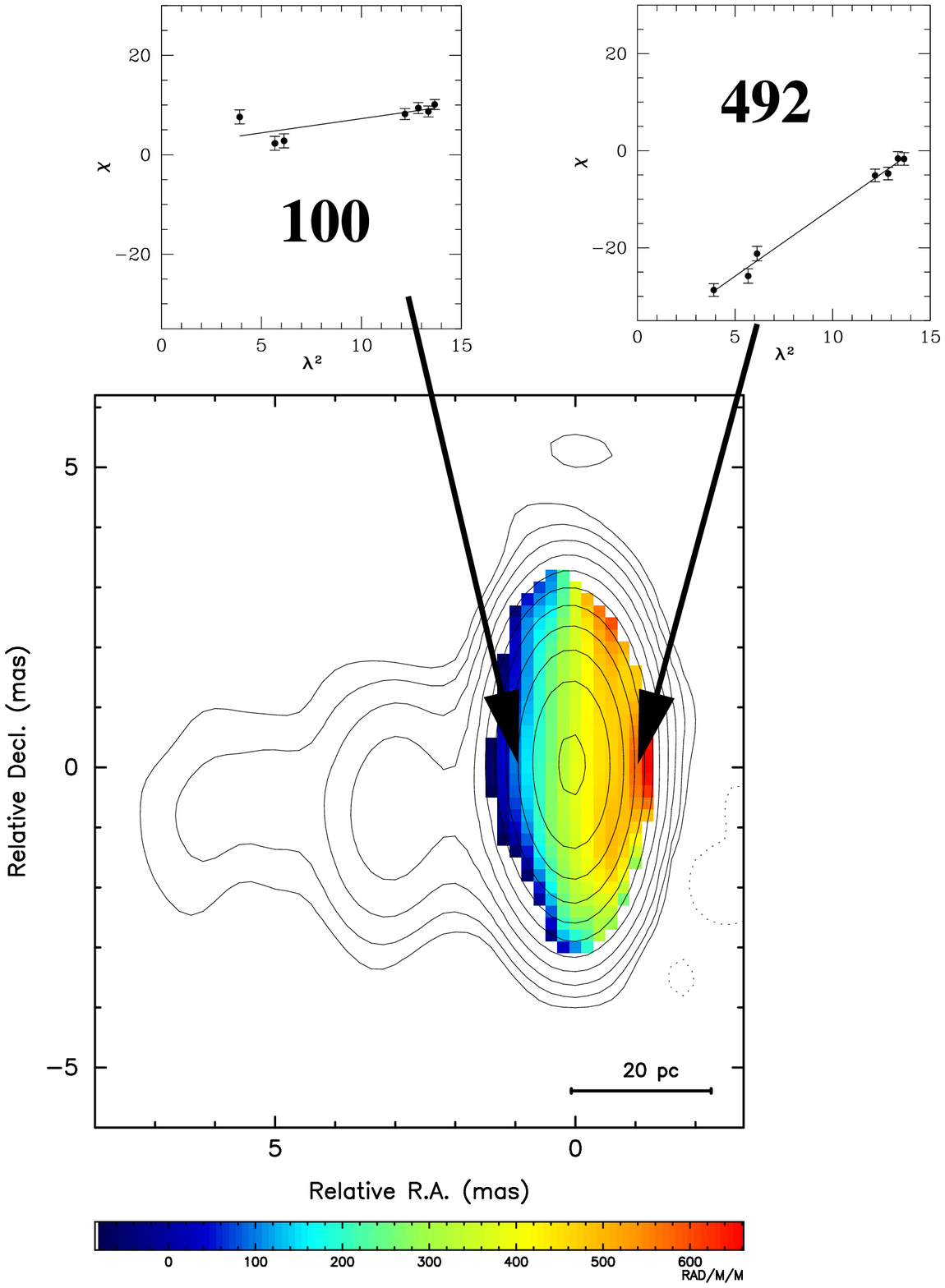}{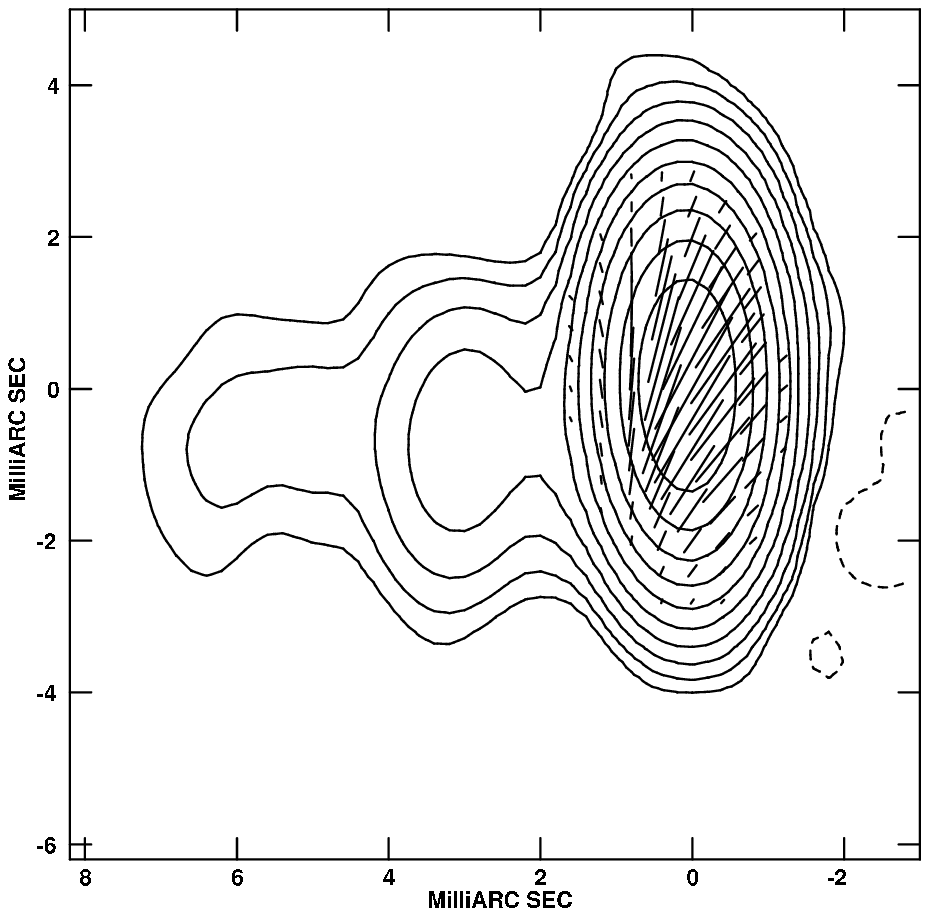}
\caption{(a) Rotation measure image (color) for 3C\,446 overlaid on
Stokes I contours at 15 GHz. The inset is a plot of EVPA $\chi$
(deg) versus \l2 (cm$^2$). (b) Electric vectors (1 mas =
200 mJy beam$^{-1}$ polarized flux density) corrected for Faraday
Rotation overlaid on Stokes I contours. Contours start at 5.1 mJy
beam$^{-1}$ and increase by factors of two.}
\label{446rm}
\end{figure}

\begin{figure}
\plotone{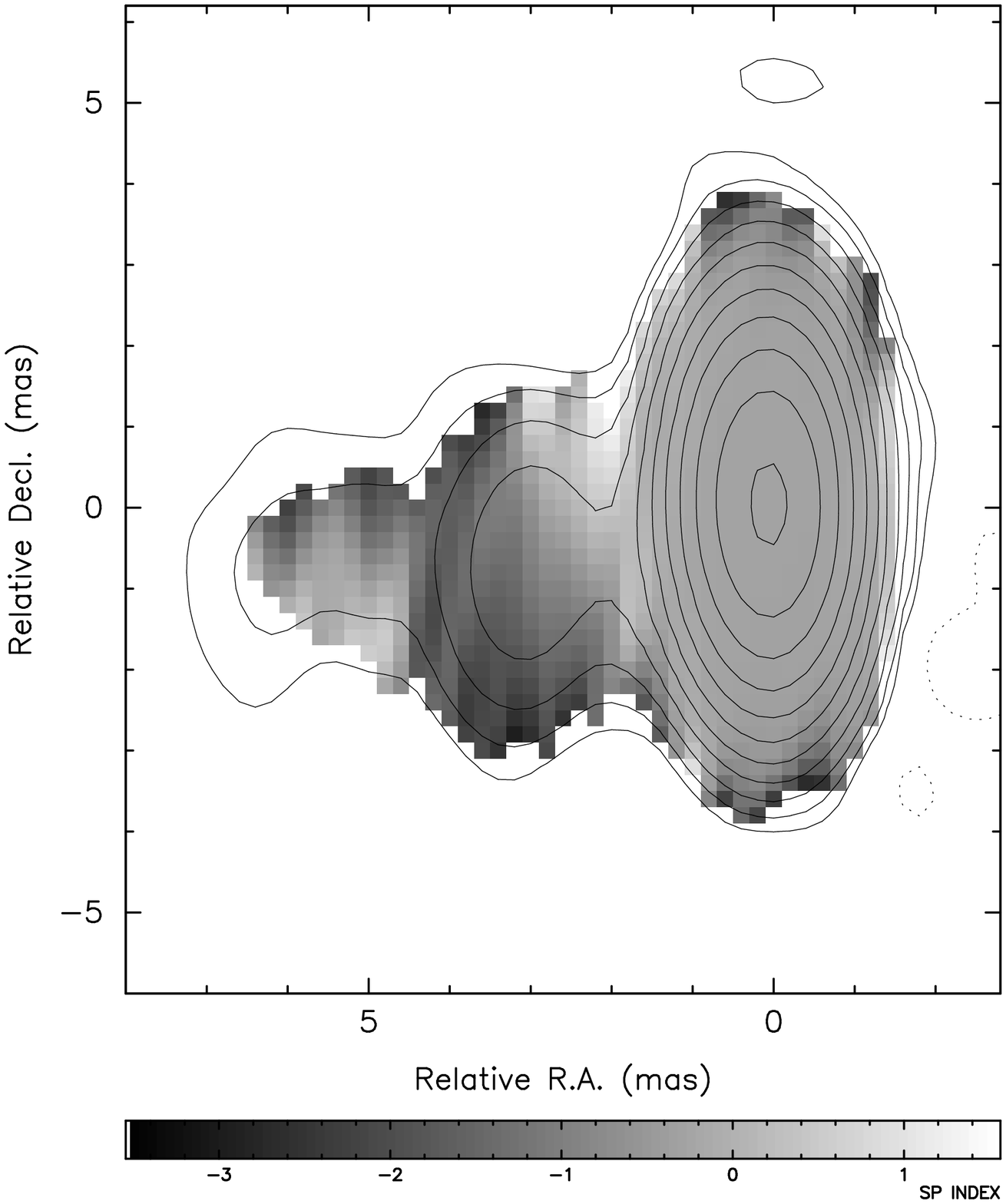}
\caption{Spectral index $\alpha_{12.1}^{8.1}$ plot for 3C\,446 overlaid on
Stokes I contours at 15 GHz. Contours start at 5.1 mJy beam$^{-1}$ and
increase by factors of two.}
\label{446si}
\end{figure}

3C\,446 could be a transition object between the quasar and BL 
Lacertae objects \citep{fal}, but the case for a quasar identification 
has also been made by \citet{breg1} and \citet{breg2}. In 
Fig.~\ref{446rm}a the RM decreases from 492 $\pm$ 23 \radm\ west of the 
core to 100 $\pm$ 22 \radm\ east of the core. This gradient in RM 
tracks a change in the RM corrected electric vector direction of 
almost 60\dg\ (Fig.~\ref{446rm}b). 3C\,446 has a flat, optically 
thick spectrum throughout its RM distribution (Fig.~\ref{446si}). 
The jet, which has no detected polarized flux, is optically thin. 

\section{Discussion} 

\begin{figure}
\plotone{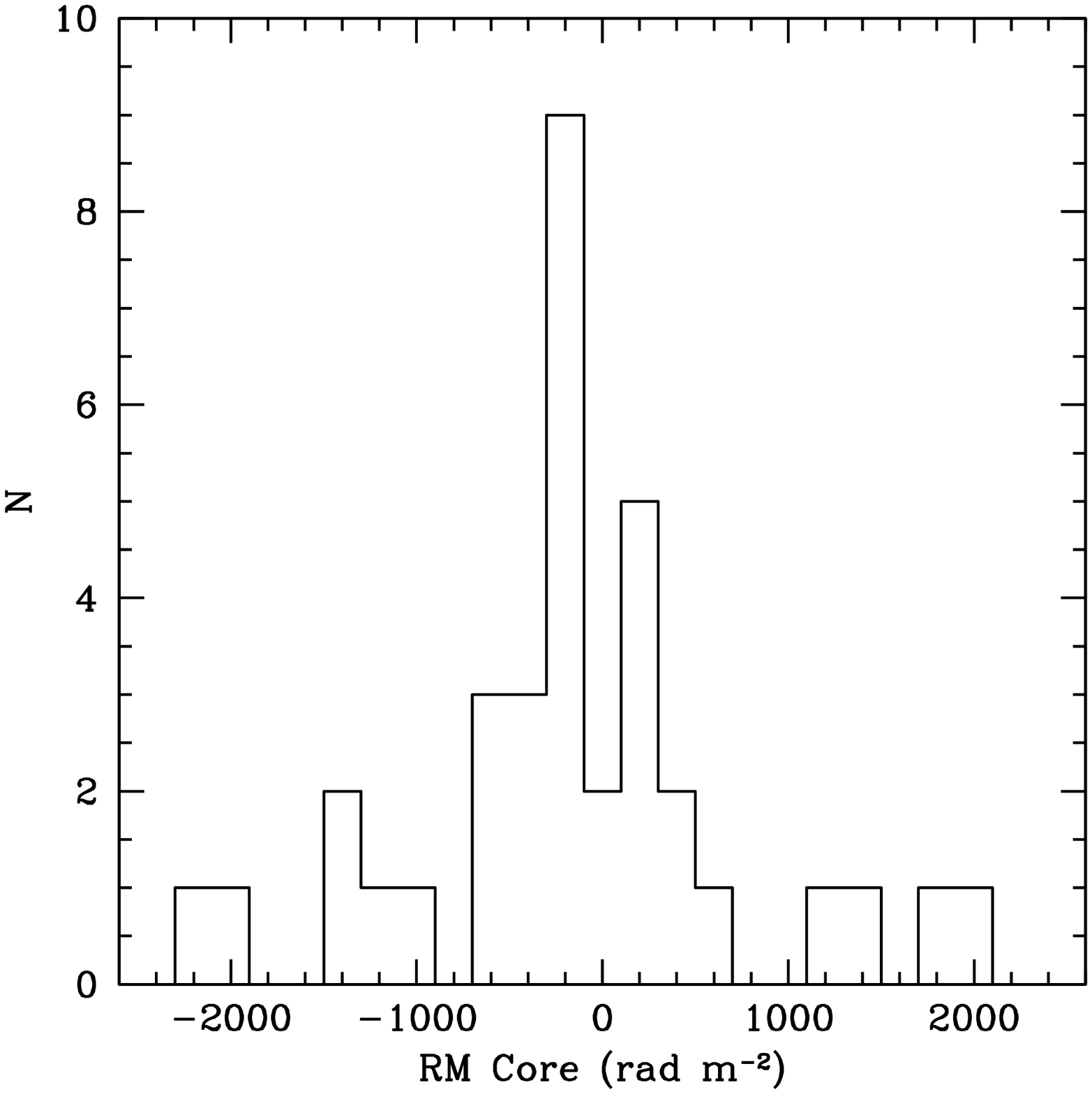}
\caption{Histogram of the RM in the core (\radm) in 200 \radm\ 
bins for the AGN presented here and in \citet{zt03}.}
\label{histo}
\end{figure}

To characterize the RM distribution of the various sources we consider
the RM value of the cores of the AGN presented here. Fig.~\ref{histo} 
shows the histogram of the observed core RM in 200 \radm\ bins. Although 
this is the RM at a single pixel, it is generally representative 
of the values found in the flat spectrum cores of the individual AGN. As 
expected there is no preference to the sign of the RM, and the mean RM observed 
is 137 \radm. The sparse sampling prevents a reliable determination of the 
distribution function, but the general appearance is consistent with a zero 
mean Gaussian distribution. To understand the magnitude of the parsec scale RM 
effect we determined the average of the absolute value of the observed core RM. 
This average absolute value, 644 \radm, is approximately twice the maximum of 
about 300 \radm\ expected on larger angular scales from the observed 
rotation measures in \citet{sim}.      

\subsection{RM and Radio Luminosity \label{zdist}}
\begin{figure}
\plotone{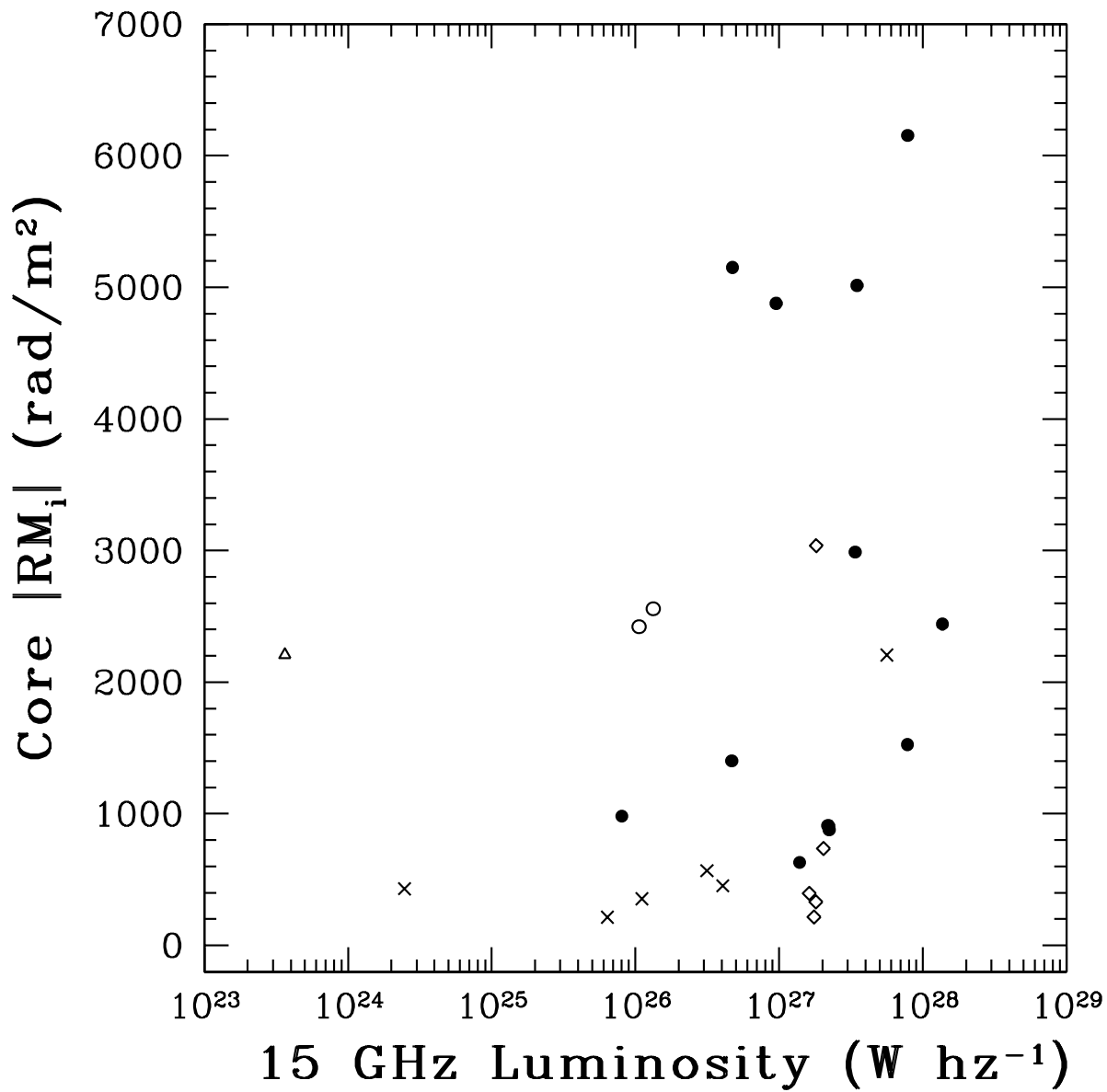}
\caption{The rest frame core rotation measure versus luminosity for the AGN 
in Table~\ref{big}. Filled circles are quasars, two open circles are the two 
epochs for 3C\,273, open diamonds are five epochs for 3C\,279, X's are BL 
Lac objects, and the open triangle is the radio galaxy 3C\,120. The luminosity 
distance was determined with $\Omega_{m}$ = 0.23, $\Omega_{vac}$ = 0.77, 
and H$_0$ = 75 km sec$^{-1}$ Mpc$^{-1}$.}
\label{lum}
\end{figure}

Our understanding of AGN is based largely on an empirical foundation
which suggests a differentiation based on luminosity \citep{lawr}.
We attempt to test for this differentiation by plotting the 
rest frame core RM versus 15 GHz radio luminosity in Fig.~\ref{lum}. The cosmology 
used was $\Omega_{m}$ = 0.23, $\Omega_{vac}$ = 0.77, and H$_0$ = 75 km sec$^{-1}$ 
Mpc$^{-1}$. We made use of E.~L. Wright's online cosmology 
calculator\footnote{http://www.astro.ucla.edu/$\sim$wright/CosmoCalc.html} 
to determine the luminosity distance, and allowed for relativistic 
beaming by using a unit solid angle. Fig.~\ref{lum} looks like 
a scatter plot, but an interesting fact emerges. The multi-epoch data for 
3C\,279 shows that the rotation measure is relatively insensitive to luminosity.
At a given radio luminosity 3C\,279 has high and low rotation measures. Whatever causes 
the change in rotation measure in the core of 3C\,279 does not require large 
changes in the radio luminosity.

The intrinsic rotation measure and radio luminosity are both redshift dependent 
properties. Therefore, a false correlation is expected in Fig.~\ref{lum}. As the plot 
resembles a scatter plot the false correlation from plotting two redshift dependent 
quantities versus each other does not appear significant. To quantify this 
we used the ASURV Revision 1.2 statistics package \citep{asurv}. 
We used the Cox and generalized Kendall's $\tau$ tests \citep{bivar} to test for 
a radio luminosity$-$intrinsic RM correlation. The Cox test gives the probability of 
no correlation at the 20\% level, and the Kendall's $\tau$ rules out a correlation 
at the 5\% level. We conclude that there is no correlation between the intrinsic RM 
and radio luminosity even though one might be expected. 

\subsection{Fractional polarization properties}
\begin{figure}
\plottwo{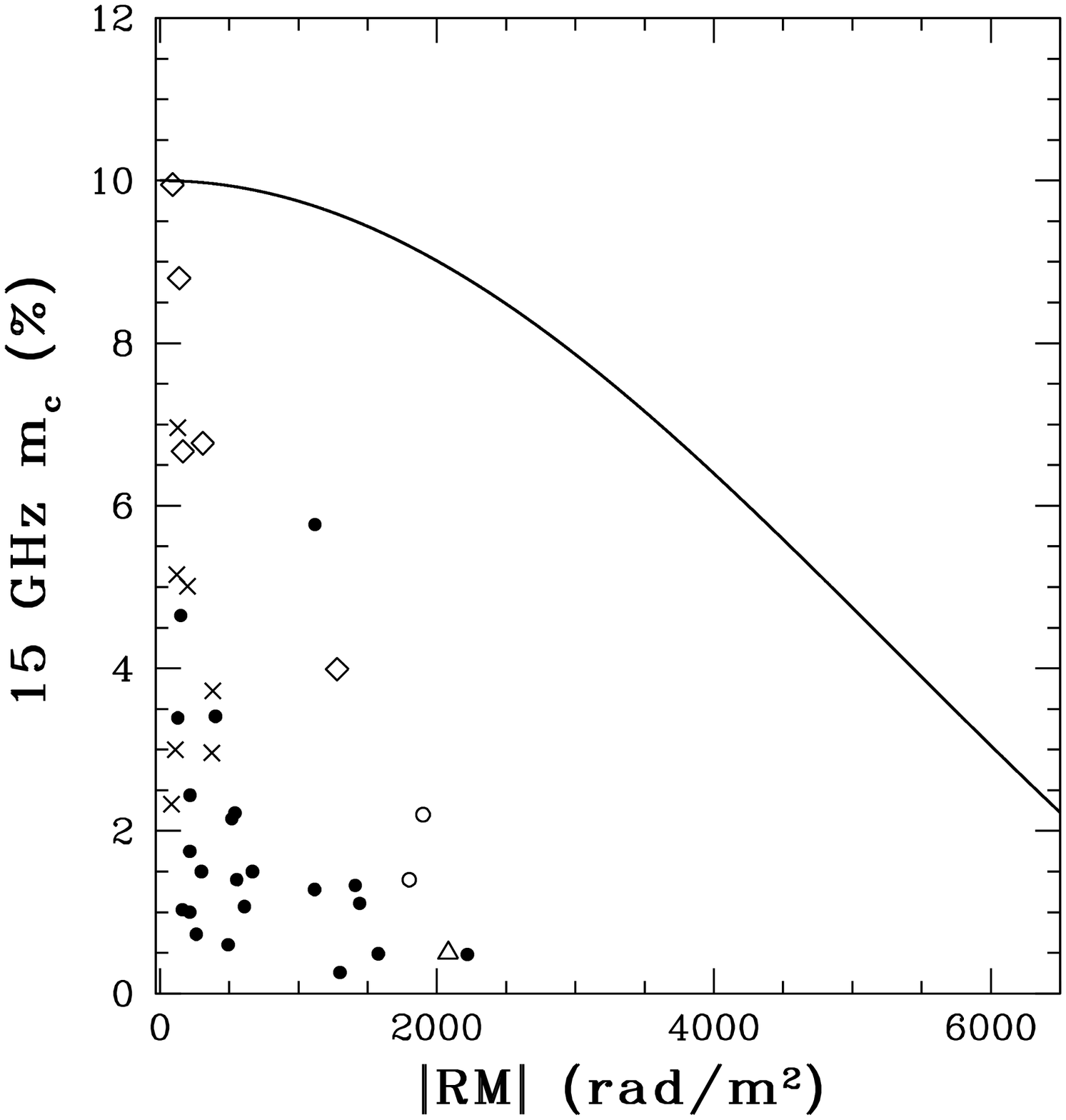}{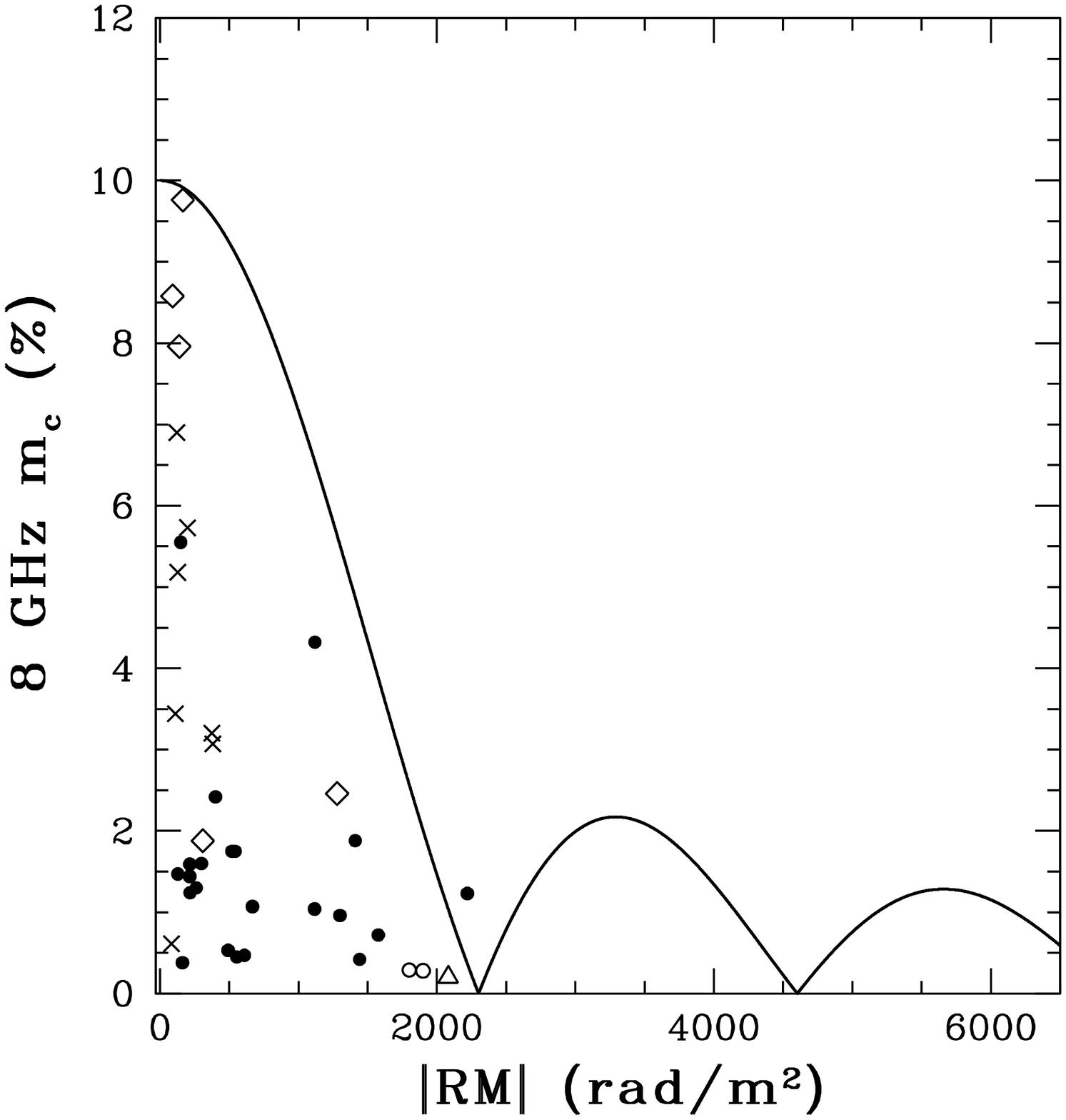}
\caption{(a) Core fractional polarization in percent at 15 GHz 
for the objects in Table~\ref{big} versus observed rotation measure.
Filled circles are quasars, open circles are two epochs of 3C\,273, 
open diamonds are 5 epochs of 3C\,279, X's are BL Lac objects, and the 
open triangle is the radio galaxy 3C\,120. 
(b) Core fractional polarization in percent at 8 GHz 
for the objects in Table~\ref{big} versus observed rotation measure. 
Symbols are the same as in a. The solid lie represents the expected beam 
depolarization from a gradient in a foreground Faraday screen using equation 
~\ref{eqdp}.}
\label{fp}
\end{figure}

Faraday rotation by a foreground screen can produce beam depolarization 
\citep{gw66}. Longer wavelengths will exhibit
this effect to a higher degree due to the \l2 nature of Faraday rotation.
Fig.~\ref{fp}a shows the 15 GHz core fractional polarization versus 
observed rotation measure for the sources in Table~\ref{big}. There is 
a lack of sources with high core fractional polarization and high observed rotation measure. 
This distinction is somewhat more pronounced at 8 GHz as seen in Fig.~\ref{fp}b.
In \citet{zt03} we noted that an RM gradient of 770 \radm\ across a beam 
is sufficient to cause substantial depolarization at 8 GHz. To more 
quantitatively account for the observed fractional polarization 
the beam depolarization can be modeled in the same manner as 
depolarization due to internal Faraday rotation \citep{gw66}. The 
observed fractional polarization is a sinc function of the rotation 
measure. By fixing \l2 and varying the RM we can plot the expected beam 
depolarization, but this requires setting an amplitude to the sinc 
function at zero RM. We set this amplitude at 10\%, in agreement with 
the maximum observed fractional polarization at 8 GHz for this small sample. 
This is similar to the maximum core fractional polarization at 5 GHz found for 106 
quasars by \citet{ptz}. In Fig.~\ref{fp}a \& b the solid line plots the 
expected beam depolarization using the equation 

\begin{equation}
\rm{m(\%)} = 10|sinc(RM\lambda^2)|
\label{eqdp}
\end{equation}

\noindent as derived by \citet{burn}. This is a simple model of a constant 
gradient across the beam. Fig.~\ref{fp}b appears 
to agree with the expected 8 GHz beam depolarization. The 15 GHz fractional 
polarization data seem to respond to the expected depolarization more strongly. 
The first null in fractional polarization for 15 GHz in this simple model is 
not expected to occur until an RM gradient of almost 8000 \radm. Yet the
fractional polarization is 2\% or less at 2000 \radm. This indicates
that the real situation is more complicated than a constant RM
gradient in a foreground screen.

\citet{trib} put forth a modification to the treatment of Burn by
considering variations in the rotation measure which are comparable to
the resolution of the telescope. His results increase the fractional
polarization as compared to the Burn model, and so would not help a
foreground gradient to explain together the 8 and 15 GHz fractional
polarization data. 

Surprisingly, the maximum core fractional polarizations of our 8 GHz 
data presented here, and at 5 GHz from \citet{ptz} are higher than the 
6\% found by \citet{ml} at 43 GHz. One might expect that the decreased 
depolarization and less blending of components at 43 GHz would 
yield higher core fractional polarizations than observed at lower 
frequencies. 

\begin{figure}
\plotone{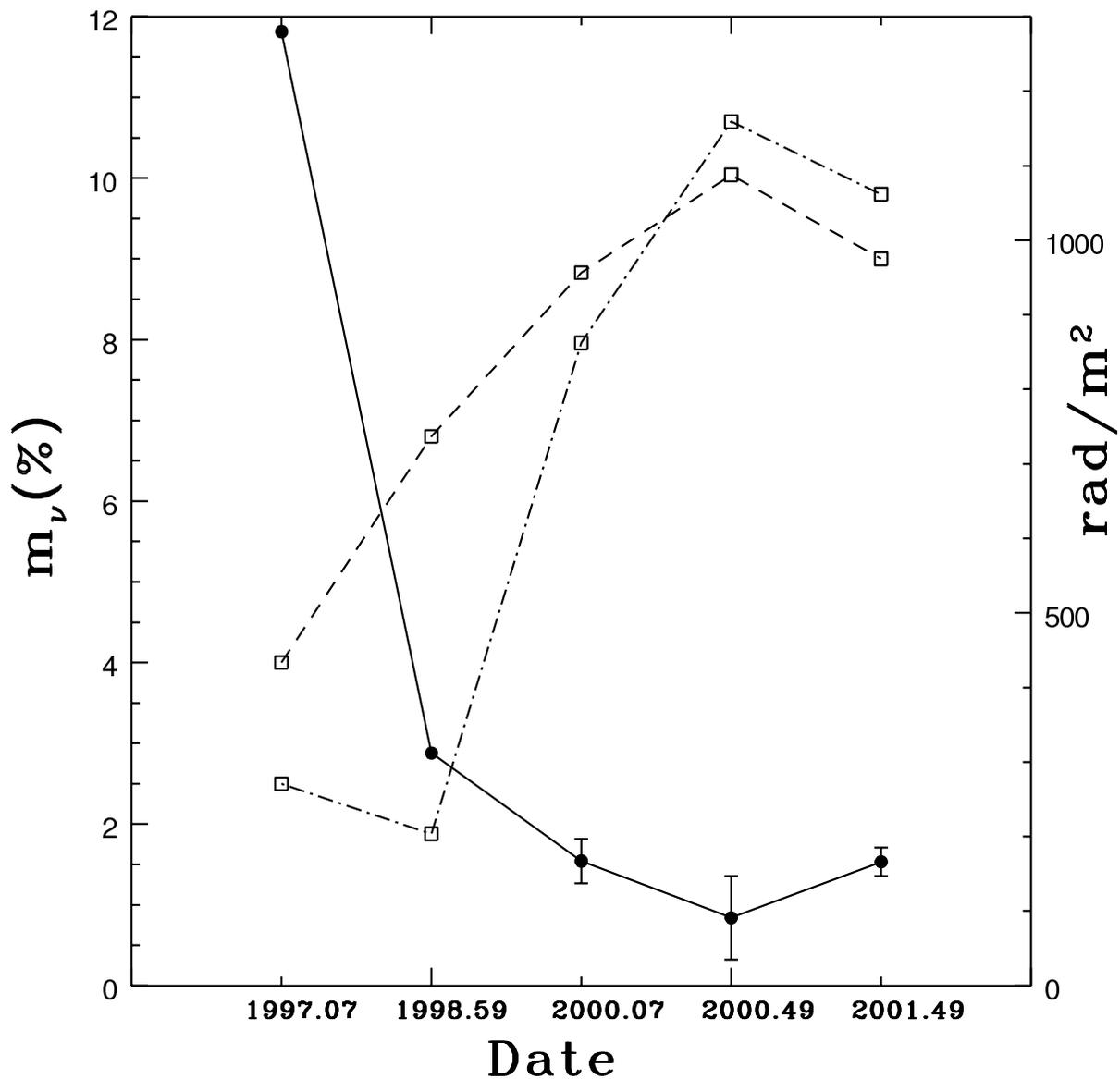}
\caption{Five year curves of the core RM and fractional polarization for the quasar 
3C\,279. The solid line shows the RM versus epoch, the dashed line the 15 GHz 
core fractional polarization (\%), and the 8 GHz core fractional polarization (\%) 
is the dash-dot line. Error bars for the fractional polarization estimates are 
approximately the size of the plotted filled circles. Errors in the RM are only 
known for the three most recent epochs.}
\label{5yr}
\end{figure}

Multi-epoch monitoring of the RM structure in 3C\,279 allows us to 
revisit the idea of a luminosity/RM correlation. Fig.~\ref{5yr} shows 
the results from five years of rotation measure data for the core of 
the quasar 3C\,279. The solid line in Fig.~\ref{5yr} shows the observed 
core rotation measure versus epoch. The 8 and 15 GHz core fractional 
polarization are shown as dashed and dash-dot lines respectively. 
What is immediately evident is the anti-correlation between the core 
fractional polarization at 8 and 15 GHz, and the observed core rotation 
measure. From Table~\ref{big} we see that the highest rotation measure 
and lowest fractional polarizations occur when the quasar has the highest 
radio luminosity. This may also be seen in the optical monitoring data of 
taken at Foggy Bottom Observatory of Colgate University \citep{bk02,bk}. From 
January 1997 to June 2001 3C\,279 brightened from an R magnitude of 15.5 
to 13.6, reaching almost to magnitude 12.5 by August 2001. Superimposed 
on this trend is considerable variability on time scales of days, 
as well as microvariability. Overinterpreting the better time-sampled 
optical light curve should be discouraged, but a relation between the 
radio and optical luminosity and the varying rotation measure deserves 
further scrutiny.

\subsection{Identification of the Faraday Screen}

Faraday rotation serves as a probe of the physical conditions 
responsible for the observed rotation, but this is only useful if 
the screen can be identified. We first consider and rule out several 
locations in order of increasing distance from the supermassive black 
hole. We then make the argument that the screen is located close to 
the relativistic jet itself.

The broad emission line region (BLR) is not a likely candidate 
for the foreground Faraday screen. The BLR is thought to be less
than a parsec with a small (1\%) volume filling factor $\epsilon$ 
\citep{ost} and cannot account for Faraday affects which appear on scales 
of tens of parsecs. Reverberation mapping in AGN provides 
similar size constraints for the BLR \citep{blr}. Additionally, 
the multi-epoch RM maps of 3C\,279 can be used to rule out 
the BLR as a source of variations in the core RM of this blazar. 
\citet{korat} have shown that the Lyman$\alpha$ line in 3C\,279 
does not track variations in the optical continuum over an eight 
year period. This implies that as the optical continuum varies any 
Faraday depth due to the BLR clouds would remain constant. Although the
sampling interval of \citet{korat} did not coincide with our  
rotation measure monitoring, it seems reasonable to accept this finding and 
disregard the BLR as a Faraday screen candidate.

Proceeding out from the center of an AGN the next viable candidate 
for the Faraday screen is the narrow emission line region (NLR), or the 
thermal gas expected to confine the NLR clouds. In \citet{zt03} and 
\citet{zt02} we ruled out the NLR clouds as a Faraday screen based on 
similar volume filling factor arguments used to eliminate the BLR clouds. 
If the NLR clouds are confined in the vicinity of the jet this eliminates 
the volume filling factor argument. 

The hot rarefied gas which confines the NLR clouds is ruled out as the observed 
RM distributions in individual sources do not exhibit a zero-mean Gaussian 
distribution \citep{zt03}. Even with this in mind, we examined the 
possibility of such a stochastic screen using the results of \citet{mm}.  
Melrose \& MacQuart predict that the variance of the Stokes parameters 
Q and U should decrease as exp($-\lambda^4$) in the presence of a stochastic 
foreground Faraday screen, while the expectation value $<Q^2 + U^2>$ should 
remain constant. This decrease in Q and U, while $<Q^2 + U^2>$ remains 
constant, is termed the polarization covariance by Melrose \& MacQuart. 
We examined polarization covariance for the quasar 1611+343 whose RM 
distribution appears in \citet{zt03}. The spatial sampling of the RM 
distribution for this quasar was fairly good, and we found that the 
variance in Q and U increased with wavelength. The polarization 
covariance remained constant, or possibly increased slightly. This 
is further evidence against a purely random Faraday screen. 

The accumulating rotation measure observations reinforce the conclusion 
of \citet{udom} that the Faraday screen cannot be located in the ISM or IGM, 
and we do not consider this suggestion further.

Essentially by process of elimination we are left to consider a Faraday screen 
in close proximity to the relativistic jets of AGN. This has important  
implications for probing the physics of relativistic jets. An exciting example 
is the suggestion by \citet{bland} that observers search for evidence of helical 
magnetic fields through observations of a gradient in the rotation measure 
transverse to a jet axis. \citet{asada} report the detection of an RM 
gradient across the jet of 3C\,273 and interpret this as evidence for the helical 
magnetic field expected by some theories and simulations. 

Interactions between the 
jets and ambient material in the centers of AGN as described by \citet{bick} 
have also been considered \citep{zt03}. A mixing layer described in \citet{zt02} 
and \citet{zt03} also has potential as a foreground Faraday screen. 
Examinations of the relatively rare \citep{ptz} broad and polarized jets 
in AGN will be required to settle the identity of the Faraday screen. 
For example, the interaction model may be tested by observing the alignment 
of magnetic vectors at the interaction site through shocks, and an increasing 
fractional polarization due to this alignment relative to regions of the jet 
upstream from the supposed interaction. 

If a turbulent mixing layer is the Faraday screen than an upper limit to the 
layer thickness is approximately a jet radius. This requirement exists to prevent 
significant deceleration of the jet due to mass entrainment \citep{dey,ros}. 
Relativistic motion in the jets of quasars and BL Lac 
objects, and 3C\,120 \citep{jlg} clearly show that deceleration has not occurred. 
Non-detection of counterjets shows that relativistic beaming is still 
substantial, and is another indicator that no significant deceleration 
has taken place. The deceleration argument limits the maximum screen 
thickness to less than the observed jet radius. The line of sight distance 
$L$ is constrained to about 10 parsecs or less. 

In \citet{zt03} an upper limit to $n_{e}$ was set at a few times 
10$^4$ cm$^{-3}$ due to the lack of apparent free$-$free absorption.
Recently published electron densities for the narrow line radio galaxy 
Cygnus~A put $n_{e}$ at 300 cm$^{-3}$ \citep{ttr}, 
and we consider this a useful lower limit. Thus, it is reasonable to set 
$n_{e}$ to 1000 cm$^{-3}$. With typical jet rotation measures of 100$-$500 
\radm\ the net line of sight B field is $\sim$ 0.1$-$0.6 $\mu$Gauss for a 1 
parsec path length. Should the same path lengths and electron densities be 
responsible for the core RMs of quasars then the field strengths will be 
approximately 1$-$4 $\mu$Gauss, for RMs of 1000$-$3000 \radm. However, the 
assumption of similar physical conditions for the screen within 10 parsecs 
of the black hole seems unlikely. A gradient in the physical conditions is 
expected as we proceed closer to the center of activity. 
      
These magnetic fields are surprisingly weak. To be in equilibrium with a 
thermal gas similar to that in the NLR (T = 10000 K, $n_{e}$ = 1000 cm$^{-3}$)
would require fields of approximately 200$\mu$G, approximately two orders of 
magnitude or more than the simple estimates here produce for the 
B fields. These weak field estimates may present a problem for a dynamically significant 
helical magnetic field. It is difficult to see how a helical 
field could be dynamically important for the relativistic jet with field 
strengths of less than 10 $\mu$G.    

\subsection{RM Properties and Optical Classification}

\begin{figure}
\plotone{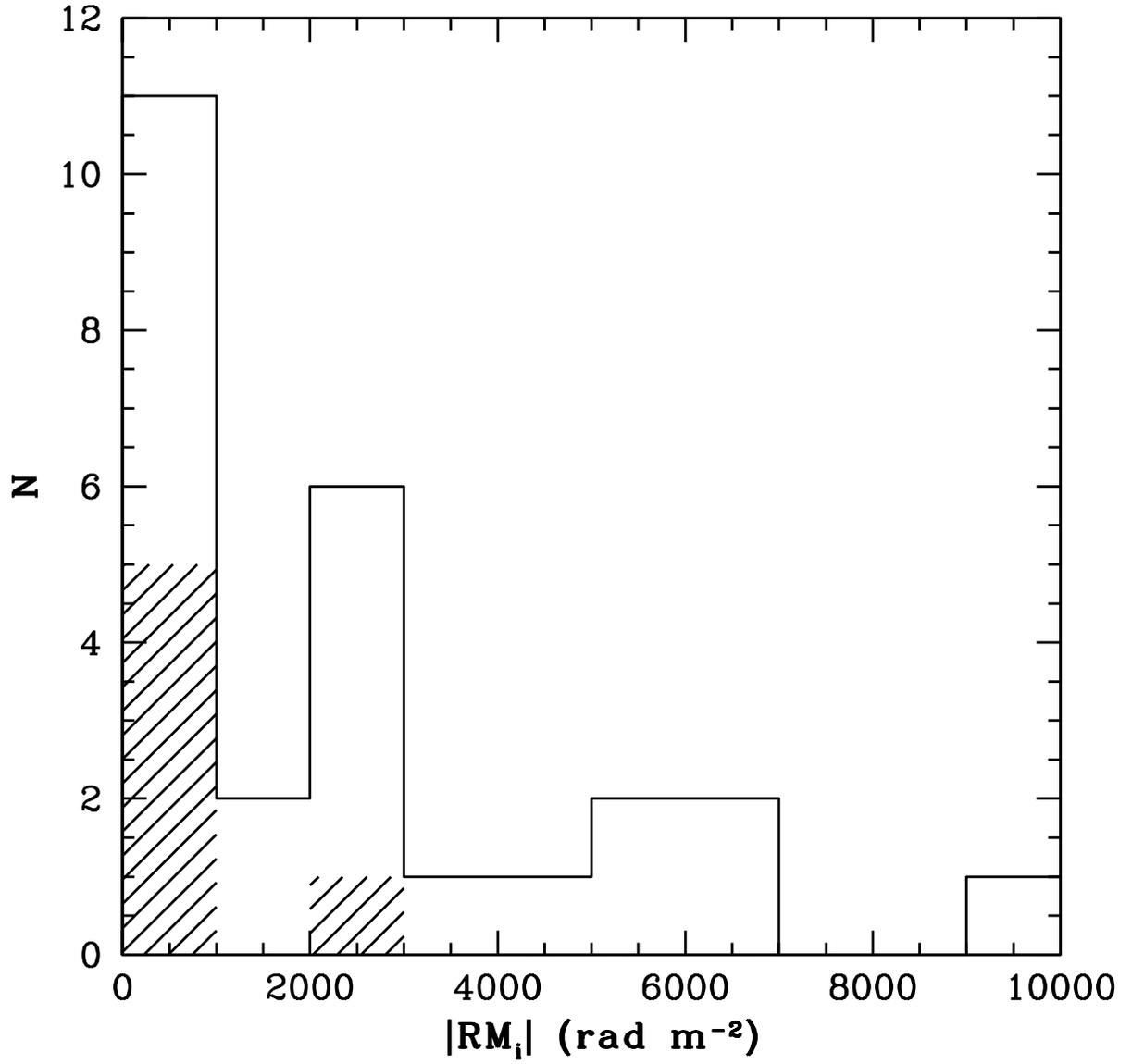}
\caption{Histogram of the rest frame core RM for quasars (open) and 
BL Lac objects (angled line).}  
\label{corehist}
\end{figure}

For some time it has been apparent that optical AGN classification correlates 
with fractional polarizations. For example, \citet{gab92} presented results which 
showed that the cores of BL Lac objects were more strongly polarized than 
quasar cores. This result was verified for a larger sample of AGN by \citet{ptz}.
Using arcsecond scale polarization data \citet{sai} noted that BL Lac objects and 
core-dominant quasars had higher fractional polarizations than either 
lobe-dominant quasars or radio galaxies. Saikia attributed this to an 
orientation effect due to an obscuring torus which depolarized the cores of 
radio galaxies and lobe-dominant quasars. Based on the high rotation measures 
found on parsec scales in quasars  \citet{tay00} predicted lower core rotation 
measures in BL Lacs as compared to quasars. This would arise if BL Lacs 
have their jets more closely aligned to the line of sight, and if the relativistic 
jet clears out the magneto-ionic gas responsible for the Faraday rotation.  
Contrary to this expectation BL Lac itself was found to have a non-negligible 
Faraday rotation \citep{rcg}.  
With this in mind we will examine whether the rotation measure properties of the 
cores and jets of BL Lacs and quasars are significantly different.
Using the values for the peak rest frame rotation measures (last column of 
Table~\ref{big}) we see that the quasars and BL Lacs appear to be different.
Table~\ref{rmcore} shows the number, the mean $\mu$, 
error of the mean , and the median rest frame 
rotation measures for quasars and BL Lac objects. For the quasars there are 26 
measurements for 21 quasars because of the multi-epoch observations of 3C\,273 and 
3C\,279. As only one radio galaxy has a core rotation measure we exclude this class 
from consideration. The quasars have a mean rest frame rotation measure three times 
that observed for the BL Lac objects. The median values, which are less affected by 
outliers, agree with this result. As expected by \citet{tay00} BL Lacs seem to have a 
systematically lower core RM compared to quasars. These are small number statistics, 
and a Kolmogorov-Smirnov test \citep{numrec} gives a probability of 0.011 that 
the BL Lac and quasar core RMs are drawn from the same parent distribution. This is 
only a 2.5$\sigma$ result. Fig.~\ref{corehist} is a histogram of the rest frame core 
RM of BL Lacs (angular line boxes) and quasars (open boxes). Clearly small number 
statistics limit our ability to distinguish any difference between the two AGN classes 
that might exist based on RM. All we can say is that there is a suggestion that quasar 
and BL Lac core rotation measures are different, and better statistics are needed to 
establish this on a firm foundation.

Some shaking to this foundation has already occurred. \citet{denn} 
report in their multi-epoch monitoring of BL Lac an observation
of a rotation measure of 6000 \radm.  This quasar-like RM further
blurs the distinction between the quasars and BL Lac's which also
exists in their optical spectral line properties \citep{lacnot}.
The BL Lac redshift distribution does not extend much
beyond a z of 1 \citep{rec} so we have only a small overlap
for quasars and BL Lacs with redshifts less than 1.  Our primarily
single epoch RM observations may certainly undersample a highly
variable phenomenon as \citet{denn} and \citet{zt01}
demonstrate.

The same cannot be said for the jets of BL Lacs and quasars. To define the jets 
we used the 8-12 GHz spectral index maps, and defined the jets to be the 
regions where the spectral index map shows the jet is optically thin 
($\alpha < -0.5$). The RM maps are blanked retaining pixels where this 
criteria for $\alpha$ is met, which enables the RM distribution for the 
predominantly optically thin jet regions to be determined. We further 
required that this ``jet'' region be at least one beamwidth from the map 
peak, the location where the core RM in Table~\ref{big} is taken. These criteria 
limited the number of sources for which we could investigate the jet RM statistics. 
Table~\ref{rmjet} presents the results of this comparison, and the smaller number 
statistics are immediately apparent. Neither the mean nor the median values 
appear significantly different. A K-S test was not performed due to the small 
numbers present in the comparison of jet RM properties. These small number statistics, 
especially in the case of the BL Lac objects, and the already noted RM variability 
of BL Lac objects (\S~\ref{zdist}), leaves these comparisons of core and jet RM 
properties suspect. A larger sample of RM observations, with good time sampling, 
is required to confirm that the jet regions are indeed similar while the 
cores appear different. 

\subsection{Breaking the \l2\ Faraday Law}

As noted in Section 3 the $\lambda^2$ law does not seem to be
universally applicable.  Both 0202+149 and 0420$-$014 are depolarized perhaps
through a superposition of components smaller than the beam size, and
no fits to a $\lambda^2$ law were possible. There are sources for
which sufficient polarized flux is detected at all frequencies and a
$\lambda^2$ law does not seem applicable. Table~\ref{reject} lists the
sources for which agreement to a $\lambda^2$ law seems unlikely based
on the reduced $\chi^2$ obtained for the RM fits. Lack of agreement to
the Faraday rotation law may result for several reasons which we now
consider.

Almost all sources have cores optically thick to
synchrotron emission as shown in the spectral index maps. This is
especially true at 8 and 12 GHz. Observations at different frequencies
see different $\tau = 1$ surfaces which may not have the same
intrinsic polarization angle. If this were the case the RM fits in the
optically thick cores should always fail to agree with the $\lambda^2$
law as the $\lambda=0$ position angles would not agree. This is not
true in general, as most sources show good agreement to the
$\lambda^2$ law even in the optically thick cores. This is especially
true for 3C 273 and 3C 279 which show good agreement to the Faraday
rotation law in their optically thick cores over several epochs
separated by months to year timescales. This is an interesting result
as it requires the different $\tau = 1$ surfaces to maintain the same
polarization angle orientation. It is known that the jets collimate
within a small distance from the black hole \citep{junor} and this
collimation may also order the magnetic field within this short
distance.  For the optically thick regions of the jet higher
frequencies see farther down the jet and closer to the black hole
\citep{blanko}. As the $\lambda^2$ law holds in the optically thick
regions, then the different $\tau = 1$ surfaces, located at different
radii from the black hole, must have the same intrinsic polarization
angle and hence magnetic field orientation.
  
\begin{figure}
\plotone{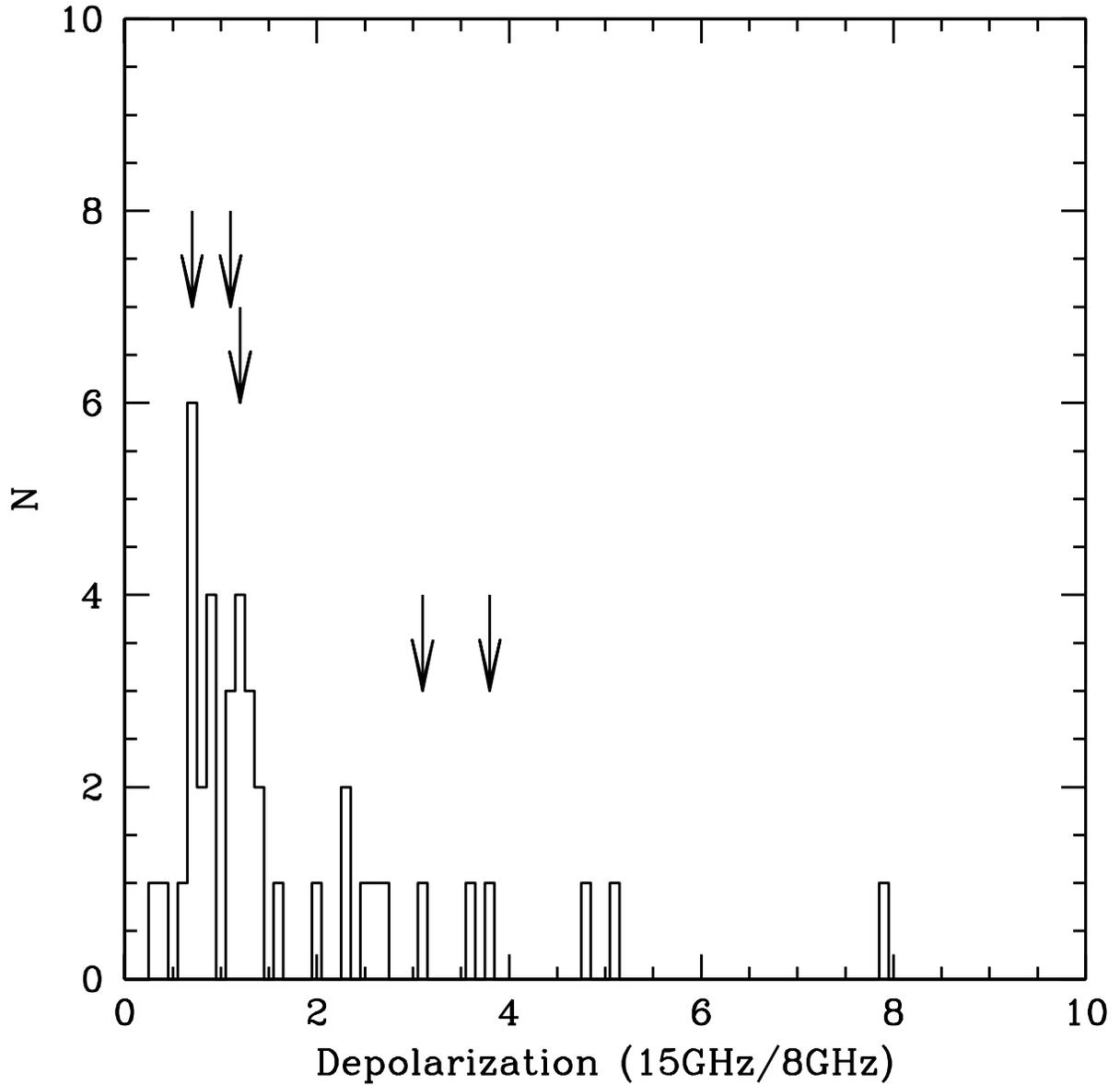}
\caption{Histogram of the depolarization, defined as ratio of core fractional 
polarization at 15 GHz to the core fractional polarization at 8 GHz for all 
sources in Table~\ref{big}. Arrows identify the positions of the five sources 
for which the \l2\ law may not apply based on the reduced $\chi^2$ as discussed 
in the text.}
\label{depol}
\end{figure}

There is no consistent observational picture for the sources which do
not show good agreement to a $\lambda^2$ law based on the reduced
$\chi^2$ of the RM fits. Comparing Table~\ref{reject} with
Fig.~\ref{lum} shows that these sources are not systematically
brighter or fainter relative to other sources in the sample.  Optical
class seems unimportant for the moment as BL Lacs and quasars appear
in proportion to their representation in the sample as a
whole. Opacity effects do not seem to be important as optically thick
sources do show good agreement to the Faraday law for most cases. We
examined depolarization as a characteristic and find that the
depolarization spans a wide range of values for these
sources. Fig.~\ref{depol} shows the depolarization as the ratio of the
15 GHz fractional polarization to the 8 GHz fractional
polarization. Arrows in Fig.~\ref{depol} show the locations of the
five sources for which a reduced $\chi^2$ is not in agreement with
that expected if the Faraday law were true. The most depolarized
source in Fig.~\ref{depol} is 3C\,273 (epoch 2000.07) which shows good
agreement to a $\lambda^2$ law even with a high depolarization
ratio. \citet{homan} report that two sources (not included in this
sample) also exhibit non-Faraday law behavior, based on variations in
polarization angles at two frequencies over several epochs.

\section{Conclusions}

The rotation measure properties for a sample of over 40 quasars, radio 
galaxies and BL Lac objects are examined. The core rotation measures in 
quasars are observed to vary from approximately 500 \radm\ to several 
thousand \radm\ within 10 parsecs of the core. Jet rotation measures are
typically 500 \radm\ or less. The cores of the seven BL Lac objects examined 
have RMs in their cores and jets similar to quasar jets. Radio galaxies 
usually have depolarized cores, and exhibit RMs in their jets varying from 
a few hundred to 10,000 \radm. A gradient in the foreground Faraday screen 
is invoked to explain the observed depolarization properties of the sample. 
The Faraday screen is likely located close to the relativistic jet, 
although its exact nature remains unclear. Observations of broad, 
polarized jets, are required to further constrain the identity of the 
Faraday screen. Net line of sight magnetic fields of 0.1$-$0.6 $\mu$Gauss 
can account for the observed jet rotation measures. If similar physical 
conditions exist in quasar cores then the field strength required is of 
order 1 $\mu$Gauss. Agreement to the \l2\ law in the optically thick 
cores of quasars and BL Lac objects requires a constant magnetic field 
orientation at different $\tau = 1$ surfaces, and thus at different radii 
from the black hole.         

\acknowledgments
Tom Balonek and Jeyhan Kartaltepe kindly provided optical monitoring 
data for 3C\,279 in advance of publication. R.T.Z gratefully acknowledges 
support from a pre-doctoral research fellowship from NRAO and from the New 
Mexico Alliance for Graduate Education and the Professiorate through NSF 
grant HRD-0086701. This research has made use of the NASA/IPAC Extragalactic 
Database (NED) which is operated by the Jet Propulsion Laboratory, 
Caltech, under contract with NASA, and NASA's Astrophysics Data System 
Abstract Service. This research has also made use of data from the
University of Michigan Radio Astronomy Observatory which is supported 
by funds from the University of Michigan.



\begin{deluxetable}{lcccccccc}
\tabletypesize{\scriptsize}
\tablecaption{T{\sc arget} S{\sc ources} \label{t1}}
\tablewidth{0pt}
\tablehead{
\colhead{Source} & \colhead{Name}   & \colhead{Identification}   &
\colhead{Magnitude\tablenotemark{a}} &
\colhead{z}  & \colhead{$S_{15}$} & \colhead{Scans} & 
\colhead{$\sigma_{15\rm{GHz}}$} & \colhead{Peak$_{15\rm{GHz}}$} \\
\colhead{(1)} & \colhead{(2)} & \colhead{(3)} & \colhead{(4)} & 
\colhead{(5)} & \colhead{(6)} & \colhead{(7)} & \colhead{(8)} 
& \colhead{(9)} }

\startdata
0202+149   &        &Q  &22.1   &0.41 &2.29 &8 &0.4 &1.52 \\
0336$-$019 &CTA26   &Q  &18.4   &0.85 &2.23 &9 &0.5 &2.01 \\
0355+508   &NRAO150 &EF &\nodata &\nodata&3.23 &7 &3.4 &6.40 \\
0458$-$020 &        &Q  &18.4   &2.29 &2.33 &9 &0.3 &0.91 \\
0552+398   &DA193   &Q  &18.0 &2.37\tablenotemark{b} &5.02 &7 &1.2 &3.02 \\
0605$-$085 &        &Q  &18.5   &0.87 &2.80 &7 &0.8 &1.10 \\
0736+017   &        &Q  &16.5   &0.19 &2.58 &7 &0.4 &1.31 \\
0748+126   &        &Q  &17.8   &0.89\tablenotemark{b} &3.25 &7 &0.3 &1.27 \\
1055+018   &        &BL &18.3   &0.89 &2.15 &11 &0.8 &4.03  \\
1253$-$055 &3C\,279 &Q  &17.8   &0.54 &21.56 &7 &2.2 &8.81 \\
1546+027   &        &Q  &18.0   &0.41 &2.83 &8 &0.4 &1.53 \\
1548+056   &        &Q  &17.7   &1.42 &4.05 &8 &1.3 &1.68 \\
1741$-$038 &        &Q  &18.6   &1.05 &4.06 &7 &2.1 &4.32 \\
1749+096   &        &BL &16.8   &0.32 &5.58 &7 &0.6 &2.54 \\
2021+317   &        &EF &\nodata &\nodata &2.02 &9 &0.3 &0.356 \\
2201+315   &        &Q  &15.5   &0.30 &3.10 &10 &0.3 &2.01 \\
2223$-$052   &3C\,446 &BL &17.2   &1.40 &3.92 &8 &1.5 &4.74 \\
\enddata

\tablenotetext{a}{Note that many sources are highly variable.}
\tablenotetext{b}{Redshift questionable, see \citet{wills}}

\tablecomments{Col. (1): B1950 source name. Col. (2): Alternate
common name. Col. (3): Optical identification from the literature 
(NED) with  Q = quasar, BL = BL Lac object, EF = empty field.
Col. (4): Optical magnitude. Col. (5): Redshift. Col. (6): 
Total flux density at 15 GHz measured by \citet{kel98}. 
Col. (7): Number of scans. Col. (8): RMS (mJy beam$^{-1}$) in 
15 GHz untapered map. Col. (9): Peak flux (Jy) in 15 GHz untapered 
map.}

\end{deluxetable}

\clearpage

\begin{deluxetable}{ccccc}
\tabletypesize{\scriptsize}
\tablecaption{\sc{EVPA Calibration using 3C\,279} \label{t2}}
\tablewidth{0pt}
\tablehead{\colhead{Telescope} & \colhead{Freq.} & \colhead{Date} & 
\colhead{Pol. Flux} & \colhead{$\chi$\tablenotemark{a}} \\
 & \colhead{GHz} & & \colhead{mJy} & \colhead{Deg.}}
\startdata
UMRAO & 8.0 & 20010620 & 2044 & 56.6  \\
 & 14.5 & 20010625 & 2058 & 60.5 \\
VLA & 5.0 & 20010624 & 1297 & 64.0 \\
 & 8.5 &20010624  & 1995 & 58.2 \\
 & 22 &20010624  & 2016 & 57.0 \\
 & 43 &20010624  & 1889 & 57.0 \\
VLBA & 8.5 & 20010621 & 1894 & 34.0 \\
 & 15.15 & 20010621 & 2147 & $-66$ \\
\enddata  
\tablenotetext{a}{$\chi$ for VLBA is before applying the EVPA 
calibration derived from the VLA data.}

\end{deluxetable}

\clearpage






\begin{deluxetable}{cc}
\tabletypesize{\scriptsize}
\tablecaption{O{\sc bservational} P{\sc arameters} \label{t3}}
\tablewidth{0pt}
\tablehead{
\colhead{Frequency} & \colhead{Bandwidth}}

\startdata
8.114, 8.209, 8.369, 8.594   &8   \\
12.115, 12.591& 16       \\
15.165 &32 \\
\enddata

\tablecomments{Frequencies in GHz, bandwidths in MHz}

\end{deluxetable}

\clearpage

\begin{deluxetable}{lccccccccccccccccc}
\tabletypesize{\scriptsize}
\rotate
\tablecaption{C{\sc ore} RM \& P{\sc olarization} P{\sc roperties} \label{big}}
\tablewidth{0pt}
\tablehead{
\colhead{} & \colhead{} & \colhead{} & \colhead{} & \multicolumn{3}{c}{8 GHz} 
& \colhead{} & \multicolumn{3}{c}{15 GHz} & \colhead{} 
& \multicolumn{2}{c}{8 GHz} & \colhead{} 
& \multicolumn{2}{c}{15 GHz} \\ 
\cline{5-7} \cline{9-11} \cline{13-14} \cline{16-17} \\
\colhead{Source} & \colhead{Name} & \colhead{ID} & \colhead{z}  & \colhead{Peak} 
& \colhead{Integ} & \colhead{PeakPOL} & \colhead{} & \colhead{Peak} & \colhead{Integ} & 
\colhead{PeakPOL} & \colhead{RM$_{0}$} & \colhead{R$_{c}$} & \colhead{m$_{c}$}
& \colhead{} & \colhead{R$_{c}$} & \colhead{m$_{c}$}  & \colhead{RM$_{i}$} \\
\colhead{(1)} & \colhead{(2)} & \colhead{(3)} & \colhead{(4)} & 
\colhead{(5)} & \colhead{(6)} & \colhead{(7)} & \colhead{} 
& \colhead{(8)} & \colhead{(9)} & \colhead{(10)} & \colhead{(11)} & \colhead{(12)} & \colhead{(13)} 
& \colhead{} & \colhead{(14)} & \colhead{(15)} & \colhead{(16)} }

\startdata
0133+476 & DA55 & Q & 0.86 & 2921 & 3103 & 55 & & 3736 & 3802 & 50 & -1410 & 0.941 & 1.88 & & 0.983 & 1.33 & -4878 \\ 
0202+149 & & Q & 0.41 & 1664 & 2016 & $<$1.9 & & 1648 & 1869 & 3 & \nodata& 0.825 & $<$1.1 & & 0.882 & 0.18 & \nodata \\ 
0212+735 & & Q & 2.37 & 2229 & 3125 & 39 & & 1844 & 2445 & 41 & -542 & 0.713 & 1.75 & & 0.754 & 2.22 & -6155 \\ 
0336$-$019 & CTA26 & Q & 0.85 & 1447 & 1919 & 15 & & 2191 & 2512 & 28 & \nodata\tablenotemark{d} & 0.754 & 1.04 & & 0.873 & 1.28 & \nodata\tablenotemark{d} \\ 
0355+508 & & EF & \nodata & 4631 & 5479 & 16 & & 7001 & 7245 & 126 & -1028 & 0.845 & 0.35 & & 0.966 & 1.80 & \nodata \\ 
0415+379 & 3C111 & G & 0.05 & 861 & 1963 &$<$2.0  & & 1537 & 2263 & $<$1.8 & \nodata& 0.439 & $<$2.3 & & 0.679 & $<$1.2 & \nodata \\ 
0420$-$014 & & Q & 0.92 & 2035 & 2377 & 7 & & 2644 & 2872 & 6 & \nodata& 0.856 & 0.34 & & 0.921 & 0.23 & \nodata \\ 
0430+052 & 3C120 & G & 0.03 & 1075 & 3307 & $<$2.1 & & 797 & 2519 & 4 & 2082 & 0.325 & $<$1.5 & & 0.316 & 3.8 & 2209 \\ 
0458$-$020 & & Q & 2.29 & 668 & 858 & 3 & & 931 & 1055 & 13 & \nodata\tablenotemark{d} & 0.779 & 0.45 & & 0.882 & 1.40 & \nodata\tablenotemark{d} \\ 
0528+134 & & Q & 2.06 & 2924 & 3479 & 11 & & 3100 & 3439 & 32 & -163 & 0.840 & 0.38 & & 0.901 & 1.03 & -1526 \\ 
0552+398 & DA193 & Q & 2.37 & 4440 & 5345 & 64 & & 3794 & 4260 & 38 & 215 & 0.831 & 1.44 & & 0.891 & 1.00 & 2442 \\ 
0605$-$085 & & Q & 0.87 & 1114 & 1674 & 27 & & 1321 & 1762 & 45 & 401 & 0.665 & 2.42 & & 0.750 & 3.41 & 1402 \\ 
0736+017 & & Q & 0.19 & 758 & 1102 & 4 & & 1330 & 1539 & 8 & \nodata\tablenotemark{d} & 0.688 & 0.53 & & 0.864 & 0.60 & \nodata\tablenotemark{d} \\ 
0748+126 & & Q & 0.89 & 955 & 1276 & 4 & & 1437 & 1689 & 16 & 1442 & 0.748 & 0.42 & & 0.851 & 1.11 & 5151 \\ 
0923+392 & & Q & 0.70 & 8367 & 10640 & 104 & & 7179 & 8959 & 175 & -218 & 0.786 & 1.24 & & 0.801 & 2.44 & -630 \\ 
1055+018 & & B & 0.89 & 2969 & 3813 & 18 & & 4117 & 4594 & 96 & \nodata\tablenotemark{d} & 0.779 & 0.61 & & 0.896 & 2.33 & \nodata\tablenotemark{d} \\ 
1226+023 & 3C273\tablenotemark{a} & Q & 0.16 & 13271 & 26828 & 38 & & 12341 & 21571 & 173 & 1800 & 0.495 & 0.29 & & 0.572 & 1.40 & 2422 \\ 
 & 3C273\tablenotemark{c} & Q & 0.16 & 9500 & 27955 & 27 & & 13500 & 27180 & 297 & -1900 & 0.340 & 0.28 & & 0.497 & 2.20 & -2557 \\ 
1228+126 & M87 & G & 0.00 & 1042 & 2180 &$<$2.5 & & 1029 & 1920 & $<$1.8 & \nodata& 0.478 & $<$2.4 & & 0.536 & $<$1.7 & \nodata \\ 
1253$-$055 & 3C279 & Q & 0.54 & 10824 & 18538 & 1056 & & 12226 & 19773 & 816 & -166 & 0.583 & 9.76 & & 0.618 & 6.67 & -394 \\ 
 & 3C279 & Q & 0.54 & 13277 & 21207 & 1139 & & 14650 & 21335 & 1457 & -91 & 0.626 & 8.58 & & 0.687 & 9.95 & -216 \\ 
 & 3C279\tablenotemark{a} & Q & 0.54 & 12860 & 20402 & 1024 & & 14070 & 21980 & 1238 & -166 & 0.630 & 7.96 & & 0.640 & 8.80 & -396 \\ 
 & 3C279\tablenotemark{b} & Q & 0.54 & 17898 & 23485 & 336 & & 19865 & 24766 & 1344 & -310 & 0.762 & 1.88 & & 0.802 & 6.77 & -735 \\ 
 & 3C279\tablenotemark{c} & Q & 0.54 & 9100 & \nodata & 224 & & 17528 & 22129 & 700 & -1280 & \nodata & 2.46 & & 0.792 & 3.99 & -3036 \\ 
1308+326 & & B & 1.00 & 843 & 1467 & 29 & & 667 & 1079 & 20 & 113 & 0.575 & 3.44 & & 0.618 & 3.00 & 452 \\ 
1546+027 & & Q & 0.41 & \nodata & \nodata & \nodata & & 1630 & 1900 & 64 & -474 & \nodata & \nodata & & 0.858 & 3.93 & -982 \\ 
1548+056 & & Q & 1.42 & 2108 & 2426 & 118 & & 2149 & 2469 & 100 & -150 & 0.869 & 5.55 & & 0.870 & 4.65 & -878 \\ 
1611+343 & DA406 & Q & 1.40 & 2630 & 4377 & 46 & & 2692 & 3880 & 58 & -519 & 0.601 & 1.75 & & 0.694 & 2.15 & -2989 \\ 
1641+399 & 3C345\tablenotemark{c} & Q & 0.59 & 4220 & \nodata & 62 & & 4920 & \nodata & 167 & -130 & \nodata & 1.47 & & \nodata & 3.39 & -329 \\ 
1741$-$038 & & Q & 1.05 & 4525 & 4890 & 72 & & 4923 & 5212 & 86 & 216 & 0.925 & 1.59 & & 0.945 & 1.75 & 908 \\ 
1749+096 & & B & 0.32 & 2347 & 2437 & 162 & & 2643 & 2702 & 136 & 122 & 0.963 & 6.90 & & 0.978 & 5.15 & 213 \\ 
1803+784 & & B & 0.68 & 1798 & 2430 & 103 & & 1716 & 2179 & 86 & -201 & 0.740 & 5.73 & & 0.788 & 5.01 & -567 \\ 
1823+568 & & B & 0.66 & 599 & 838 & 31 & & 647 & 829 & 45 & -128 & 0.715 & 5.18 & & 0.780 & 6.96 & -353 \\ 
1828+487 & 3C380\tablenotemark{c} & Q & 0.69 & 650 & \nodata & 8 & & 1250 & \nodata & 6 & -2220 & \nodata & 1.23 & & \nodata & 0.48 & -6341 \\ 
1901+319 & 3C\,395\tablenotemark{b} & Q & 0.64 & 834 & \nodata & 13 & & 890 & \nodata & 13 & 300 & \nodata & 1.6 & & \nodata & 1.5 & 807 \\
1928+738\tablenotemark{b} & & Q & 0.30 & 1970 & \nodata & 19 & & 2310 & \nodata & 6 & -1300 & \nodata & 0.96 & & \nodata & 0.26 & -2197 \\ 
2005+403 & & Q & 1.74 & 750 & 2154 & 8 & & 1403 & 2327 & 21 & 654 & 0.348 & 1.07 & & 0.603 & 1.50 & 4911 \\ 
2021+317 & & EF & \nodata & 437 & 721 & 18 & & 476 & 683 & 16 & -31 & 0.606 & 4.12 & & 0.697 & 3.36 & \nodata \\ 
2021+614 & & G & 0.23 & 1764 & 3016 &$<$2.0  & & 1559 & 2174 & $<$1.5 & \nodata& 0.585 & $<$0.11 & & 0.717 & $<$0.1 & \nodata \\ 
2134+004\tablenotemark{b} & & Q & 1.93 & 3240 & \nodata & 140 & & 3170 & \nodata & 183 & 1120 & \nodata & 4.32 & & \nodata & 5.77 & 9615 \\ 
2200+420 & BL Lac & B & 0.07 & 1969 & 3277 & 63 & & 2029 & 2982 & 60 & -376 & 0.601 & 3.20 & & 0.680 & 2.96 & -430 \\ 
2201+315 & & Q & 0.30 & 1111 & 1779 & 8 & & 2262 & 2746 & 11 & \nodata\tablenotemark{d} & 0.625 & 0.72 & & 0.824 & 0.49 & \nodata\tablenotemark{d} \\ 
2223$-$052 & 3C446 & B & 1.40 & 6184 & 7227 & 190 & & 5779 & 6446 & 215 & 383 & 0.856 & 3.07 & & 0.897 & 3.72 & 2206 \\ 
2230+114 & CTA102\tablenotemark{b} & Q & 1.04 & 2740 & \nodata & 13 & & 4490 & \nodata & 48 & -610 & \nodata & 0.47 & & \nodata & 1.07 & -2539 \\ 
2251+158 & 3C454.3 & Q & 0.86 & 5084 & 9382 & 66 & & 5509 & 8442 & 40 & -263 & 0.542 & 1.30 & & 0.653 & 0.73 & -910 \\ 
\enddata
\tablenotetext{a}{\citet{zt01}}
\tablenotetext{b}{\citet{tay00}}
\tablenotetext{c}{\citet{tay98}}
\tablenotetext{d}{Agreement to \l2\ law ruled out based on reduced $\chi^2$.}
\tablecomments{This table is also available in the electronic edition of
the Journal.  The printed edition contains only a sample. Col. (1): B1950 source 
name. Col. (2): Alternate common name. Col. (3): Optical identification from 
the literature (NED) with  Q = quasar, BL = BL Lac object, EF = empty field. 
Col. (5): 8.11 GHz Peak flux density (mJy beam$^{-1}$). Col. (6) 8.11 GHz 
Sum of CLEAN components (mJy). Col. (7) 8.11 GHz polarized flux density 
(mJy beam$^{-1}$) at location of peak.  Col. (8-10): Same as for 5$-$7, for
15.1 GHz. Col. (11): Observed core RM (\radm\ ). Col. (12): 8.11 GHz core 
dominance. Col. (13): 8.11 GHz core fractional polarization (\%). 
Col. (14$-$15): Same as 12$-$13, for 15.1 GHz. Col. (16): Core rest 
frame RM (\radm\ ).   
 }

\end{deluxetable}

\clearpage

\begin{deluxetable}{lcccc}
\tablecaption{\sc Rest Frame Core RM Properties \label{rmcore}}
\tablewidth{0pt}
\tablehead{\colhead{Type} & \colhead{Number} & \colhead{$\mu$} & \colhead{$\sigma_{\mu}$} 
& \colhead{Median} \\
\colhead{} & \colhead{} & \colhead{\radm} & \colhead{\radm} & \colhead{\radm}}
\startdata
Quasars & 26 & 2515 & 106 & 1862 \\
BL Lacs & 6  & 704 & 274 & 441 \\
\enddata
\end{deluxetable}

\clearpage

\begin{deluxetable}{lcccc}
\tablecaption{\sc Rest Frame Jet RM Properties \label{rmjet}}
\tablewidth{0pt}
\tablehead{\colhead{Type} & \colhead{Number} & \colhead{$\mu$} & \colhead{$\sigma_{\mu}$} 
& \colhead{Median} \\
\colhead{} & \colhead{} & \colhead{\radm} & \colhead{\radm} & \colhead{\radm}}
\startdata
Quasars & 12 & 600 & 43 & 458 \\
BL Lacs & 4  & 330 & 20 & 264 \\
\enddata
\end{deluxetable}

\clearpage

\begin{deluxetable}{lccc}
\tablecaption{\sc Properties of \l2\ Law Breakers\label{reject}}
\tablewidth{0pt}
\tablehead{\colhead{Source} & \colhead{Type} & \colhead{15 GHz Lum} & \colhead{Depol} \\
\colhead{} & \colhead{} & \colhead{W hz$^{-1}$} & \colhead{}}
\startdata
0336$-$019 & Q  & 6.3$\times$10$^{26}$ & 1.23 \\
0458$-$020 & Q  & 3.1$\times$10$^{27}$ & 3.11 \\
0736+017   & Q  & 1.1$\times$10$^{25}$ & 1.13 \\
1055+018   & B  & 5.6$\times$10$^{25}$ & 3.82 \\
2201+315   & Q  & 1.3$\times$10$^{27}$ & 0.68 \\
\enddata
\end{deluxetable}

\clearpage

\clearpage

\clearpage

\clearpage
 
\clearpage

\clearpage
 
\clearpage

\clearpage
 
\clearpage

\clearpage
 
\clearpage

\clearpage
 
\clearpage

\clearpage
 
\clearpage

\clearpage
 
\clearpage

\clearpage
 
\clearpage

\clearpage
 
\clearpage

\clearpage
 
\clearpage

\clearpage
 
\clearpage

\clearpage
 
\clearpage

\clearpage
 
\clearpage

\clearpage
 
\clearpage

\clearpage
 
\clearpage

\clearpage
 
\clearpage

\clearpage

\clearpage

\clearpage
 
\clearpage

\clearpage

\clearpage

\end{document}